\documentclass{aa}

\usepackage[]{aalongtable}
\usepackage{scalefnt}
\usepackage{pdflscape}
\usepackage{enumerate}
\usepackage{float}
\usepackage{graphicx}
\usepackage{subfig}
\begin{document}
%\input{psfig}

%			Definitions: new commands
%===============================================================================

%===============================================================================

% Next 5 lines define \simless and \simgreat: "less than or approximately
% equal to" and "greater than or approximately equal to".
\newbox\grsign \setbox\grsign=\hbox{$>$} \newdimen\grdimen \grdimen=\ht\grsign
\newbox\simlessbox \newbox\simgreatbox
\setbox\simgreatbox=\hbox{\raise.5ex\hbox{$>$}\llap
     {\lower.5ex\hbox{$\sim$}}}\ht1=\grdimen\dp1=0pt
\setbox\simlessbox=\hbox{\raise.5ex\hbox{$<$}\llap
     {\lower.5ex\hbox{$\sim$}}}\ht2=\grdimen\dp2=0pt
\def\simgreat{\mathrel{\copy\simgreatbox}}
\def\simless{\mathrel{\copy\simlessbox}}
% Next lines define "approximately proportional to"
\newbox\simppropto
\setbox\simppropto=\hbox{\raise.5ex\hbox{$\sim$}\llap
     {\lower.5ex\hbox{$\propto$}}}\ht2=\grdimen\dp2=0pt
\def\simpropto{\mathrel{\copy\simppropto}}

\title{Iron-peak elements Sc, V, Mn, Cu, and Zn in Galactic
 bulge globular clusters \thanks{Observations 
collected both  at the European  Southern  Observatory,  Paranal  and La 
Silla,  Chile  (ESO programes 65.L-0340 (HP1), 65.L-0371, 67.D-0489 and 69.D-0582,
88.D-0398A (N6522), 93.D-0123A (N6558), 93.D-0124A (HP1))} }

%\subtitle{Analysis of four giants in NGC 6553}
%
\author{
H. Ernandes\inst{1}
\and
B. Barbuy\inst{1}
\and
A. Alves-Brito\inst{2}
\and
A. Fria\c ca\inst {1}
\and
C. Siqueira-Mello\inst{1}
\and
D. M. Allen\inst{1}
}
\offprints{H. Ernandes}
\institute{
Universidade de S\~ao Paulo, IAG, Rua do Mat\~ao 1226,
Cidade Universit\'aria, S\~ao Paulo 05508-900, Brazil
\and
Universidade Federal do Rio Grande do Sul, Departamento de Astronomia,
CP 15051, Porto Alegre 91501-970, Brazil
%\and
%Observat\'orio do Valongo, Universidade Federal do Rio de Janeiro, Rio de
%Janeiro, Brazil \\
%e-mail: dinah@astro.iag.usp.br
}

\date{Received ; accepted }
% \abstract{}{}{}{}{} 
% 5 {} token are mandatory
\abstract{}
{Globular clusters are tracers of the history
of star formation and chemical enrichment in the early Galaxy.
 Their abundance pattern can help understanding their chemical
enrichment processes. In particular,
the iron-peak elements have been relatively little studied so far
in the Galactic bulge.}%  Given that
% the nucleosynthesis of these elements is complex, such observations 
%can give clues on the supernovae that produce them. Chemical tagging based
% on these elements may also be discriminators between different stellar
%populations. }
{ The main aim of this work is to verify the strength of abundances of iron-peak elements for chemical tagging in view of identifying different stellar populations. Besides, the nucleosynthesis processes that build these elements are
complex, therefore observational data can help constraining theoretical
models, as well as give hints on the kinds of supernovae that enriched
the gas before these stars formed.  }
{The abundances of iron-peak elements are derived for the sample
 clusters, and compared with bulge field, and thick disk stars.
We derive abundances of the iron-peak elements Sc, V, Mn, Cu, and Zn 
in individual stars of five bulge globular clusters
(NGC~6528, NGC~6553, NGC~6522, NGC~6558, HP~1), and of the reference thick disk/inner halo cluster 
 47 Tucanae 
 (NGC 104). High resolution spectra were obtained with the UVES spectrograph
 at the Very Large Telescope over the years.}
 %The analysis is carried out using MARCS model atmospheres, and the abundances of the studied elements is obtained by comparing observed and synthetic spectra}
{The sample globular clusters studied span metallicities in the range -1.2$\simless$[Fe/H]$\simless$0.0. 
 V and Sc appear to vary in lockstep with Fe,  indicating that they are produced in the same
supernovae as Fe. % the data might show a decrease for [Fe/H]$\simgreat$-0.3. 
 We find that  Mn  is deficient in metal-poor stars, confirming that it is
underproduced in massive stars; Mn-over-Fe steadily increases at the higher
metallicities due to a metallicity-dependent enrichment by supernovae of type
Ia. Cu behaves as a secondary element, indicating its production in a weak-s
process in massive stars.  Zn has an alpha-like behaviour at low metallicities,
 which can be explained in terms of nucleosynthesis in hypernovae. 
At the metal-rich end, Zn 
decreases with increasing metallicity, similarly to the alpha-elements.}
% There is a trend for the cluster stars to be deficient relative to field bulge stars
%for Sc, V, and Zn, that could be attributed to noise in the spectra, in particular
%at the metal-rich end.  }
% conclusions
%{We conclude that predictions of nucleosynthesis in massive stars
%reproduce rather well the abundances of these elements, and that
%classical chemodynamical evolution models apply suitably to abundances
%of bulge globular clusters.}

\maketitle

\section{Introduction} 
 Chemical tagging is expected to be a key discriminator of
chemical evolution processes and stellar populations (e.g. Freeman
\& Bland-Hawthorn 2002). Among the different groups of chemical
elements, alpha-elements, light odd-Z elements and heavy elements are
more often studied. The iron-peak elements are less extensively
derived, probably because they are formed in more complex processes,
and require atomic data, such as hyperfine structure  for the odd-Z
elements, not always
available. 

The majority of the iron-peak elements show solar abundance ratios in most 
objects in the metallicity range of bulge stars ([Fe/H$\simgreat$ $-$1.5).
  There are exceptions, such as the LMC,
where Ni, Co, Cr  vary in lockstep with Fe, but they
 are deficient relative to Fe [Ni,Co,Cr/Fe]$\sim$-0.15
(Pomp\'eia et al. 2008).
The elements Sc,  Mn, Cu and Zn however show different
trends relative to Fe (e.g. Nissen et al. 2000; Ishigaki et al. 2013). 
 In particular, Zn is found to be enhanced in metal-poor halo
 and thick disk stars
in the Milky Way (e.g. Cayrel et al. 2004, Ishigaki et al. 2013), 
and in  dwarf spheroidal galaxies (Sk\'ulad\'ottir et al. 2017).
Mn is deficient in metal-poor stars, and increases with metallicity for
[Fe/H]$\simgreat$$-$1.0
(e.g. Cayrel et al. 2004; Ishigaki et al. 2013) for halo and
thick disk stars and McWilliam et al. 2003; Sobeck et al. 2006;
Barbuy et al. 2013, Schultheis et al. 2017) for bulge stars. 
The same applies to Cu in field halo, thick disk and bulge stars 
(Ishigaki et al. 2013; Johnson et al. 2014).

The iron-peak elements include elements with atomic numbers in the range
21 $\leq$ Z $\leq$ 32, from scandium to germanium.
%Sc, Ti, V, Cr, Mn, Fe, Co, Ni, Cu, Zn, Ga, Ge.
Sc with Z=21 is a transition element between the so-called alpha-elements and the iron-peak elements. 
 They are produced in complex nucleosynthesis processes, 
such that they can be subdivided in two
groups, the lower iron group: 21 $\leq$ Z $\leq$ 26 %, 45 $\leq$A$\leq$ 56,
 including  Scandium (Sc), Titanium (Ti),
 Vanadium (V), Chromium (Cr), Manganese (Mn), Iron (Fe); 
and the upper iron Group: 27 $\leq$ Z $\leq$
32 %, or 59 $\leq$A$\leq$ 72,
 which includes
 Cobalt (Co), Nickel (Ni), Copper (Cu), Zinc (Zn), Gallium (Ga) 
and Germanium (Ge) 
(Woosley \& Weaver 1995, hereafter WW95, Woosley et al. 2002, 
Woosley, private communication).
The lower iron group elements are produced in explosive oxygen
burning at temperatures 3$\times$10$^9$ $<$T$<$4$\times$10$^9$ K,
explosive Si burning at  4$\times$10$^9$ $<$T$<$5$\times$10$^9$ K,
or nuclear statistical equilibrium for T$>$5$\times$10$^9$
(WW95, Nomoto et al. 2013).  The upper iron group elements 
% 57$\leq$A$\leq$ 66 (up to $^{66}$Zn, 
%but excluding $^{67,68}$Zn, $^{69}$Ga, $^{70,71}$Ge)
 are produced in two processes: neutron capture on iron group
nuclei during helium burning and later burning stages, and the 
alpha-rich freezeout from material heated to $>$5$\times$10$^9$K in the
deepest layers. The amount of each element ejected at the supernova event depends on the amount of mass that
falls back.

There are very few analyses of the odd-Z iron-peak elements
Sc, V, Mn, Co, Cu in bulge stars.
This may be due to the need to consider hyperfine structure for them
 and to having only a few reliable lines. 
McWilliam et al. (2003) derived Mn abundances in 8 bulge field stars.
Barbuy et al. (2013, 2015) have derived abundances of Mn, and Zn,
for 56 bulge field stars, based on FLAMES-UVES spectra from the
Zoccali et al. (2006) sample.
Johnson et al. (2014) have derived abundances
of the iron-peak elements Cr, Co, Ni, Cu in stars located in
 bulge field stars using
FLAMES-GIRAFFE data by Zoccali et al. (2008), comprising
 205 stars in the (+5\fdg25,-3\fdg02) field near the globular cluster 
NGC~6553, and 109 stars in the (0,-12$^{\circ}$) field.
More recently, abundances of iron-peak elements were presented
for bulge metal-poor stars by Howes et al. (2014, 2015, 2016),
Casey \& Schlaufman (2015), and Koch et al. (2016).
The stars from Howes et al. (2015) that had determined
their orbital parameters  
were selected here as bulge members (their Table 6).
As far as we know these data are all that is available presently
as concerns bulge stars.
 Reviews on chemical abundances in the Galactic bulge
can be found in McWilliam (2016), and Barbuy et al. (2018).

In the present work we analyse individual stars in  the
reference globular cluster 
 47 Tucanae ([Fe/H]\footnote{We adopted here the usual
spectroscopic notation that [A/B] = log(N$_{\rm A}$/N$_{\rm B}$)$_{\star}$ $-$
log(N$_{\rm A}$/N$_{\rm B}$)$_{\odot}$ and $\epsilon$(A) = log(N$_{\rm
A}$/N$_{\rm B}$) + 12 for each elements A and B.} = $-$0.67, Alves-Brito et
al. 2005),
and the bulge globular clusters
 NGC 6553 ([Fe/H] = $-$0.20, Alves-Brito et al. 2006),
 NGC 6528 ([Fe/H] = $-$0.11, Zoccali et al. 2004),
 NGC 6522 ([Fe/H] = $-$0.95, Barbuy et al. 2014),
 HP~1 ([Fe/H] = $-$1.00, Barbuy et al. 2006, 2016),
and NGC 6558  ([Fe/H] = $-$1.00, Barbuy et al. 2017).
In previous work we have derived the abundances of
the $\alpha$-elements (O, Mg, Ca, Si, Ti), odd-Z (Na, Al), 
s-process (Ba, La, Zr), and
r-process (Eu) elements in these clusters.
 In the present work we derive the abundances of the
 iron-peak elements Sc, V, Mn, Cu, Zn.

Star clusters are tracers of the formation history of different
components of galaxies. Globular clusters are probably the earliest 
objects to have formed, and they trace the formation of
the halo and bulge of our Galaxy (Hansen et al. 2013;
Kruijssen 2015; Renzini 2017).
As to whether the field stars and globular clusters can be 
identified as having a same origin, has been a matter of debate in
the literature. The detection of abundance anomalies in field stars
similar to the anomalies found in globular cluster stars
(Gratton et al. 2012) has been used to conclude that at least
a fraction of the field stars have their origin in the clusters
(Kraft 1983; Martell et al. 2011; Schiavon et al. 2017).

The observations are briefly described in Sect. 2.  Line parameters are reported
in Sect. 3. Abundance analysis is described in Sect. 4.
Results and discussions are presented in Sect. 5.
Conclusions are drawn in Sect. 6.

%===============================================================================
%				Observations
%===============================================================================

\section{Observations}

The sample consists of 28 red giant stars, including five in 47 Tucanae
(Alves-Brito et al. 2005), four in NGC
6553 (Alves-Brito et al. 2006), three in NGC 6528 (Zoccali et al. 2004), 
eight  in HP~1 (Barbuy et al. 2006, 2016), 
 four  in NGC 6522 (Barbuy et al. 2009,
2014), and four in NGC 6558 (Barbuy et al. 2017), all observed with UVES at the
8.2 m Kueyen ESO telescope.
The wavelength coverage is 4800-6800 {\rm \AA}.
The red portion of the spectrum (5800-6800 {\rm \AA}) was obtained with the
ESO CCD \# 20, an MIT backside illuminated, of 4096x2048 pixels, and pixel
size  15x15$\mu$m.
The blue portion of the spectrum (4800-5800 {\rm \AA}) uses ESO Marlene EEV
CCD-44, backside illuminated, 4102x2048 pixels, and pixel size  15x15$\mu$m. 
With the UVES standard setup 580, the  UVES resolution is
  R $\sim$ 45 000 for a 1 arcsec slit width, while R $\sim$ 55 000 
for a slit of 0.8 arcsec. The pixel scale is 0.0147 {\rm \AA}/pix.
  The log of  observations is given in Table \ref{seeing}.
 
%Typical signal-to-noise ratio obtained by considering average values at
%different wavelengths varies from {\bf 30 $\leq$ S/N $\leq$ 280} 
%per pixel in the program stars. 

\begin{figure}[h]
\includegraphics[scale=0.45]{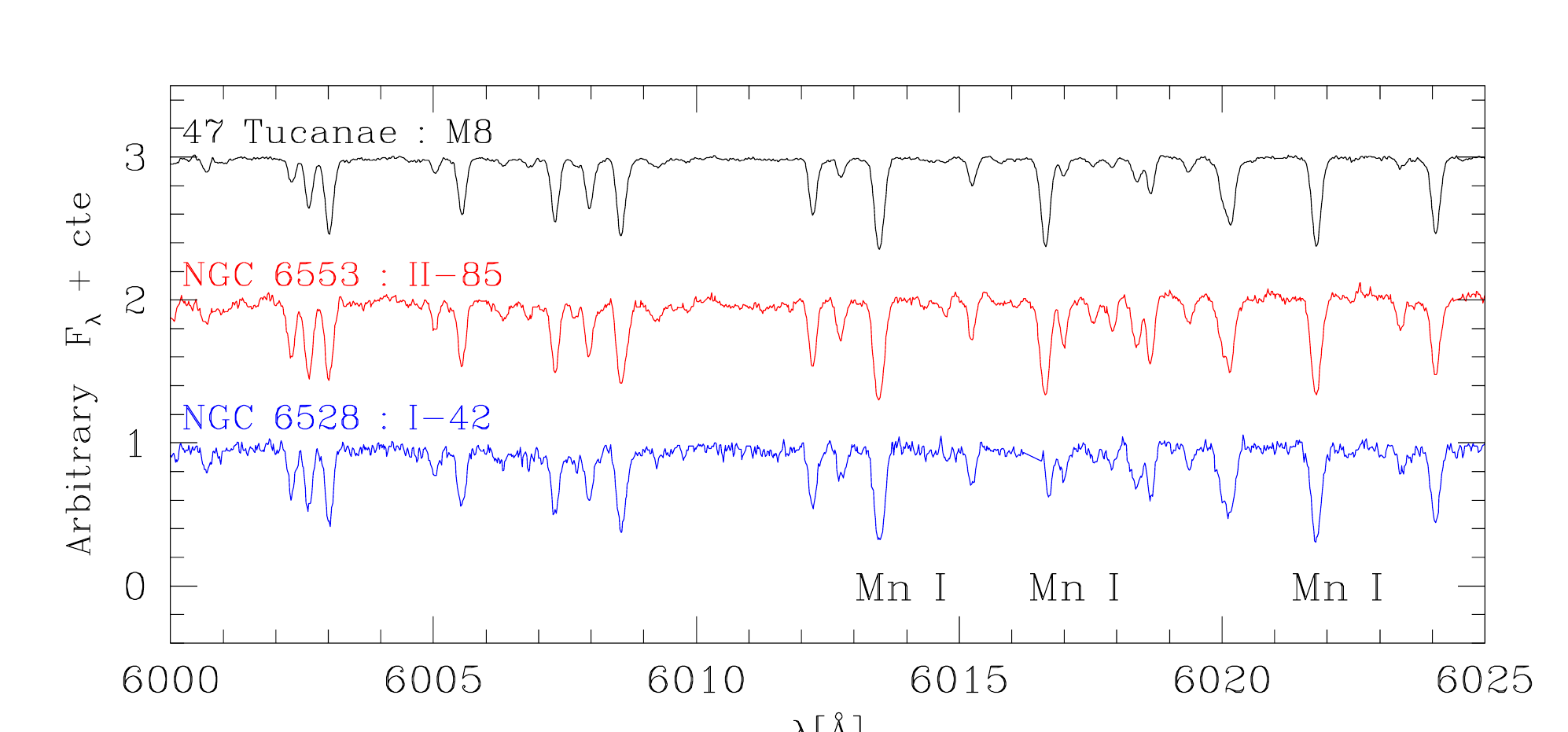}
\includegraphics[scale=0.45]{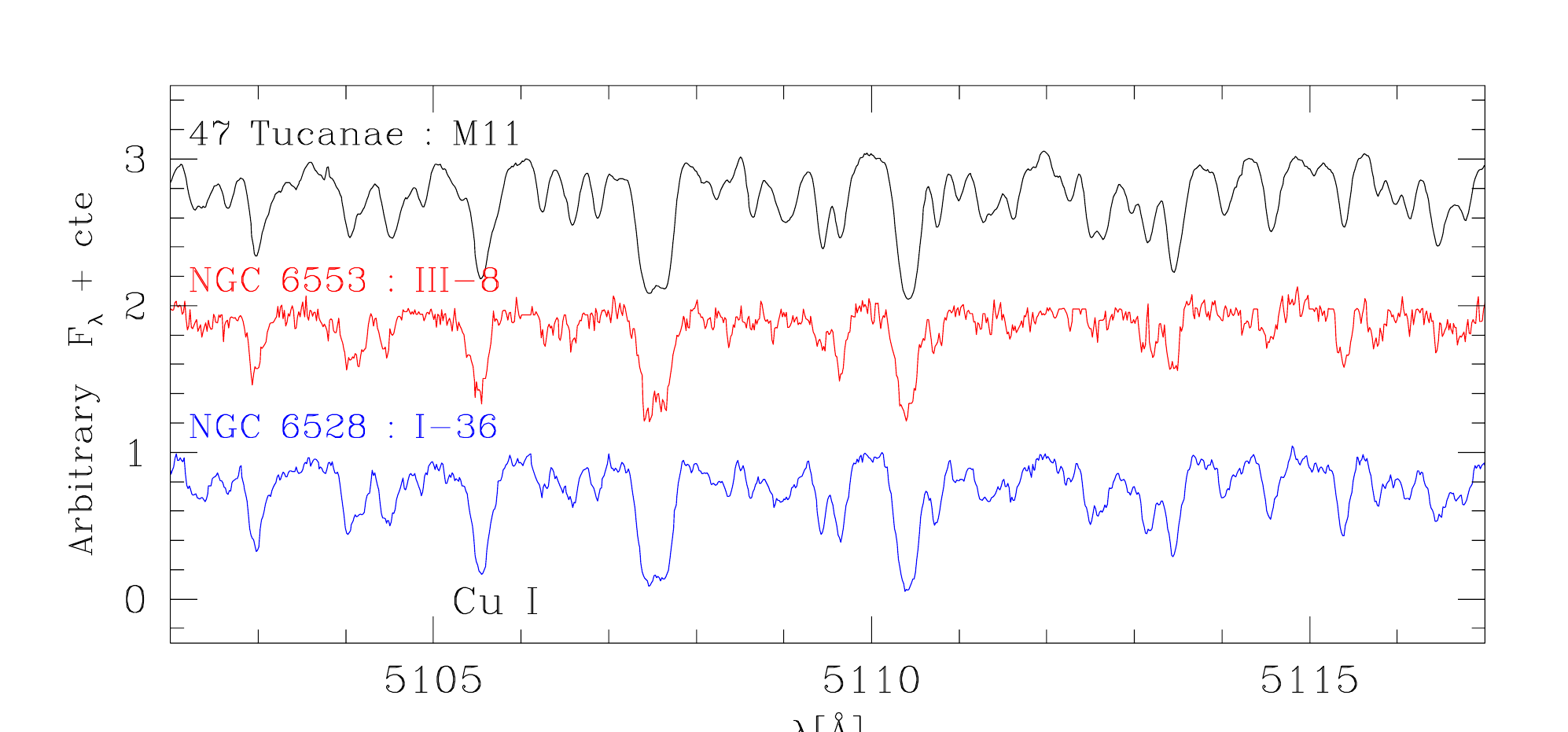}
\includegraphics[scale=0.45]{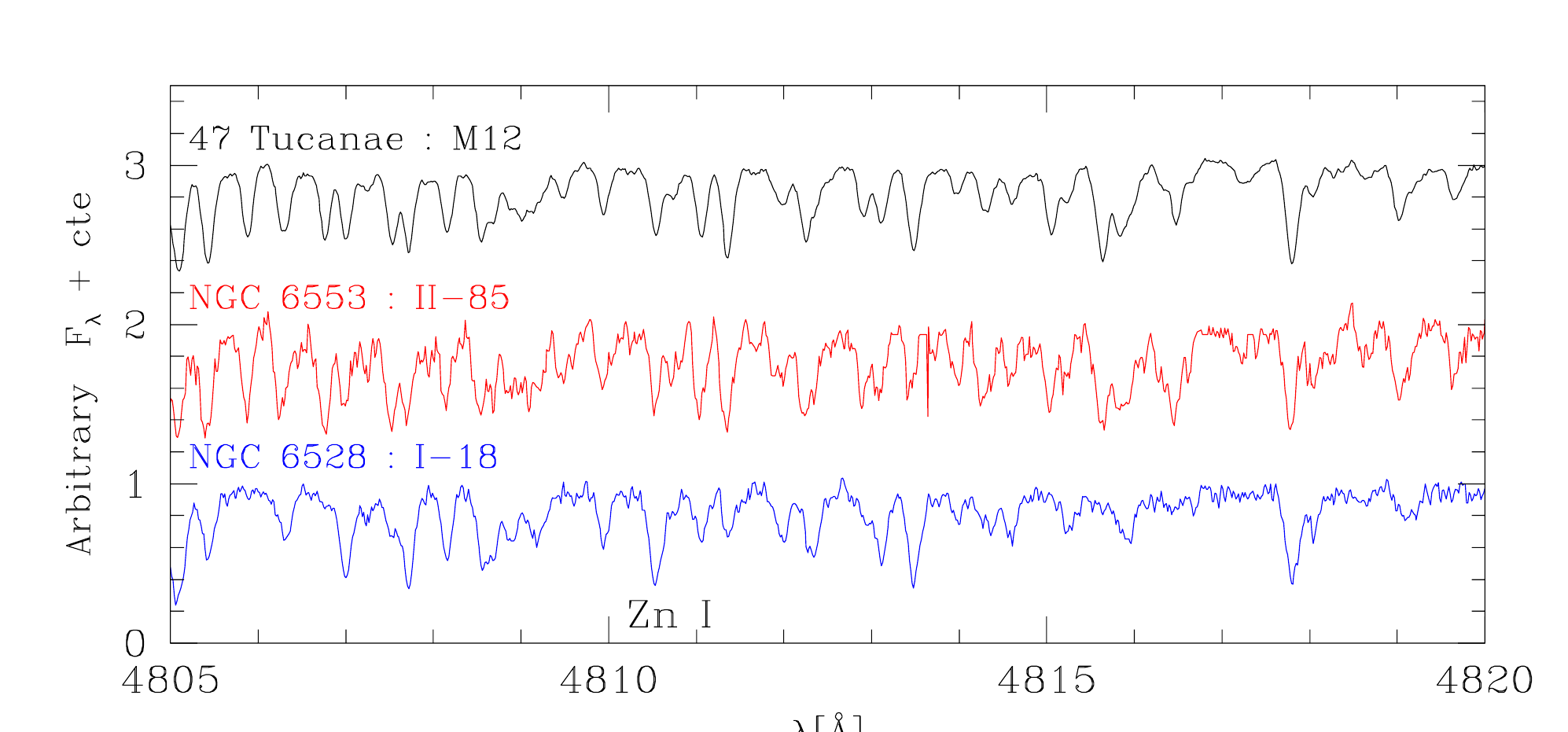}
\caption{Comparison of the spectra for each globular cluster in the sample. The
features are marked for \ion{Mn}{I} ({\it top}), \ion{Cu}{I} ({\it middle}), and \ion{Zn}{I} lines
({\it bottom}).}
\label{espectros} 
\end{figure}

Reductions are described in the references given above, in all cases
including bias and inter-order background subtraction, flatfield correction,
extraction and wavelength calibration (Ballester et al. 2000).
Fig. \ref{espectros} shows the spectra for some of the program stars around
the features studied.

\begin{table}[h]
\caption{Log of spectroscopic observations of globular clusters, 47 Tucanae, NGC 6553, NGC 6528, HP 1, NGC 6558.} 
\label{seeing} 
\centering                  
\begin{tabular}{c c c } 
\hline\hline             
Star  &  Slit width & (S/N)/px  \\ 
\hline  
\multicolumn{3}{c}{\hbox{ 47 Tucanae}}  \\
\hline  
M8 & 0.8$\arcsec$ & 280  \\    
M11 & 0.8$\arcsec$ & 241 \\ 
M12 & 0.8$\arcsec$ & 247  \\   
M21 & 0.8$\arcsec$ & 213  \\          
M25 & 0.8$\arcsec$ & 258 \\  
\hline  
\multicolumn{3}{c}{\hbox{NGC 6553}} \\
\hline  
II-64 & 0.8$\arcsec$ & 110  \\ 
II-85 & 0.8$\arcsec$ & 200  \\ 
III-8 & 0.8$\arcsec$ & 170 \\ 
267092 & 0.8$\arcsec$ & 110 \\
\hline  
\multicolumn{3}{c}{\hbox{NGC 6528}} \\
\hline  
I-18 & 1.0$\arcsec$ & 40  \\ 
I-36 & 1.0$\arcsec$ & 40  \\ 
II-42 & 0.8$\arcsec$ & 30  \\  
\hline  
\multicolumn{3}{c}{\hbox{ HP 1 }}\\
\hline  
2 & 0.8$\arcsec$ & 70 \\    
3 & 0.8$\arcsec$ & 45 \\ 
2115 & 0.8$\arcsec$ & $>$ 200 \\   
2461 & 0.8$\arcsec$ & $>$ 200 \\          
2939 & 0.8$\arcsec$ & $>$ 200 \\ 
3514 & 0.8$\arcsec$  & $>$ 200 \\
5037 & 0.8$\arcsec$ & $>$ 200 \\ 
5485 & 0.8$\arcsec$ & $>$ 200 \\ 
\hline  
\multicolumn{3}{c}{\hbox{NGC 6522}} \\
\hline  
B-107 & 0.9$\arcsec$  & 180 \\
B-122 & 0.9$\arcsec$  & 170 \\
B-128 & 0.9$\arcsec$  & 180 \\
B-130 & 0.9$\arcsec$  & 210 \\
\hline  
\multicolumn{3}{c}{\hbox{NGC 6558}} \\
\hline  
283  & 1.0$\arcsec$ & 130 \\
364  & 1.0$\arcsec$ & 150 \\
1072 & 1.0$\arcsec$ & 190 \\
1160 & 1.0$\arcsec$ & 170 \\
\hline   
\hline                          
\end{tabular}
%\tablebib{
\end{table}

%===============================================================================
%				Atmospheric Parameters
%===============================================================================

\section{Line parameters: hyperfine structure, oscillator strengths, and solar
abundances.}
To settle suitable values of oscillator strengths and central wavelengths, the studied lines were checked by using high-resolution spectra of the Sun  (using the same instrument settings as the present sample of spectra\footnote{http://www.eso.org/\-observing/\-dfo/\-quality/\-UVES/\-pipeline/solar{$_{-}$}spectrum.html} ), Arcturus (Hinkle et al. 2000) and the metal-rich giant star 
$\mu$ Leo (Lecureur et al. 2007).
%and abundances of the iron-peak elements in Arcturus and $\mu$ Leo.
We adopted the stellar parameters
 effective temperature (T$_{\rm eff}$), surface gravity (log~g), 
metallicity ([Fe/H]) and microturbulent velocity (v$_{\rm t}$)  of
 (4275 K, 1.55, -0.54, 1.65 km.s$^{-1}$) for Arcturus from 
Mel\'endez et al. (2003), and (4540 K, 2.3, +0.30, 1.3 km.s$^{-1}$) for $\mu$ Leo from Lecureur et al. (2007).
In Table \ref{adopted_ab}, we present the adopted abundances for the Sun, 
Arcturus and $\mu$ Leo. 

Oscillator strengths for \ion{Sc}{I}, \ion{Sc}{II}, \ion{V}{I}, and \ion{Cu}{I} reported in  
Table \ref{sclines2}, are
 from Kurucz (1993)\footnote{http://kurucz.harvard.edu/atoms.html}, 
NIST (Martin et al. 2002)\footnote{http://physics.nist.gov/PhysRefData/ASD/lines$_-$form.html},
VALD3 (Piskunov et al. 1995)\footnote{http://vald.astro.univie.ac.at},
 literature values, and adopted final values.

\begin{table}[h]
\caption{Adopted abundances for the Sun, Arcturus, and $\mu$Leo.
References:
[1]: Grevesse et al. (1996);% Grevesse \& Sauval (1998); 
[2]: Allende Prieto et al. (2001); 
[3]: Ram\'irez \& Allende Prieto (2011); 
[4]: Mel\'endez et al. (2003)
[5]: this work; 
[6]: Barbuy et al. (2015); 
[7]: Fulbright et al. (2007); 
[8]: Smith et al. (2013); 
[9]: Gratton \& Sneden (1990);
[10]: Steffen et al. (2015).}   
\label{adopted_ab} 
\centering                  
\begin{tabular}{c c c c} 
\hline\hline             
El. & A(X)$_{\odot}$ & A(X)$_{Arcturus}$ & A(X)$_{\mu Leo}$ \\ 
\hline  
Fe & 7.50 [1] & 6.96 [4]  & 7.80 [6] \\                
C  & 8.55 [1] & 8.32 [3] & 8.55 [6] \\   
N  & 7.97 [1] & 7.68 [4] & 8.83 [6]\\   
O  & 8.76 [10] & 8.66 [4] & 8.97 [6]\\   
Na & 6.33 [1] & 5.82 [3] & 7.06 [7]\\   
Mg & 7.58 [1] & 7.47 [3] & 7.85 [8]\\   
Al & 6.47 [1] & 6.26 [3] & 6.90 [8]\\   
Si & 7.55 [1] & 7.30 [3] & 7.76 [8]\\   
K  & 5.12 [1] & 4.99 [3] & 5.63 [8]\\   
Ca & 6.36 [1] & 5.94 [3] & 6.62 [8]\\   
Sc & 3.17 [1] & 2.81 [3] & 3.34 [9]\\   
Ti & 5.02 [1] & 4.66 [3] & 5.40 [8]\\   
V  & 4.00 [1] & 3.58 [3] & 4.18 [8]\\   
Cr & 5.67 [1] & 4.99 [3] & 6.14 [8]\\   
Mn & 5.39 [1] & 4.74 [3] & 5.79 [8]\\   
Co & 4.92 [1] & 4.71 [3] & 5.23 [8]\\   
Ni & 6.25 [1] & 5.73 [4] & 6.60 [8]\\   
Cu & 4.21 [1] & 3.67 [5] & 4.46 [5]\\   
Zn & 4.60 [1] & 4.06 [6] & 4.80 [6]\\   
\hline                          
\end{tabular}
%\tablebib{
\end{table}

\subsection{Scandium and Vanadium}

$^{45}$Sc is the unique species of Sc, and the
 V abundance corresponds to  99.75\% of $^{51}$V
and 0.25\% of $^{50}$V (Asplund et al. 2009), 
therefore we adopted $^{51}$V as unique isotope.
We selected \ion{Sc}{I}, \ion{Sc}{II}, and
\ion{V}{I} lines that showed to be strong enough to
be detected in red giants. Hyperfine structure (HFS) was taken into account,
by applying the code made available by McWilliam et al. (2013), 
together with the A, and B constants
reported in Table \ref{sclines1}. 
V and Sc have a nuclear spin I = 7/2. 
 Some lines that were blended in the sample stars,
or affected by telluric lines, were discarded.
 This applies to the lines \ion{V}{I}
 4831.640, 4851.480, 4875.480, 4932.030, 5627.640, 5670.850, 6216.370,
and 6285.160 {\rm \AA}. \\ \\ \\

\subsection{Copper}

\begin{figure}[h]
\includegraphics[scale=0.45]{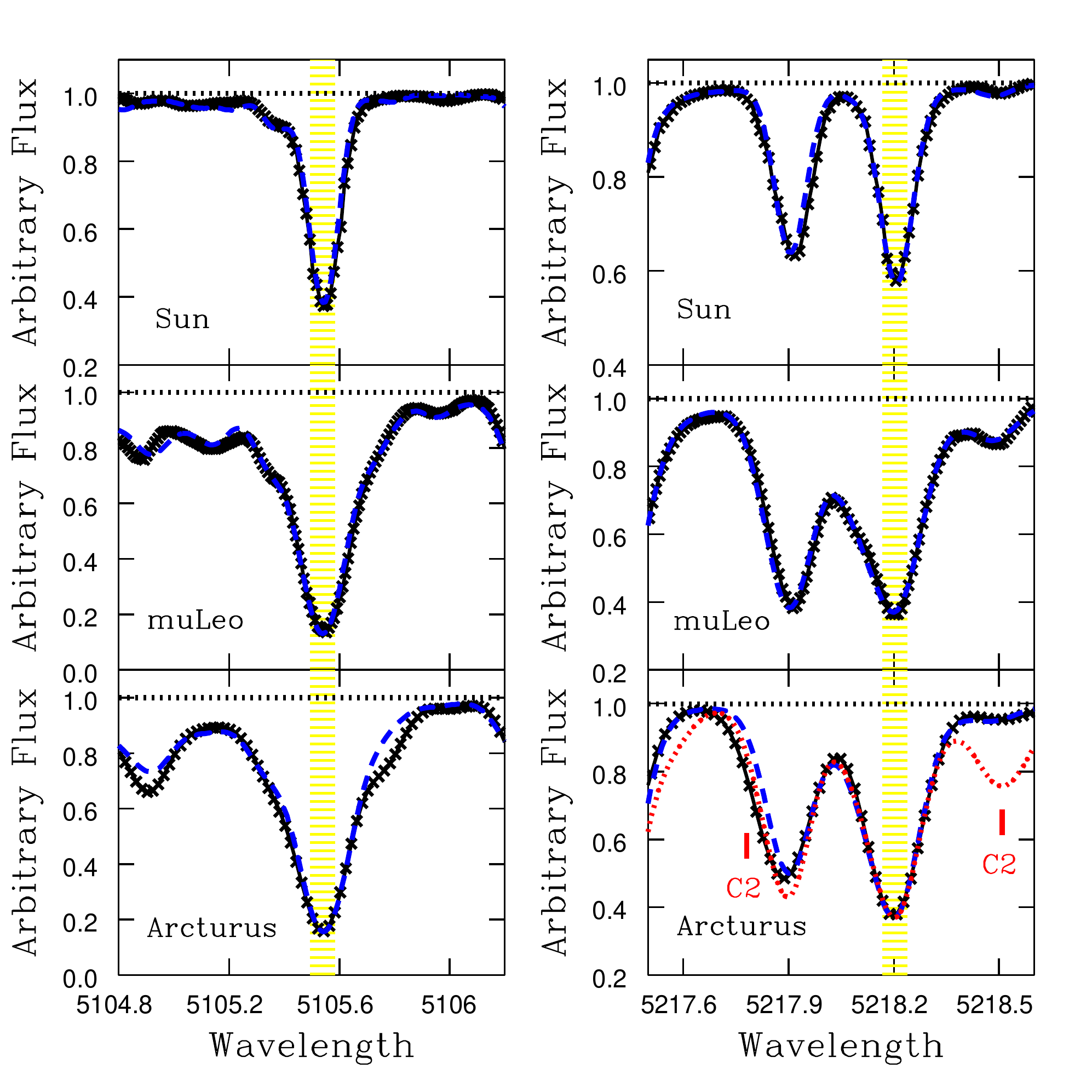}
\caption{Fittings on solar, Arcturus, and $\mu$Leo spectra 
for the \ion{Cu}{I} lines at 5105 {\rm \AA} and 
5218 {\rm \AA} (yellow lines). Observations (black crosses) 
are compared with synthetic spectra computed using 
the adopted abundaces (dashed blue lines). 
%The Arcturus synthetic spectrum with A(C)~$=8.80$ is also shown (dotted red line).
}
\label{figura_Cu} 
\end{figure}

The isotopic fractions of 0.6894 for $^{63}$Cu and
 0.3106 for $^{65}$Cu (Asplund et al. 2009) are considered. 
Copper abundances were derived from the  \ion{Cu}{I} lines at 
5105.50 {\rm \AA} and 5218.20 {\rm \AA}. The 5782 {\rm \AA} line is
 not available in the UVES spectra analysed, which
 cover the wavelengths 4780-5775 {\rm \AA} and 5817-6821 {\rm \AA},
 therefore with a gap of about 40 {\rm \AA} in the
 range 5775-5817 {\rm \AA}.
Oscillator strengths of the \ion{Cu}{I} lines were selected in 
the literature from Kurucz (1993), Bielski (1975), 
NIST or VALD, and the final  adopted values are reported
 in Table \ref{sclines2}. 

In Table \ref{sclines1} the magnetic dipole A-factor and
the electric quadrupole B-factor constants were adopted
from Kurucz (1993), 
and Biehl (1976), in order to compute  HFS. For the 5218 {\rm \AA} line
the constants for the 4d 2D level are not available.
According to R. Kur\'ucz (private communication), the upper level should
have much smaller HFS than the lower, because its wavefunction is further
away from the nucleus, and setting its splitting to 0.0 is acceptable.
The HFS components for the \ion{Cu}{I} lines and
corresponding oscillator strengths are reported in Table
 \ref{hfsCu}.

%In any case, for the   \ion{Cu}{I} 5218 {\rm \AA} line, the
%hyperfine structure line component patterns were 
%adopted from Allen \& Porto de Mello (2011), 
%which in turn were adopted from Steffen (1985). 

The \ion{Cu}{I}~5105 {\rm \AA} and \ion{Cu}{I}~5218 {\rm \AA} lines in the 
solar spectrum were fitted adopting A(Cu)$_{\odot}=4.21$
 (Grevesse et al. 1996) cf. Table \ref{adopted_ab}.
%and the oscillator strength log gf~$=-1.52$.
The adopted or derived
abundances for each of the reference stars are also presented in Table 
\ref{adopted_ab}, corresponding to 
 [Cu/Fe]=0.0 and ~+0.05 in Arcturus and $\mu$Leo, respectively.

%XXXXXX The blend of lines located at  around 5217.4, on the
% left to the \ion{Cu}{I}~5218  {\rm \AA} is well fitted assuming
%[Yb/Fe]=-0.2 and +0.2 in Arcturus and  $\mu$Leo, respectively.
 
Figure \ref{figura_Cu} shows the fits to the solar, Arcturus, 
and $\mu$Leo spectra for the Cu lines.
For the \ion{Cu}{I}~5218 {\rm \AA} in Arcturus,
 an asymmetry remained in the blue wing of the 
Fe profile close to the \ion{Cu}{I} line.
% which is 
%caused by C$_{2}$ lines. Fig. \ref{figura_Cu} shows a synthetic spectrum 
%computed with A(C)~$=8.80$ (dotted red line), increased to better 
%reproduce the profile. {\bf check CCC!! However, the inspection of other spectral 
%regions indicates A(C)~$=8.32$} (cf. Table \ref{adopted_ab}),  as a better
% C abundance. 
Consequently, the \ion{Cu}{I}~5218 {\rm \AA} line was used in the 
sample stars with caution.\\

%Central wavelengths
%and oscillator strengths are given in Table \ref{lines2}.

\subsection{Manganese and Zinc}

Manganese has one isotope $^{55}$Mn and, for zinc,
$^{64,66,68}$Zn are the dominant species
 with 48.63/27.90/18.75\% fractions (Asplund et al. 2009).
For these elements a splitting in isotopes was not considered.
Manganese abundances were derived from the \ion{Mn}{I} triplet lines at
 6013.513,  6016.640, 6021.800 {\rm \AA}.
% and the MnI 5394.67  $\rm \AA$ line.
 The line list of HFS components
are given in Barbuy et al. (2013). For zinc we used the
\ion{Zn}{I} 4810.529 and 6362.339 {\rm \AA} lines 
as detailed in Barbuy et al. (2015).

\section{Abundance Analysis}

\subsection{Atmospheric parameters and abundance derivation}

The adopted effective stellar atmospheric parameters for all program stars were derived in previous
work
 (Zoccali et al. 2004; Alves-Brito et al. 2005; Alves-Brito et al. 2006;
Barbuy et al. 2006, 2014, 2016, 2017 (in prep.)).

The method described in the original papers follow standard procedures:
\begin{enumerate}[i]

\item  Colours V-I, V-K, and J-K, corrected by the reddening values
reported in Table \ref{distance}, were used together with
colour-temperature calibrations by Alonso et al. (1999, 2001),
and/or Houdashelt et al. (2000). 

\item Gravities of the sample stars were obtained adopting the classical
relation below, and final log g values were obtained from
ionization equilibrium of Fe I and Fe II lines.

\begin{equation}
\log g_*=4.44+4\log \frac{T_*}{T_{\odot}}+0.4(M_{\rm bol*}-M_{\rm bol{\circ}})+\log \frac{M_*}{M_{\odot}} 
\end{equation}

We adopted T$_{\odot}=5770$~K and M$_{\rm bol \odot}=4.75$ for the Sun and 
M$_{*}$=0.80 to 0.88~M$_{\odot}$ for the red giant branch (RGB) stars. 

 In Table \ref{distance} are reported the distance moduli
assumed for each sample cluster and corresponding references.

\begin{table}%[h]
\caption{Reddening and distance moduli adopted.
References: 1 Harris (1996); 2 Zoccali et al. (2001a);
3 Zoccali et al. (2004); 4 Barbuy et al. (1998); 
5 Guarnieri et al. (1998); 6 Barbuy et al. (2006);
7 Ortolani et al. (2007, 2011); 8 Barbuy et al. (2009);
9 Terndrup (1988); 10 Rossi et al. (2015). } 
\label{distance} 
\centering                  
\begin{tabular}{c c c c c } 
\hline\hline             
Cluster  &  \hbox{E(B-V)} & Ref. \hbox{$(m-M)_{0}$} & Ref. \\ 
\hline  
\hbox{47 Tucanae} & 0.04 & 1 & 13.09  & 2   \\ 
\hbox{NGC~6528} & 0.46 & 3 & 14.45 & 4   \\ 
\hbox{NGC~6553} & 0.70 & 5 & 13.54  & 4  \\ 
\hbox{HP~1} & 1.12 & 6 & 14.15 & 7  \\ 
\hbox{NGC~6522}  & 0.45 & 8 & 13.91  & 4  \\ 
\hbox{NGC~6558} & 0.38 & 9 & 14.43  & 10   \\ 
\hline   
\hline                          
\end{tabular}
%\tablebib{
\end{table}

\item The initial photometric temperatures and gravites were 
used to compute the excitation and ionization equilibrium.
 Effective temperatures were then checked by imposing excitation
equilibrium for FeI and FeII lines of different excitation potential,
and gravities were checked against ionization equilibrium.

\item Microturbulent velocity v$_{\rm t}$ was determined by canceling any
trend in a FeI abundance versus equivalent width diagram.

\item Finally, the metallicities for the sample were derived using a set of
equivalent widths of \ion{Fe}{I} and \ion{Fe}{II} lines.

\end{enumerate}

Table \ref{atmos} summarizes the final atmospheric parameters obtained for the
program stars. In this Table we also present carbon, nitrogen and oxygen abundance ratios, derived from the C$_{2}$(0,1)
 A$^3$$\Pi$-X$^3$$\Pi$ bandhead   % d$^{3}\Pi_g$-a$^{3}\Pi_u$  bandhead
at 5635.3 {\rm \AA}, CN(5,1)  A$^{2}\Pi$-X$^{2}\Sigma$  6332.18  and the forbidden [OI] 6300.311 {\rm \AA} lines.

%{\bf FAZER} A fit to these lines in 47Tuc:m11 is shown in Fig. 
%\ref{cno}.

Elemental abundances were obtained through line-by-line spectrum synthesis
calculations. The calculations of synthetic spectra were carried out using 
the code described in Barbuy et al. (2003),
and Coelho et al. (2005). Atomic lines are as described in Sect. 3.2, and
molecular lines of  CN  A$^2$$\Pi$-X$^2$$\Sigma$, C$_2$  Swan 
A$^3$$\Pi$-X$^3$$\Pi$ and TiO A$^3$$\Phi$-X$^3$$\Delta$ $\gamma$ and
B$^3$$\Pi$-X$^3$$\Delta$ $\gamma$' systems are taken into account.
The atmospheric models were obtained by interpolation in the grid
of MARCS LTE models (Gustafsson et al. 2008), adopting
 their spherical and mildly CN-cycled 
([C/Fe]$=-0.13$, [N/Fe]$=+0.31$) subgrid. These models
consider [$\alpha$/Fe]=+0.20 for [Fe/H]=-0.50 and 
[$\alpha$/Fe]=+0.40 for [Fe/H]$\leq-$1.00.

\begin{figure*}[!h]
\centering
\subfloat[\ion{Sc}{I} 6604.601 {\rm \AA} line.]{\includegraphics[width = 2.55in]{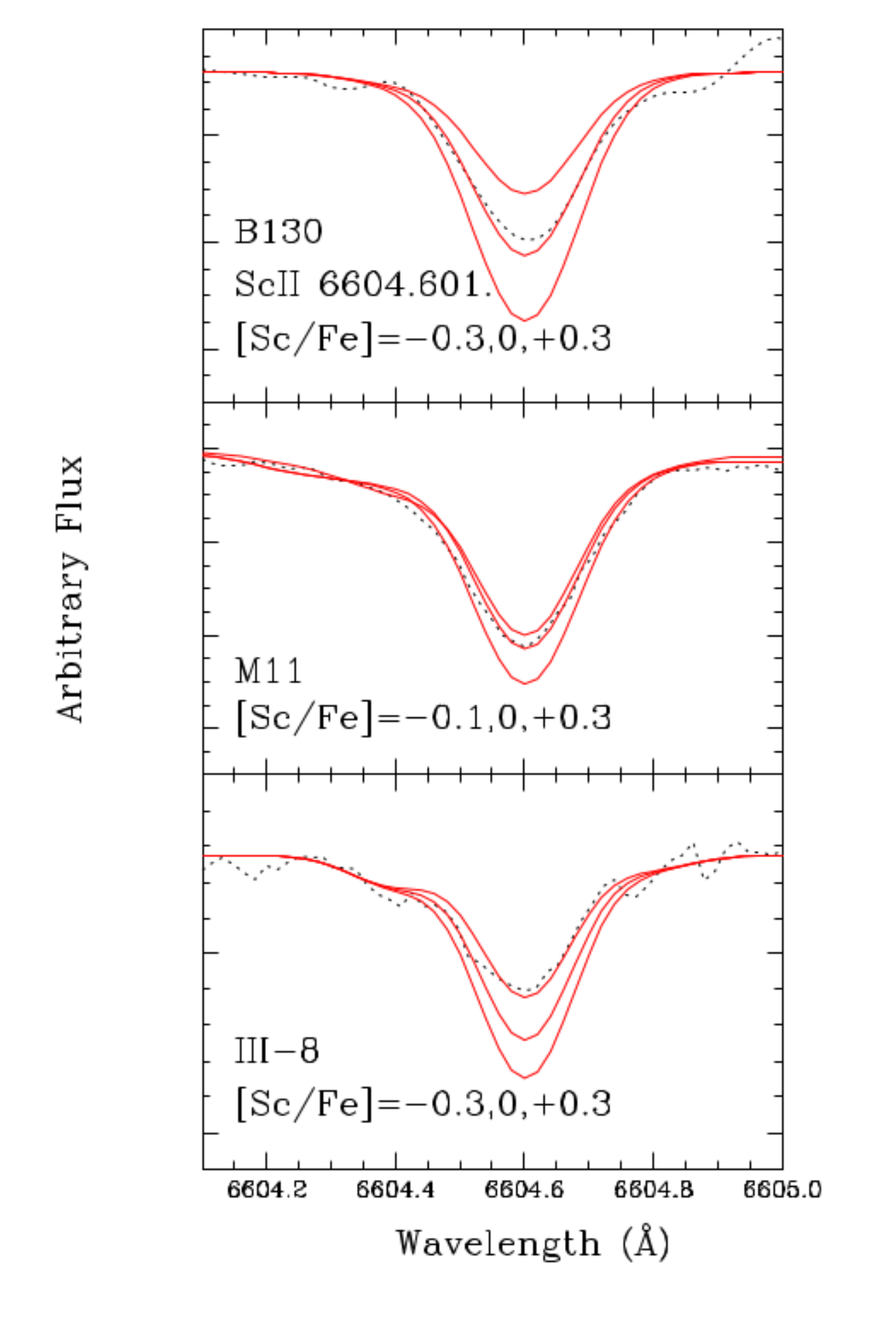}}
\subfloat[\ion{V}{I} 5703.560 {\rm \AA} line.]{\includegraphics[width = 2.38in]{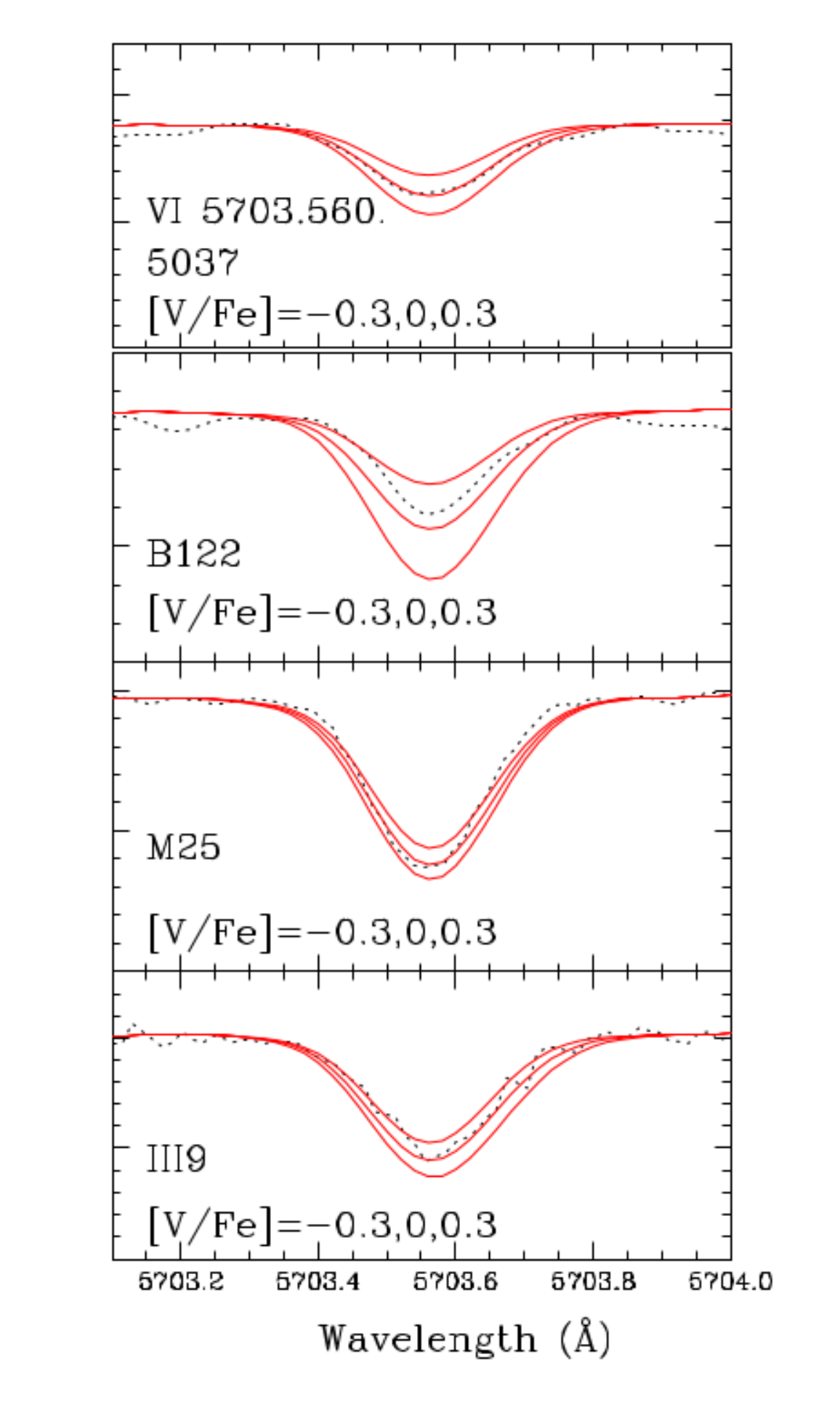}}
\subfloat[\ion{Mn}{I} 6013.513 {\rm \AA} line.]{\includegraphics[width = 2.38in]{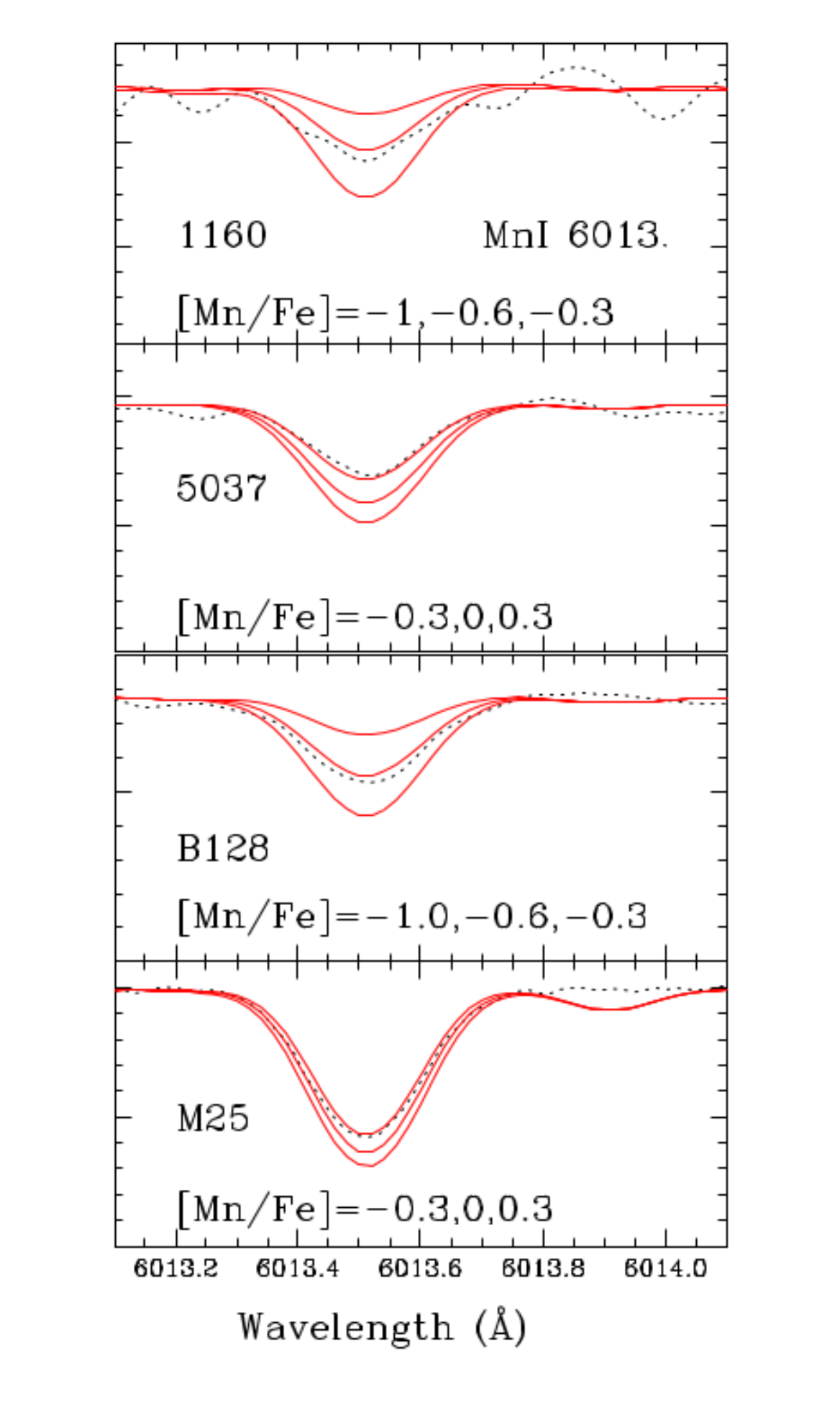}}
%\subfloat[\ion{Cu}{I} 5105.537 {\rm \AA} line.]{\includegraphics[width = 1.38in]{Culinescut.pdf}}
%\subfloat[\ion{Zn}{I} 4810.529 {\rm \AA} line.]{\includegraphics[width = 1.38in]{Znlinescut.pdf}}
\caption{Fits of best lines of \ion{Sc}{I}, \ion{V}{I}, and \ion{Mn}{I}
% \ion{Cu}{I} and \ion{Zn}{I}
 for some sample stars.}
\label{spectrasc1}
\end{figure*}

\begin{figure*}[!h]
\centering
%\subfloat[\ion{Sc}{I} 6604.601 {\rm \AA} line.]{\includegraphics[width = 1.55in]{Sclinescut.pdf}}
%\subfloat[\ion{V}{I} 5703.560 {\rm \AA} line.]{\includegraphics[width = 1.38in]{Vlinescut.pdf}}
%\subfloat[\ion{Mn}{I} 6013.513 {\rm \AA} line.]{\includegraphics[width = 1.38in]{Mnlinescut.pdf}}
\subfloat[\ion{Cu}{I} 5105.537 {\rm \AA} line.]{\includegraphics[width = 2.38in]{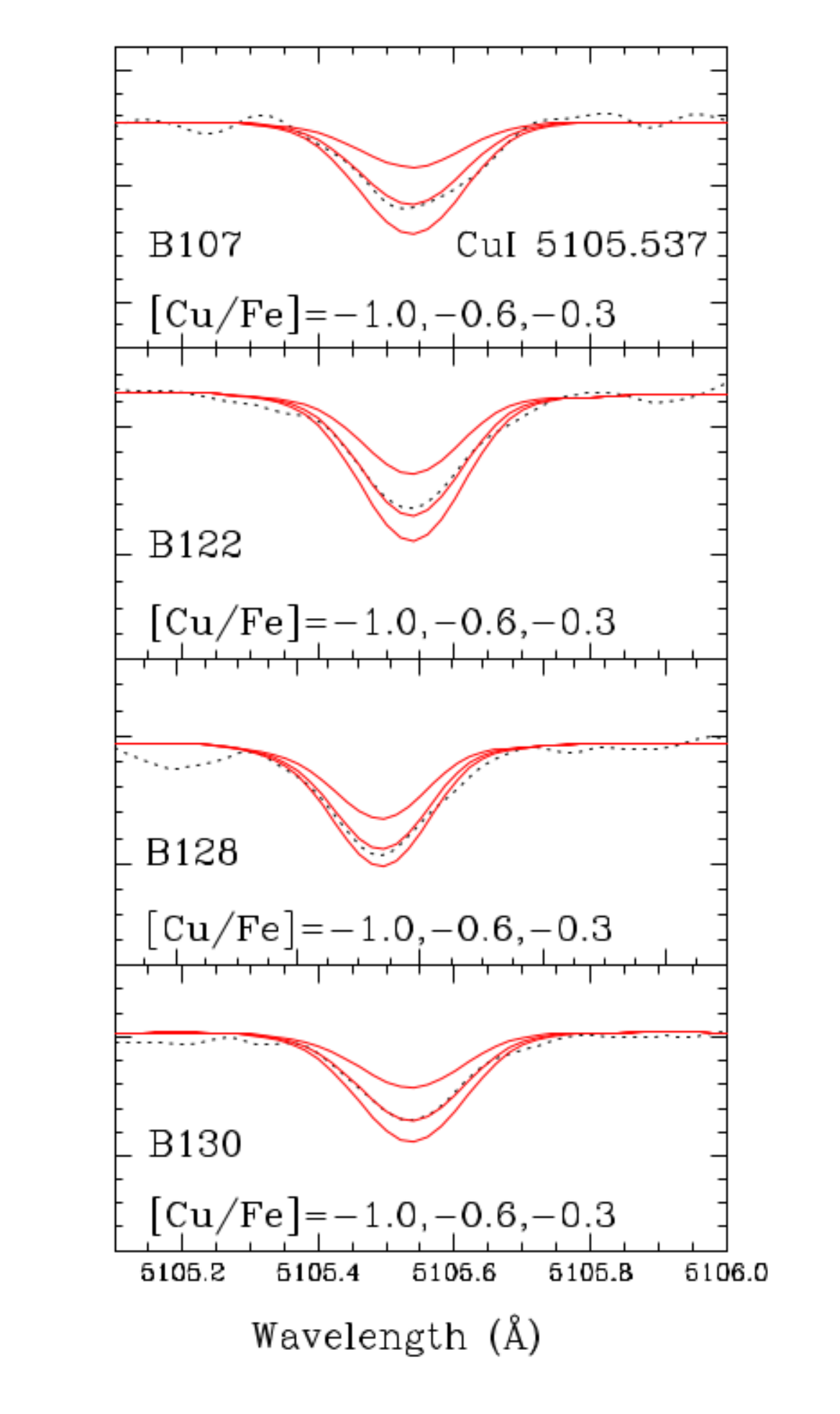}}
\subfloat[\ion{Zn}{I} 4810.529 {\rm \AA} line.]{\includegraphics[width = 2.38in]{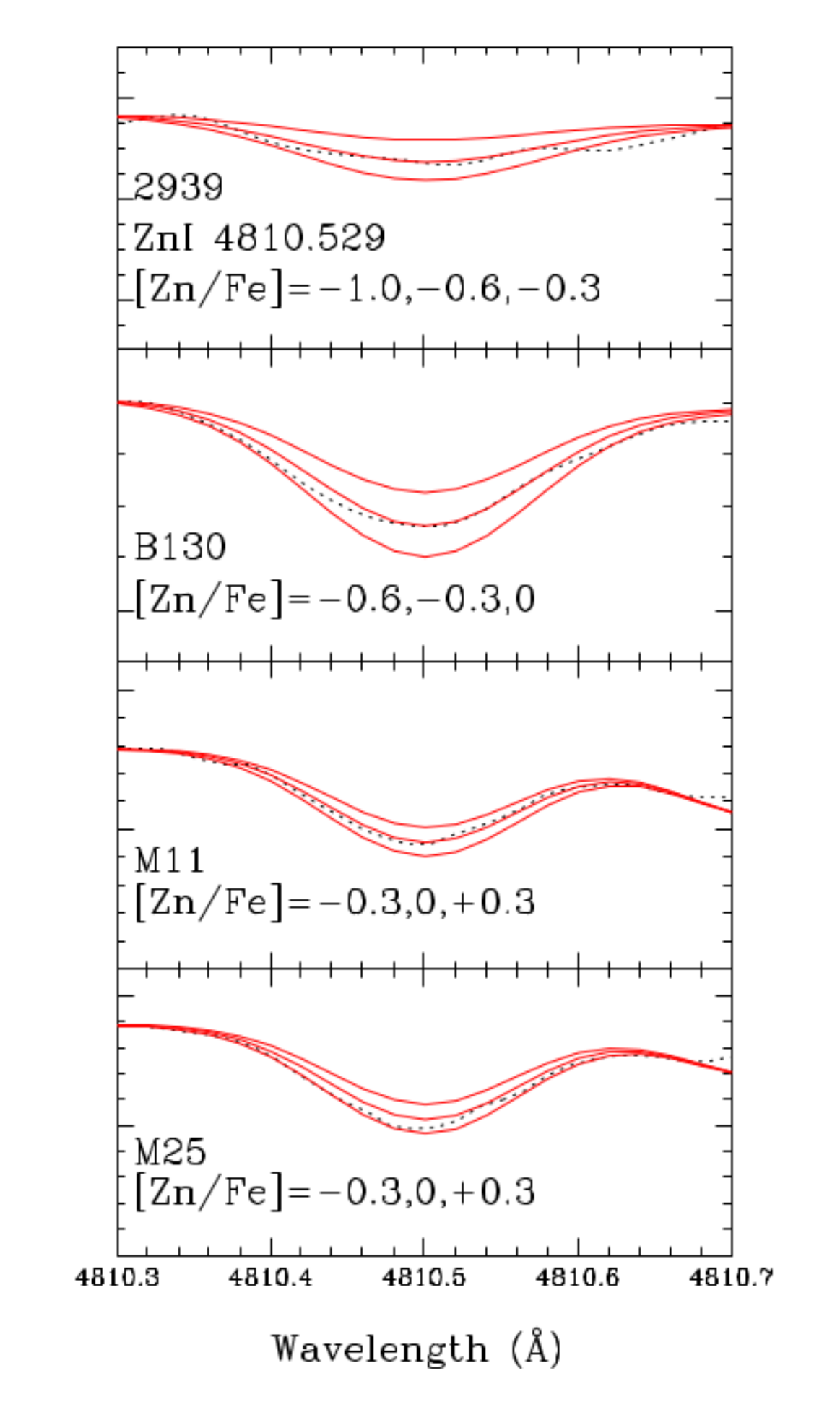}}
\caption{Fits of best lines of \ion{Cu}{I} and \ion{Zn}{I} for some sample stars.}
\label{spectrasc2}
\end{figure*}

In Figures \ref{spectrasc1}a,b,c and \ref{spectrasc2}a,b
are shown examples of fitting of synthetic spectra to
the observed lines. 

%===============================================================================
\begin{table*}[h]
\begin{flushleft}
%\centering
\caption{Atmospheric parameters adopted.}             
\label{atmos}      
\centering          
\begin{tabular}{lcccccccccccc}     % 12 columns 
\noalign{\smallskip}
\hline\hline    
\noalign{\smallskip}
\noalign{\vskip 0.1cm} 
Star & T$_{\rm eff}$ [K] & logg & [FeI/H] &
[FeII/H]  & [Fe/H]  & v$_{\rm t}$ [kms$^{-1}$] & [C/Fe] & [N/Fe]  & [O/Fe] & \\        
(1) & (2)  & (3)  & (4)  & (5)  & (6) & (7) & (8) & (9) & (10) \\                    
\noalign{\vskip 0.1cm}
\noalign{\hrule\vskip 0.1cm}
\noalign{\vskip 0.1cm}    
\multicolumn{10}{c}{\hbox{\bf 47 Tucanae}} \\
\noalign{\vskip 0.1cm}
\noalign{\hrule\vskip 0.1cm}
\noalign{\vskip 0.1cm} 
M8  & 4086 & 1.48 & $-$0.62 & $-$0.65 &$-$0.64 &$+$1.42& $+$0.20 & $+$0.50 & $+$0.45 & \\ 
M11 & 3945 & 1.20 & $-$0.62 & $-$0.62 &$-$0.62 &$+$1.49& $+$0.00 & $+$0.50 & $+$0.25 & \\ 
M12 & 4047 & 1.45 & $-$0.63 & $-$0.68 &$-$0.66 &$+$1.45& $+$0.00 & $+$0.50 & $+$0.45 & \\ 
M21 & 5100 & 2.46 & $-$0.77 & $-$0.82 &$-$0.80 &$+$1.42& $+$0.20 & $+$0.50 & $+$0.30 & \\ 
M25 & 4200 & 1.65 & $-$0.64 & $-$0.67 &$-$0.66 &$+$1.37& $-$0.10 & $+$0.20 & $+$0.35 & \\ 
\noalign{\smallskip}
\noalign{\vskip 0.1cm}
\noalign{\hrule\vskip 0.1cm}
\noalign{\vskip 0.1cm}
\multicolumn{10}{c}{\hbox{\bf NGC 6553}}\\
\noalign{\vskip 0.1cm}
\noalign{\hrule\vskip 0.1cm}
\noalign{\vskip 0.1cm} 
II-64  &  4500 & 2.20 & $-0.20$ & $-0.20$&$-$0.20 &$+$1.45 &$+$0.00 &$+$0.50 &$+$0.45 &  \\ 
II-85  &  3800 & 1.10 & $-0.23$ & $-0.29$&$-$0.26 &$+$1.38 &$+$0.00 &$+$0.50 &$+$0.30 &  \\
III-8  &  4600 & 2.40 & $-0.17$ & $-0.17$&$-$0.17 &$+$1.40 &$+$0.00 &$+$0.50 &$+$0.30 & \\
267092 &  4600 & 2.50 & $-0.21$ & $-0.22$&$-$0.22 &$+$1.50 &$+$0.00 &$+$1.00  & --- &  \\
\noalign{\vskip 0.1cm}                   
\noalign{\hrule\vskip 0.1cm}
\noalign{\vskip 0.1cm}  
\multicolumn{10}{c}{\hbox{\bf NGC 6528}} \\
\noalign{\vskip 0.1cm}
\noalign{\hrule\vskip 0.1cm}
\noalign{\vskip 0.1cm}  
I-18 & 4700 & 2.00 & $-0.05$ & $-0.11$&$-$0.08  &$+$1.50 &$-$0.20 &$+$0.30 &$+$0.30 & \\ 
I-36 & 4200 & 1.50 & $-0.13$ & $-0.09$&$-$0.11  &$+$1.50 &$-$0.30 &$+$0.80 &$+$0.00 & \\ 
I-42 & 4100 & 1.60 & $-0.14$ & $-0.08$&$-$0.11  &$+$1.20 &$+$0.00  &$+$0.20 &$+$0.05 & \\ 
\noalign{\vskip 0.1cm}                
\noalign{\hrule\vskip 0.1cm}          
\noalign{\vskip 0.1cm}  
\multicolumn{10}{c}{\hbox{\bf HP~1}} \\
\noalign{\vskip 0.1cm}
\noalign{\hrule\vskip 0.1cm}
\noalign{\vskip 0.1cm}  
HP~1-2 & 4630 & 1.70 & $-1.02$ & $-$0.97& $-$1.00  & $+$1.60 &$+$0.00 &$+$0.20 &$+$0.30 &\\ %calculado em NEWHP1/PFANT06
HP~1-3 & 4450 & 1.75 & $-0.99$ & $-$0.95& $-$0.97  &$+$ 1.40 &$+$0.00 &$+$0.20 &$+$0.30 & \\  % idem
 2115  & 4530 & 2.00 & $-0.98$  & $-$1.02 & $-$1.00 & $+$ 1.45 &$+$0.00 &$+$0.70 &$+$0.40 & \\
 2461  & 4780 & 2.05 & $-$1.13 & $-$1.09 & $-$1.11 & $+$ 1.90 &$+$ 0.00 &$+$0.50 &$+$0.50 & \\
 2939  & 4525 & 2.00 & $-$1.07 & $-$1.07 & $-$1.07 & $+$ 1.65 & $+$0.00 &$+$0.50 &$+$0.50 &  \\
 3514  & 4560 & 1.80 & $-$1.18 & $-$1.19 & $-$1.18 & $+$ 2.00 & $+$0.00 &$+$0.80 &$+$0.40 &  \\
 5037  & 4570 & 2.15 & $-$0.98 & $-$1.03 & $-$1.00 & $+$ 1.20 & $+$0.00 &$+$0.50 &$+$0.35 &  \\
 5485  & 4920 & 2.07 & $-$1.18 & $-$1.18 & $-$1.18 & $+$ 1.80 & $+$0.00 &$+$0.50 &$+$0.40 &  \\
\noalign{\vskip 0.1cm}
\noalign{\hrule\vskip 0.1cm}
\noalign{\vskip 0.1cm}  
\multicolumn{10}{c}{\hbox{\bf NGC 6522}} \\
\noalign{\vskip 0.1cm}
\noalign{\hrule\vskip 0.1cm}
\noalign{\vskip 0.1cm}  
B-107 & 4990 & 2.00  & $-1.11$ & $-1.14$&$-$1.13 &$+$ 1.40 &$+$0.00 &$+$0.70 &$+$0.30 & \\
B-122 & 4900 & 2.70  & $-0.80$ & $-0.82$&$-$0.81 & $+$1.55 &$-$0.20 &$+$0.70 &$+$0.40 & \\
B-128 & 4800 & 2.50  & $-0.81$ & $-0.82$&$-$0.82 & $+$1.25 &$+$0.10 &$+$0.70 &$+$0.50 & \\ 
B-130 & 4850 & 2.20  & $-1.03$ & $-1.04$&$-$1.04 & $+$1.45 &$+$0.10 &$+$0.70 &$+$0.50 & \\
\noalign{\vskip 0.1cm}                   
\noalign{\hrule\vskip 0.1cm}
\noalign{\vskip 0.1cm}  
\multicolumn{10}{c}{\hbox{\bf NGC~6558}} \\
\noalign{\vskip 0.1cm}
\noalign{\hrule\vskip 0.1cm}
\noalign{\vskip 0.1cm}  
%B11  & 4650 & 2.20  & $-1.04$ & $-0.92$& 1.50 & & & & \\
%B64  & 4850 & 2.60  & $-0.94$ & $-1.00$& 1.20 & & & & \\ 
%B73  & 4700 & 2.30  & $-0.92$ & $-0.83$& 1.30 & & & & \\
%B117 & 5000 & 1.30  & $-1.14$ & $-1.06$& 1.30 & & & &  \\ 
%F42  & 3800 & 0.50  & $-1.05$ & $-1.65$& 1.65 & & & & \\
%F97  & 4820 & 2.30  & $-0.97$ & $-0.94$& 1.30 & & & & \\ 
 283 & 4840 & 2.50  & $-$1.14 & $-$1.16  &$-$1.15  & $+$1.05 &$+$0.10: &$+$0.70 &$+$0.50 & \\
 364 &  4880 &  2.35  &  $-$1.18 &  $-$1.13 &  $-$1.15  & $+$1.90 &$+$0.10 &$+$0.80 &$+$0.20 & \\
1160 & 4890 & 2.35  & $-$1.03 & $-$1.04  &$-$1.04  & $+$0.73 &$+$0.20 &$+$1.00 &$+$0.50 & \\
1072 & 4850 & 2.60  & $-$1.20 & $-$1.26  &$-$1.23  & $+$1.10 &$+$0.10 &$+$1.00 &$+$0.55 & \\ 
\hline                               
\hline                  
\end{tabular}
\end{flushleft}
\end{table*}

%\begin{figure}
%\centering
%\psfig{file=47tucm11cno.eps,angle=0.,width=9.0 cm}
%\caption{CNO in 47Tuc:M11: Fits to the Swan C$_2$ (0,1) bandhead at
%5635 {\rm \AA}, the red CN (5,1) bandhead
%at 6332.18 {\rm \AA}, and the forbidden oxygen [OI]6300.311 {\rm \AA}
% line.
%}
%\label{cno} 
%\end{figure}

%\begin{figure}[!h]
%\centering
%\psfig{file=Sclines.eps,angle=0.,width=9.0 cm}
%\caption{Fits to the \ion{Sc}{I} 6604.601 {\rm \AA} line.
%}
%\label{spectrasc}
%\end{figure}

%\begin{figure}[!h]
%\centering
%\psfig{file=Vlines.eps,angle=0.,width=9.0 cm}
%\caption{Fits to the \ion{V}{I}5703.560 {\rm \AA} line.
%}
%\label{spectrav}
%\end{figure}

%\begin{figure}[!h]
%\centering
%\psfig{file=Mnlines.eps,angle=0.,width=9.0 cm}
%\caption{Fits to the \ion{Mn}{I} 6013.513 {\rm \AA} line.}
%\label{spectramn} 
%\end{figure}

%\begin{figure}[!h]
%\centering
%\psfig{file=Culines.eps,angle=0.,width=9.0 cm}
%\caption{Fits to the \ion{Cu}{I} 5105.537 {\rm \AA} line.
%}
%\label{spectracu} 
%\end{figure}

%\begin{figure}[!h]
%\centering
%\psfig{file=Znlines.eps,angle=0.,width=9.0 cm}
%\caption{Fits to the \ion{Zn}{I} 4810.529 {\rm \AA} line.
%}
%\label{spectrazn} 
%\end{figure}

%===============================================================================
%				Main Results
%===============================================================================

\subsection{Uncertainties}
The final adopted atmospheric parameters for
all program stars were based on Fe I and Fe II lines
in the papers cited above, and we have adopted their
estimated uncertainties in the atmospheric parameters, i.e., $\pm$ 100 K for
temperature, $\pm$ 0.20 for surface gravity,
 $\pm$ 0.10 dex for [Fe/H]
 and $\pm$ 0.20 kms$^{-1}$ for
microturbulent velocity.
In Table \ref{error} the final uncertainties in the abundances
of the iron-peak elements studied are reported  for
the metal-poor star HP-1:2115, and the metal-rich star NGC 6528:I36.
Given that the stellar parameters are correlated among them,
 the covariance will 
be non-zero.
 Since we have taken into account only the diagonal terms of the
 covariance matrix, these errors are overestimated.

\begin{figure}[!h]
\centering
\includegraphics[width = 3.7in]{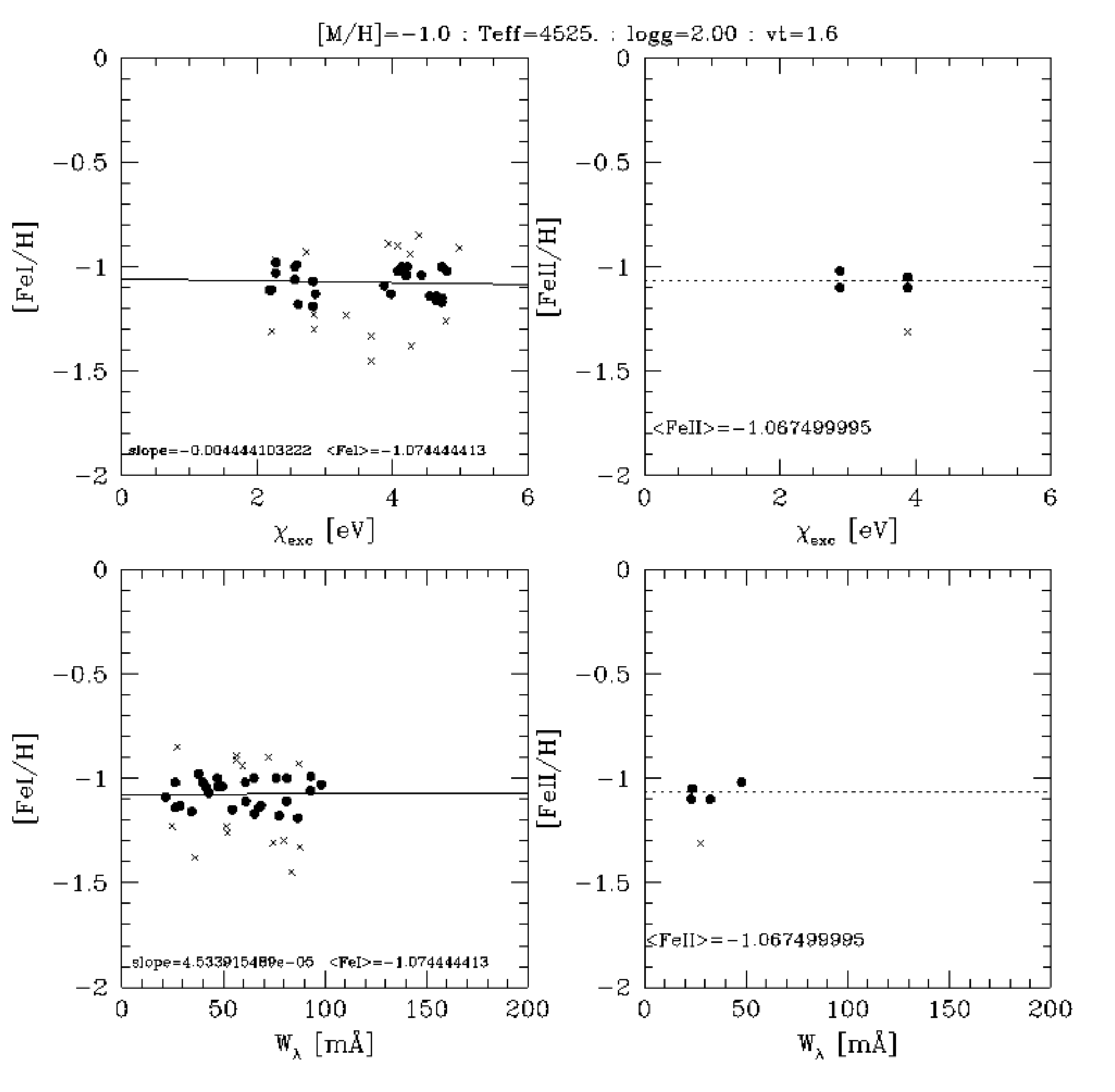}
\caption{Excitation and ionization equilibria of \ion{Fe}{I} and \ion{Fe}{II} lines for star 2939 
 in LTE.}
% b) applying NLTE abundance correction; c) applying NLTE gravity correction.  }
\label{abon2a}
\end{figure}

\begin{figure*}[!h]
\centering
%\subfloat[LTE.]{\includegraphics[width = 2.4in]{abon2939final.pdf}}
\subfloat[NLTE abundance corrected.]{\includegraphics[width = 3.7in]{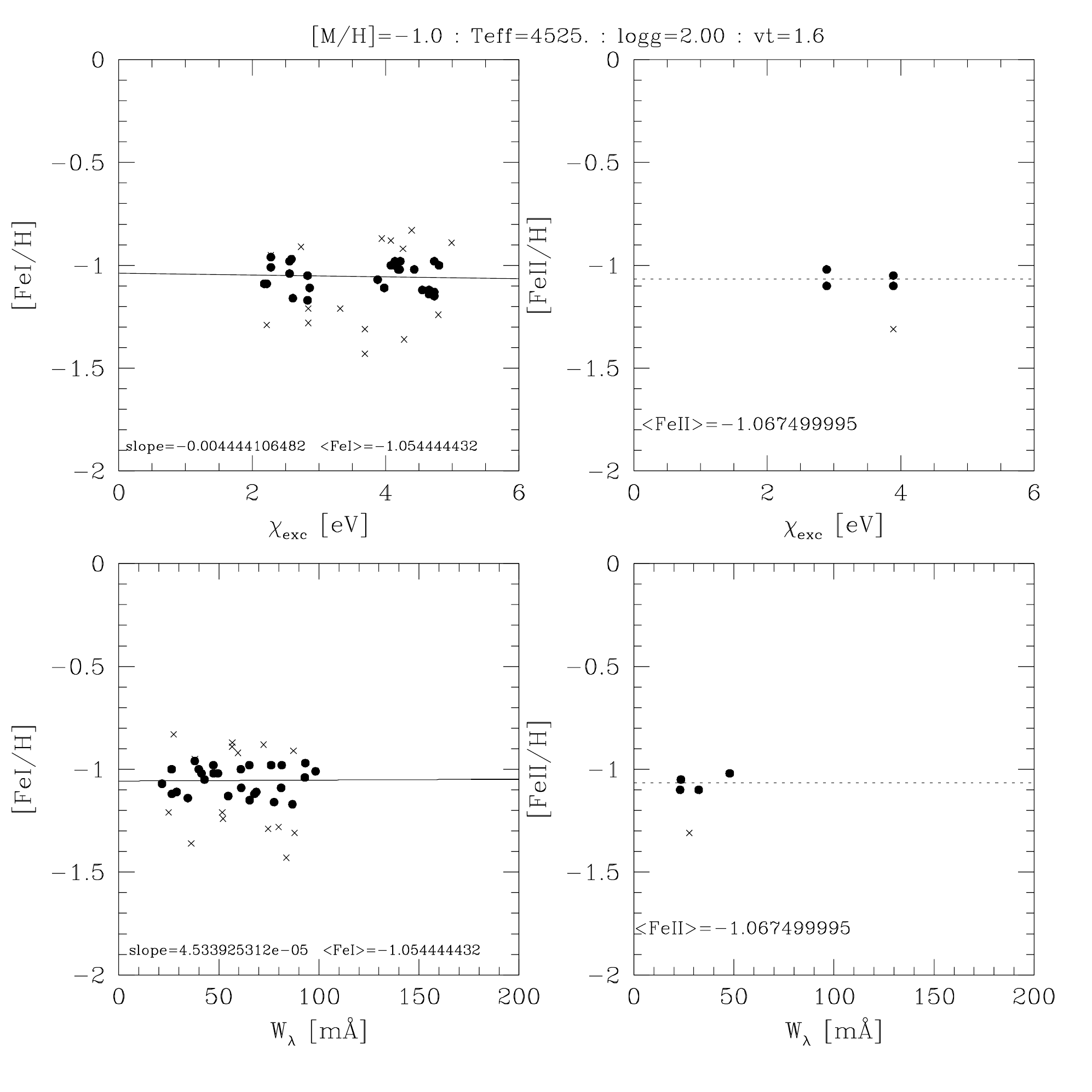}}
\subfloat[NLTE gravity corrected.]{\includegraphics[width = 3.7in]{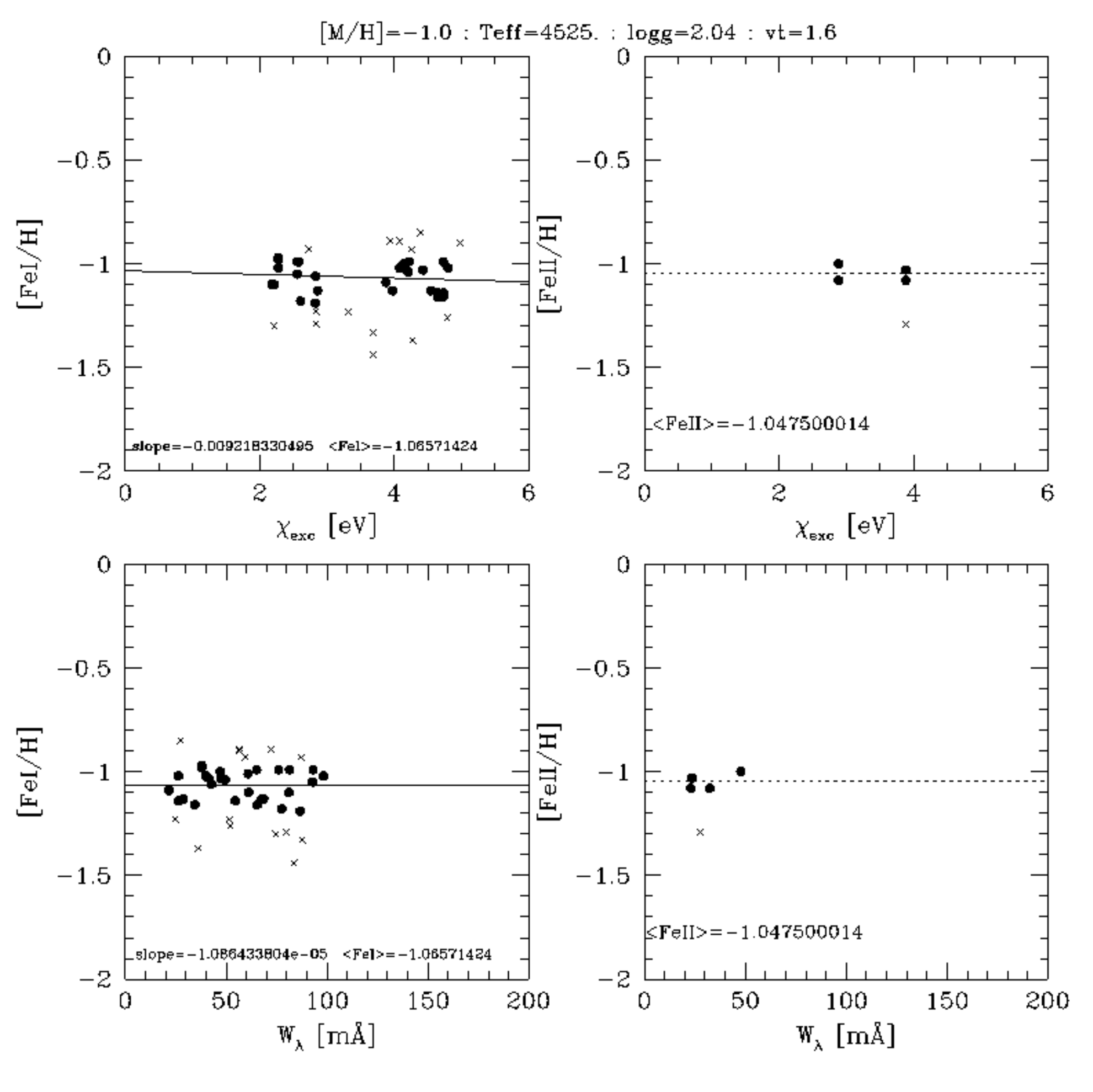}}
\caption{Excitation and ionization equilibria of \ion{Fe}{I} and \ion{Fe}{II} lines for star 2939. 
a) applying NLTE abundance correction; b) applying NLTE gravity correction.  }
\label{abon2b}
\end{figure*}

 In order to further inspect the errors in stellar parameters,
we applied NLTE corrections to abundances as given in Lind et al. (2012),
and following suggestions given in Bergemann et al. (2013).
For star HP1-2939 as an example,
we show the LTE excitation and ionization potential plots in
Fig. \ref{abon2a}, restricting to lines with excitation potential
$\chi_{ex}\geq$ 2.0 eV.
 %then we have ran the LTE excitation, and
%ionization equilibrium, 
We then applied
a) the NLTE abundance
correction to each  \ion{Fe}{I}  line, and ran the
excitation and ionizatin equilibrium once more. This is
shown in Fig. \ref{abon2b}a (left panel).
The result is a negligible change in metallicity from \ion{Fe}{I} lines
of about 0.015 dex.

b) the NLTE correction on gravity log g, amounting to 
$\Delta$log g=0.04 - see Fig. \ref{abon2b}b (right panel). 
It can be seen that the difference in 
metallicity between \ion{Fe}{I} and  \ion{Fe}{II} increased
to 0.02 instead of the previous 0.01 difference.

c) the NLTE correction on temperature is neglibigle (6 K)

d) the NLTE on microturbulence velocity is also negligible 
at these metallicities.

As shown in Fig. 6 by Bergemann et al. (2013), the effects are
not pronounced for stars of metallicity [Fe/H]$\simgreat$ $-$1.0,therefore
they can be neglected for the present sample stars, 
given the other larger errors.

We also have to take into account errors on S/N and equivalent widths.
These errors are given by the Cayrel (1988) formula 
(see also Cayrel et al. 2004)
%\begin{center}
%\begin{equation}
$\sigma={1.5 \over S/N} \sqrt{FWHM*\delta{x}}$.
%\label{cayrel}
%\end{equation}
% \end{center}
Adopting a mean FWHM = 12.5 pixels, or 0.184 {\rm \AA}.
 The CCD pixel size is 15 $\mu$m, or $\delta{x}$ = 0.0147 {\rm \AA}
in the spectra. By assuming a mean
S/N=100, we derive an error $\Delta$EW $\sim$ 0.8 m{\rm \AA}
(note that this formula neglects the uncertainty in 
the continuum placement).

In order to take the S/N and fitting error into account, 
we adopt $\delta_{noise} = \frac{\sigma}{\sqrt{N-1}}$.
These are reported in Table \ref{meanabundance}.

The final error is given by the equation %\ref{error}.
%\begin{center}
%\begin{equation}
$\delta_{[X/Fe]} = \sqrt{\delta_{noise}^2 +\delta_{parameters}^2}$
%\label{error}
%\end{equation}
% \end{center}
\noindent where for the error on stellar parameters the values given in Table 
\ref{error} for the metal-poor stars HP1:2115 are applied  
to the stars more metal-poor than [Fe/H]$<$$-$0.5, and
those for NGC~6528:I36 to the metal-rich stars.

\begin{table*}[!h]
\begin{flushleft}
\centering
\caption{Sensitivity of abundances to changes of
$\Delta$T$_{\rm eff}$ = 100 K,
$\Delta$log g = +0.20, $\Delta$[Fe/H]=+0.1, and $\Delta$v$_{\rm t}$ = 0.20 km s$^{-1}$. In the last
column the corresponding total error is given.}
\label{error}
\begin{tabular}{lcccccccc}
\hline\hline
\noalign{\smallskip}
\hbox{Species} & \hbox{$\Delta$T} & \hbox{$\Delta$ $\log$ g} & \hbox{$\Delta$[Fe/H]}
 & \hbox{ $\Delta$v$_{t}$} & \hbox{($\sum$x$^{2}$)$^{1/2}$} \\ 
 & \hbox{(100 K)} & \hbox{(+0.20 dex) }  & \hbox{(+0.10 dex) } & \hbox{(+0.20 kms$^{-1}$}) & \hbox{}  &  \\
\hbox{(1)} & \hbox{(2)} & \hbox{(3)} & \hbox{(4)} & \hbox{(5)}  & \hbox{(6)} \\
\noalign{\smallskip}
\noalign{\smallskip}
\noalign{\vskip 0.1cm}
\noalign{\hrule\vskip 0.1cm}
\noalign{\vskip 0.1cm}
\multicolumn{6}{c}{\hbox{\bf HP~1 : 2115}}\\
\noalign{\vskip 0.1cm}
\noalign{\hrule\vskip 0.1cm}
%\hbox{[Fe/H]}     & +0.06     & +0.03 & & $-$0.07  & +0.10 \\
\hbox{[Sc/Fe]}    & $-$0.02   & +0.10 & $-$0.010 & $-$0.01  & +0.10   \\
\hbox{[V/Fe]}     & +0.15     & +0.01 & $-$0.015 & $-$0.01  & +0.15  \\
\hbox{[MnI/Fe]}   & +0.01     & +0.01 & $-$0.010  & $-$0.02 & +0.03   \\
\hbox{[CuI/Fe]}   & +0.10     & +0.02 & $-$0.005 & $-$0.10  & +0.14    \\
\hbox{[ZnI/Fe]}   & $-$0.05   & +0.10 & $-$0.030 & $-$0.01  &  +0.12    \\
\noalign{\hrule\vskip 0.1cm}
\multicolumn{6}{c}{\hbox{\bf NGC 6528 : I-36}}\\
\noalign{\vskip 0.1cm}
\noalign{\hrule\vskip 0.1cm}
%\hbox{[Fe/H]}     & $- $    & $- $   & &$ -  $  & $-$ \\
\hbox{[Sc/Fe]}    & $-$0.02  & +0.10 &  $-$0.030  & $-$0.05  &+0.12 \\
\hbox{[V/Fe]}     & +0.05    & +0.01 &  $-$0.020  & $-$0.12  & +0.13 \\
\hbox{[MnI/Fe]}   & +0.01    & +0.01 &  $-$0.015  & $-$0.10  & +0.10 \\
\hbox{[CuI/Fe]}   & +0.02    & +0.01 &  $-$0.010  & $-$0.12  & +0.12 \\
\hbox{[ZnI/Fe]}   & +0.07    & +0.07 &  $-$0.050  & $-$0.15  & +0.19 \\
\hline
\end{tabular}
\end{flushleft}
\end{table*}

\section{Results and discussions}

Very few abundances are available for iron-peak elements in bulge stars.
In this Section, we present results
and discuss the available chemical evolution models, and
associated nucleosynthesis of the studied species.
%The main isotopic observable species are:
%$^{45}$Sc (100\%), $^{48}$Ti (73.72\%), $^{51}$V (99.75\%), 
%$^{52}$Cr (83.789\%), $^{55}$Mn (100\%), $^{56}$Fe (91.754\%),
%$^{59}$Co (100\%), $^{58,60}$Ni (68.08,26.22\%), $^{63,65}$Cu (69.17,30.83 \%),
%$^{64,66,68}$Zn (48.63,27.90,18.75\%) (Asplund et al. 2009). 
We have included all chemical evolution
models available for the Galactic bulge for these elements.

 The abundances of \ion{Sc}{I}, \ion{Sc}{II}, \ion{V}{I}, \ion{Mn}{I},
  \ion{Cu}{I}, and \ion{Zn}{I} for each sample star are listed in Table
\ref{abonds}. 
In Figures  \ref{plotsc}, \ref{plotvvv}, \ref{plotmn}, \ref{plotcu}, 
\ref{plotzn} we plot the element-to-iron ratio versus the metallicity [Fe/H].

\subsection {Scandium, Vanadium and Manganese}

Sc is intermediate between the alpha-elements and the iron-peak elements.
 $^{45}$Sc is produced  in central He burning and in C-burning shell,
in a so-called weak-s process, and 
 during neon burning
and  as the radioactive progenitor $^{45}$Ti in explosive oxygen and
silicon burning (WW95,LC03).
% For Samland et al. (1998) 
%45Sc is synthesized during explosive oxygen and neon
%burning.
V, Cr, Mn are mainly produced in incomplete explosive Si burning 
in outer layers of massive stars (WW95, Limongi \& Chieffi 2003, 
hereafter LC03). 

\begin{figure*}[!h]
%\centering
\includegraphics[scale=1.0]{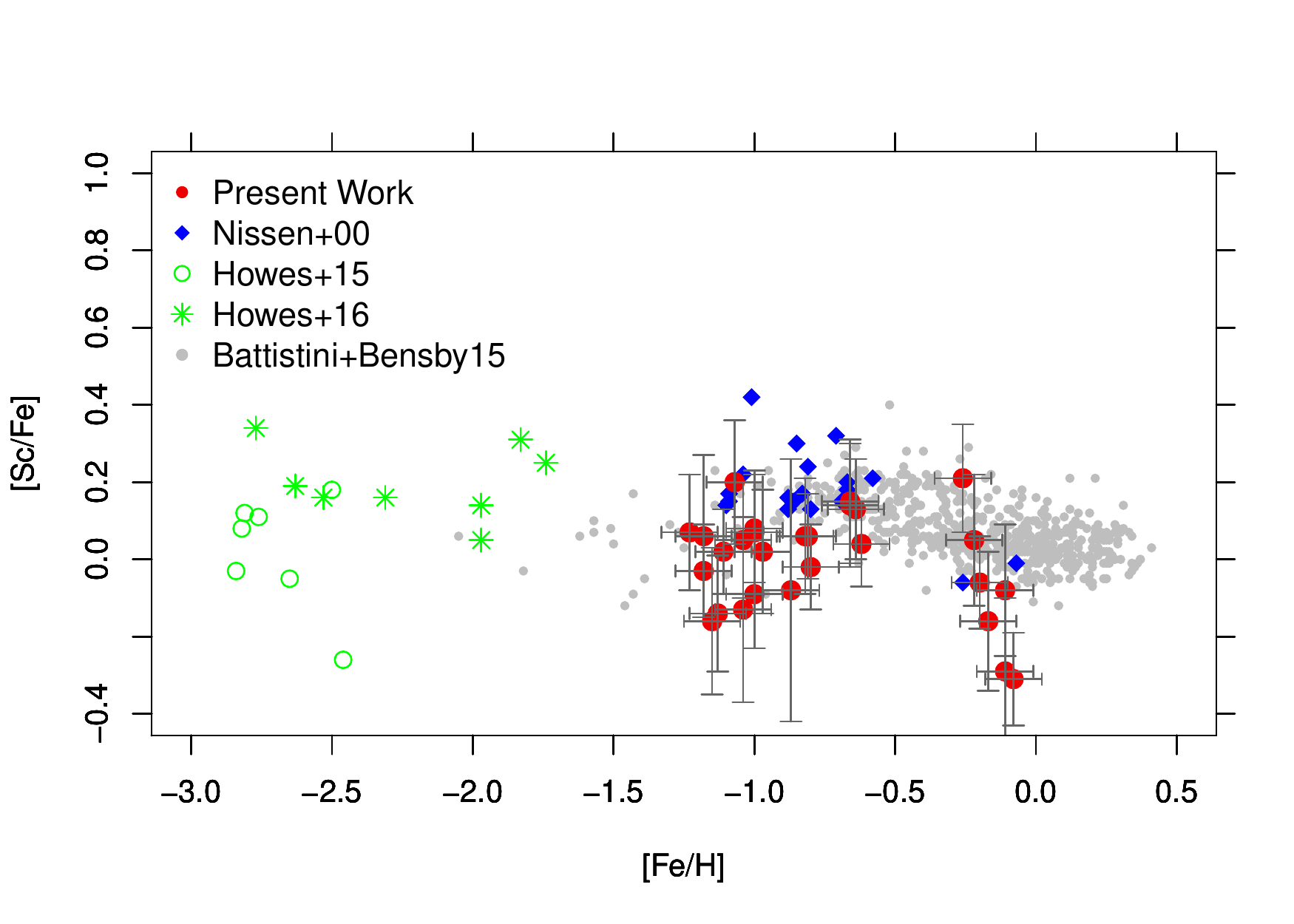}

\caption{[Sc/Fe] vs. [Fe/H]. Symbols: present work (filled red circles),
Bensby et al. (2017) (black filled circles),
Nissen et al. (2000) (filled blue diamonds), Howes et al. (2015) (open
green circles), Howes et al. (2016) (green crosses);
Battistini \& Bensby (2015) (grey filled circles).
 Error bars on [Sc/Fe] are indicated. Erros on [Fe/H] can be assumed
as constant, of the order of $\Delta$[Fe/H]=$\pm$0.17dex.}

\label{plotsc}
\end{figure*}

Fig. \ref{plotsc} compares the present [Sc/Fe] values with
 metal-poor bulge stars by Howes (2015, 2016) for 
 thick disk and halo stars by  Nissen et al. (2000) and Ishigaki et al. (2013)
and thin and thick disk stars by Battistini \& Bensby (2015). 
 The data show a considerable spread, but it is possible
to interpret the metal-poor side from Howes et al. and Ishigaki et al. 
 as somewhat enhanced with [Sc/Fe]$\sim$0.2.
Fishlock et al. (2017) confirm the findings by Nissen et al. (2000), that high-
and low-alpha halo stars show high and low Sc abundances respectively.
Data by Nissen et al. (2000) tend to show a trend of 
decreasing [Sc/Fe] with increasing metallicity.
Contrarily to the Sc enhancement in the more metal-poor stars
from Howes et al. (2015, 2016), the present results on
moderately metal-poor bulge clusters 
do not show a significant Sc enhancement. Our results are lower than
Nissen et al. (2000)'s values, and  fit well the level of [Sc/Fe] values
 by Battistini \& Bensby (2015) for thin and thick disk stars. 
The metal-rich globular clusters NGC 6528 and NGC 6553 show
a spread in Sc abundances at [Fe/H]$\sim$-0.2, that might
be considered as a decrease with increasing metallicity at the
high metallicity end.

\begin{figure*}[!]
\centering
\includegraphics[scale=1.0]{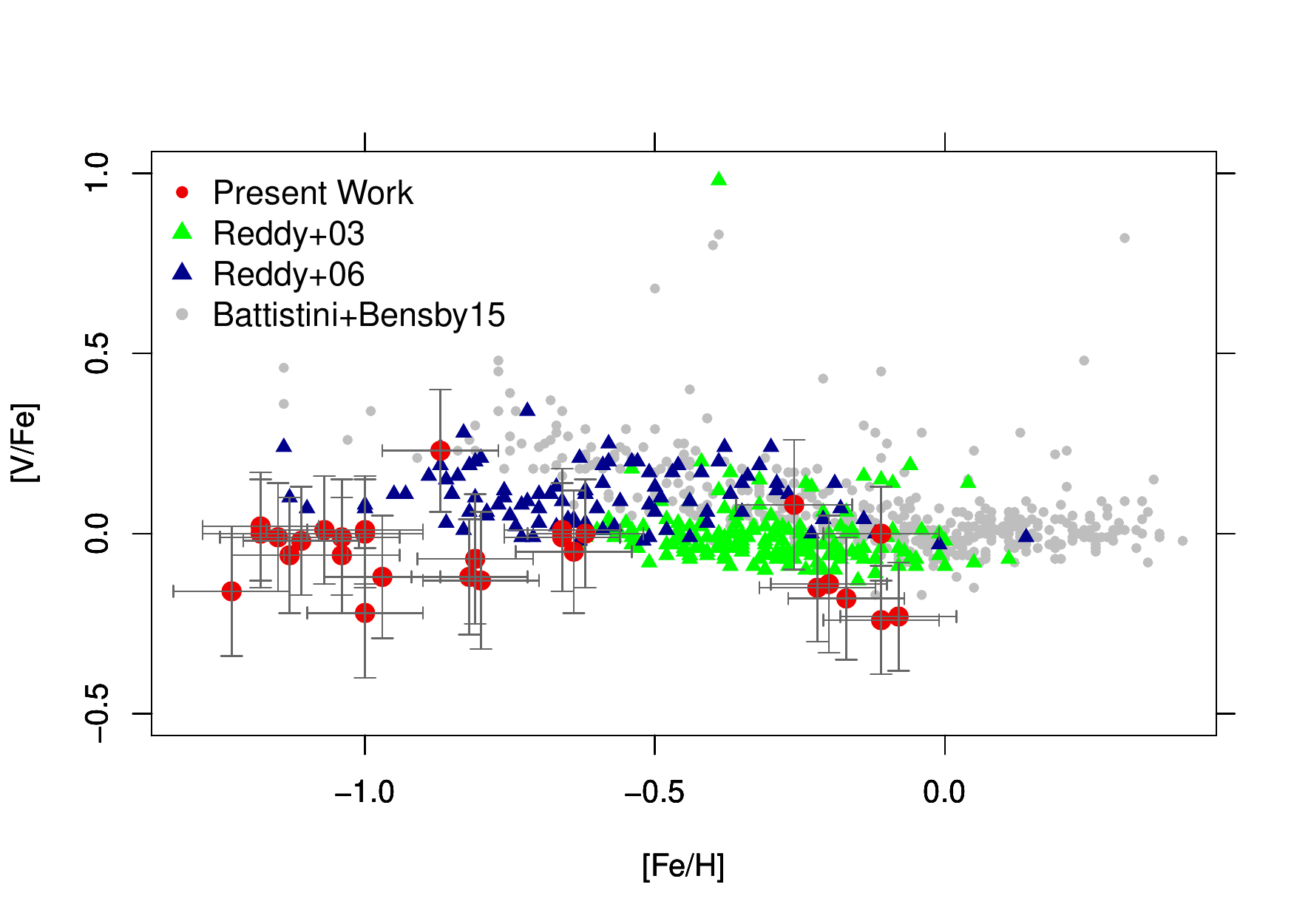}
\caption{[V/Fe] vs. [Fe/H].
Symbols: present work (filled red circles);
Reddy et al. (2003) (green filled triangles),
Reddy et al. (2006) (navyblue filled triangles)
Battistini \& Bensby (2015) (grey filled circles). Errors are assumed as in Fig. \ref{plotsc}.} 
\label{plotvvv} 
\end{figure*}

Figure \ref{plotvvv} shows that V varies in lockstep with
Fe. There are no V abundances for bulge stars other than the present
data. The thin and thick disk data from Reddy et al. (2003, 2006)
are overplotted. The thick disk V abundances from Reddy et al. (2006) appear
to be enhanced with respect to thin disk stars (Reddy et al. 2003),
as well as to the present results, whereas they
seem to be at the same level as thin and thick disk stars
by Battistini \& Bensby (2015). For the more metal-rich
stars the bulge globular cluster stars tend to decrease with
increasing metallicity, which could be due to enrichment
in Fe by SNIa. Due to uncertainties, the spread in the data
do now allow to derive further conclusions from V abundances.

Kobayashi et al. (2006, hereafter K06) have shown that Sc,
 and V yields are underabundant 
by 1 dex based on previous nucleosynthesis prescriptions.
 Umeda \& Nomoto (2005, hereafter UN05), Kobayashi et al. (2006),
Nomoto et al. (2013) have introduced a low-density model,
 during explosive burning, enhancing Sc abundance through
 the alpha-rich freezeout. Yoshida et al. (2008) applied a $\nu$-process 
to Si explosive nucleosynthesis, producing larger
amounts of Sc, V, and Mn production by a factor of 10. 
Fr\"ohlich et al. (2006) showed that a delayed neutrino mechanism 
     leading to an electron fraction value of Y$_{\rm e}$ $\simgreat$ 0.5 
in the innermost region gives larger
production of Sc, Ti and Zn.
In conclusion, given that the available models do not reproduce 
the observations (Kobayashi et al. 2006) for both Sc and V, due to low yields
from nucleosynthesis yields, these models are not overplotted
to the data.

\begin{figure*}[!]
\centering
\includegraphics[scale=1.0]{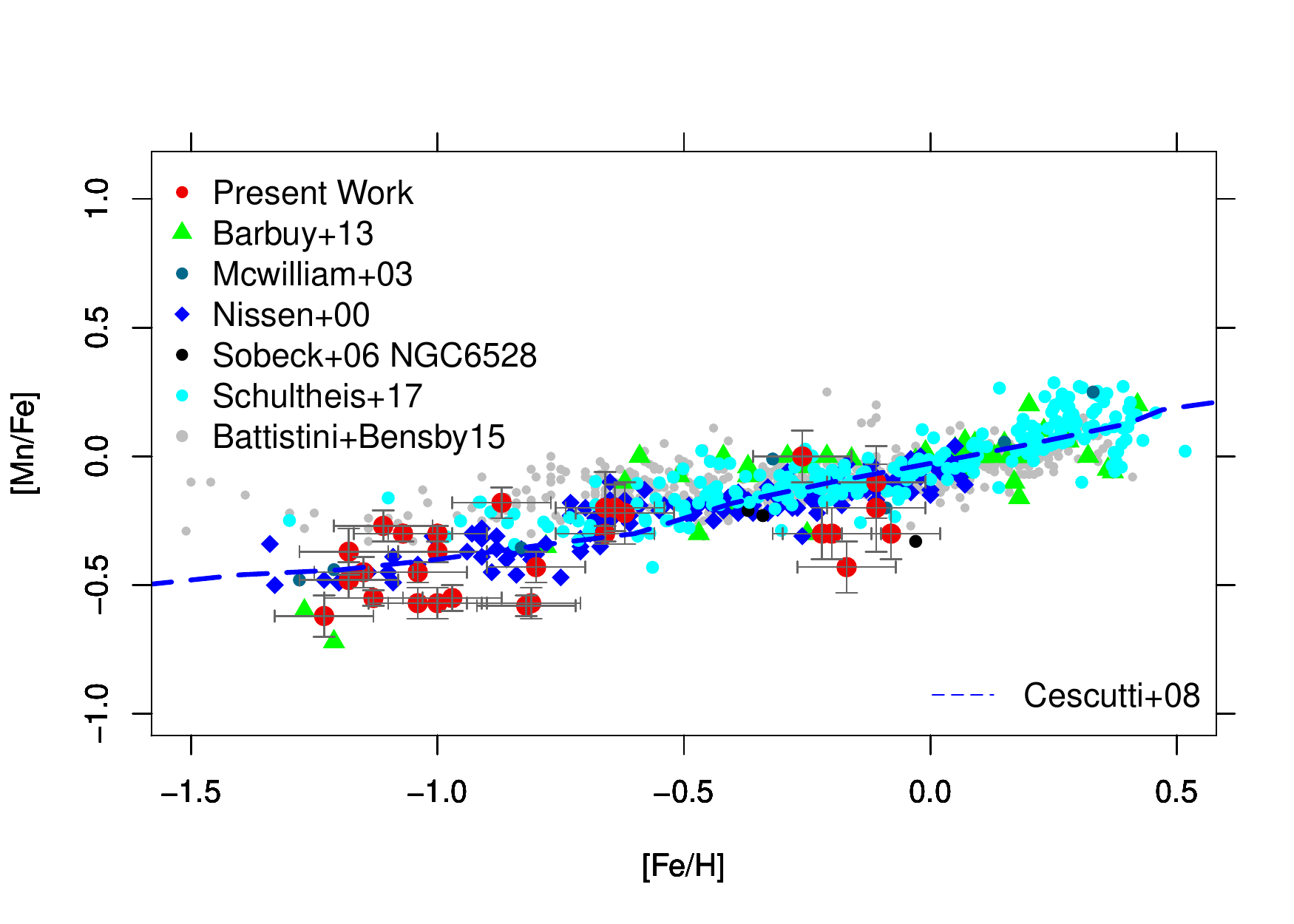}
\caption{[Mn/Fe] vs. [Fe/H] for the sample stars and literature data:
the present sample (red filled circles), Nissen et al. (2000) (blue filled diamonds),
Sobeck et al. (2006) (black filled circles), McWilliam et al. (2003) (deep sky blue filled
circles), Barbuy et al. (2013) (green filled triangles), 
Schultheis et al. (2017) 
(blue filled circles), and
Battistini \& Bensby (2015) (grey filled circles).
Chemical evolution models
by Cescutti et al. (2008)
(blue dashed line); are overplotted.
 Errors are assumed as in Fig. \ref{plotsc}.
}
\label{plotmn} 
\end{figure*}

Figure  \ref{plotmn} shows [Mn/Fe] vs. [Fe/H] for the
present results, together with previous results
 in Galactic bulge stars measured by McWilliam et al. (2003), 
Barbuy et al. (2013), and Schultheis et al. (2017), and
results for thin and thick disk stars by 
Battistini \& Bensby (2015). It is important
to note that NLTE corrections in the range of parameters of the
present data are small (Bergemann \& Gehren 2008).
The only available
bulge chemical evolution models by Cescutti et al. (2008) and
Kobayashi et al. (2006) are overplotted.
%Barbuy et al. (2013) compared the data to models
%by Cescutti et al. (2008) 
%where metallicity dependent models of
%both SN II and SN Ia were adopted.
Cescutti et al. (2008) computed models for Mn enrichment in the Galactic bulge,
adopting a star formation rate 20 times faster than in the solar neighbourhood,
and a flatter IMF. Their preferred model adopts metallicity dependent yields
from WW95 for massive stars, and Iwamoto et al. (1999) for intermediate
mass stars.
 K06 produced a grid of yields, 
including both SNII and hypernovae,
and further have built  chemical evolution models for Galaxy components, 
including the bulge.

The present Mn abundances in globular cluster
 stars follow the trend of field stars,
i.e., with [Mn/Fe]$\sim$-0.5 at [Fe/H]$\sim$-1.5, increasing
steadily with increasing metallicity, and they are well reproduced by
the models. %The metallicity-dependent yields in the
% Cescutti et al. (2008) model appear to fit the data most suitably.

\begin{figure*}[!]
\centering
\includegraphics[scale=1.0]{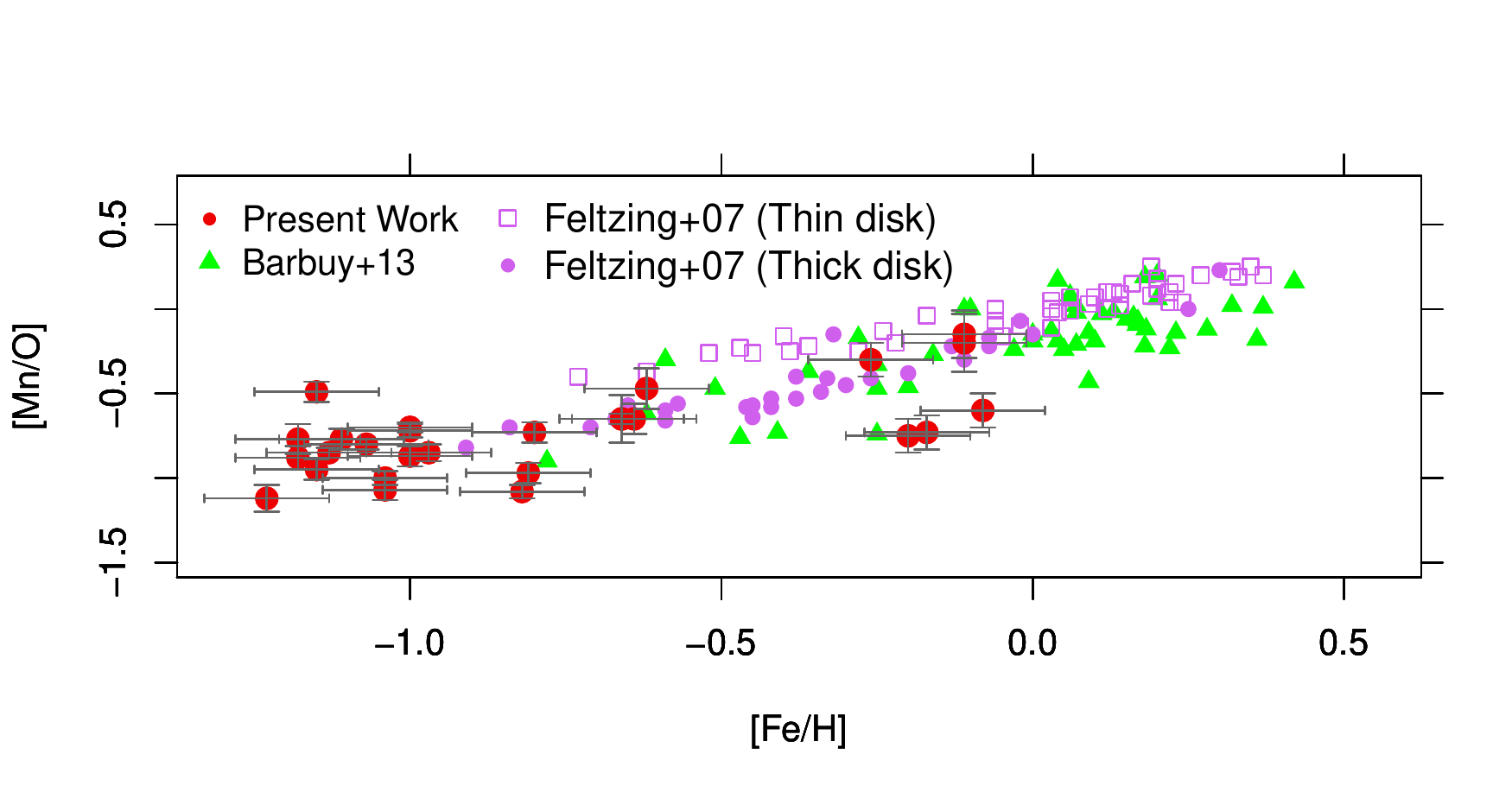}
\caption{[Mn/O] vs. [Fe/H]. Symbols: present work (red filled circles), Barbuy et al. (2013) (green filled triangles), Feltzing et al. (2007) thin disk (violet open squares), Feltzing et al. (2007) thick disk (violet filled circles).
 Errors are assumed as in Fig. \ref{plotsc}.
}
\label{MnOx} 
\end{figure*}

Finally, Fig. \ref{MnOx} shows [Mn/O] vs. [Fe/H], 
 revealing  differences
 between thin, thick and bulge stars, as previously pointed out by
Feltzing et al. (2007) and Barbuy et al. (2013). This is  
of great importance since [Mn/O] can be used as a discriminator
between different stellar populations, that otherwise have a similar
behaviour.

\subsection{Copper}

$^{63,65}$Cu are mainly produced
through neutron-capture during core He burning and convective
 shell carbon burning, therefore Cu may be classified as produced 
in a weak s-process component (LC03).
 Some primary $^{65}$Cu is also made as $^{65}$Zn 
in explosive nucleosynthesis through alpha-rich freezeout 
(WW95, Pignatari et al. 2010).  Cu is not significantly produced in
SNIa, nor in AGB stars  or through the r-process (Pignatari et al. 2010).

Johnson et al. (2014) derived copper abundances for a large
sample of bulge red giants. Their results show a low Cu abundance ratio
 at low metallicities that increases with increasing [Fe/H]. 
For supersolar metallicities, [Cu/Fe] values appear to be enhanced relative
to other stellar populations.

\begin{figure*}[!]
\centering
\includegraphics[scale=1.0]{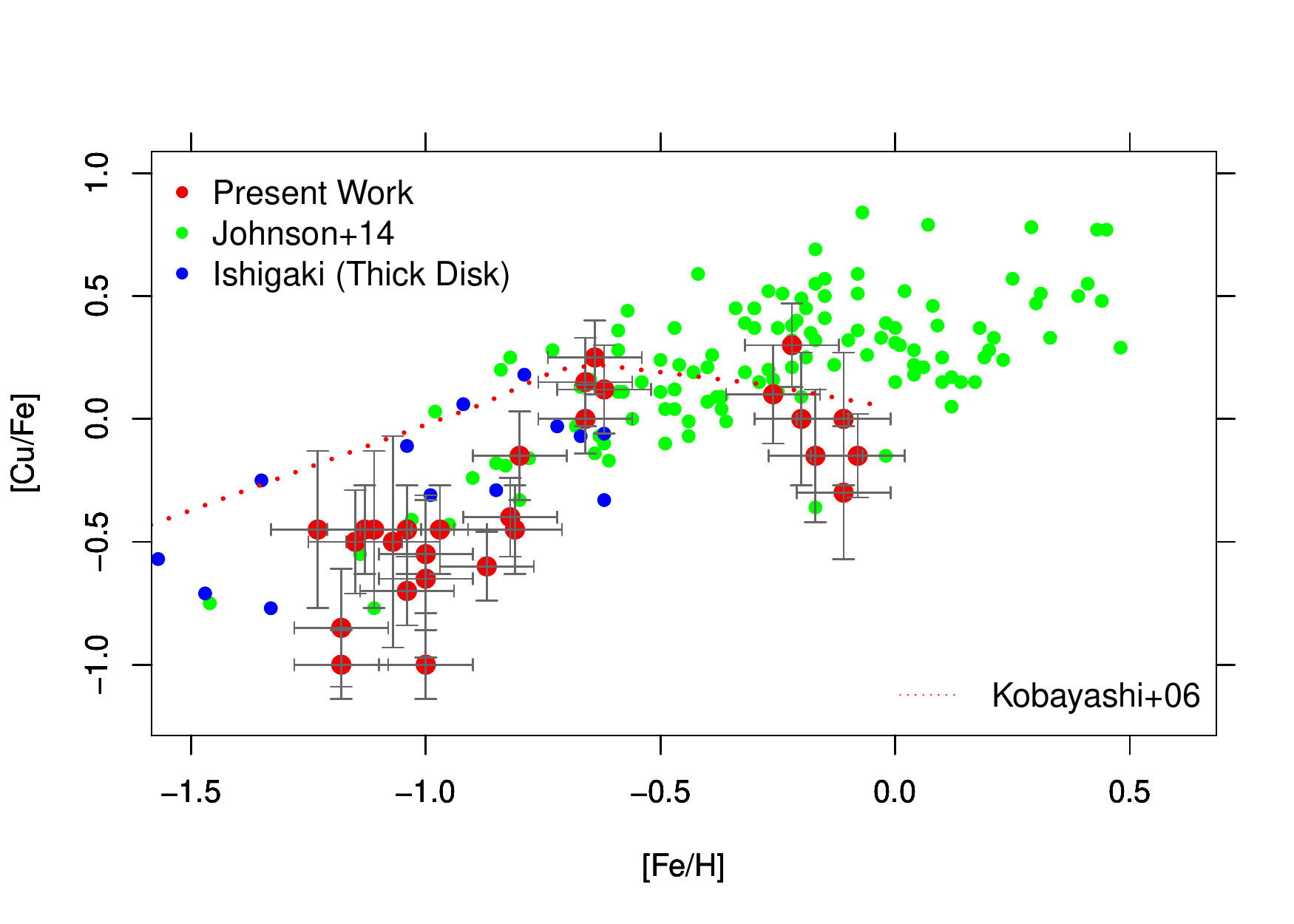}
\caption{[Cu/Fe] vs. [Fe/H] for the present sample (red filled circles),
Johnson et al. (2014) (green filled circles), and Ishigaki et al. (2013) (blue filled circles).
 Chemical evolution models by
and Kobayashi et al. (2006) (red dotted lines) are overplotted.
 Errors are assumed as in Fig. \ref{plotsc}.}
\label{plotcu} 
\end{figure*}

The present results are plotted in Fig. \ref{plotcu}, together with data from
Johnson et al. (2014) for the bulge, and Ishigaki et al. (2013) for the thick
disk. Our results tend to be less enhanced than those by Johnson et al. (2014).
 There is good agreement between the data and the models by
 by Kobayashi et al. (2006).
The metal-poor clusters show very low [Cu/Fe].

\begin{figure*}[!]
\centering
\includegraphics[scale=1.0]{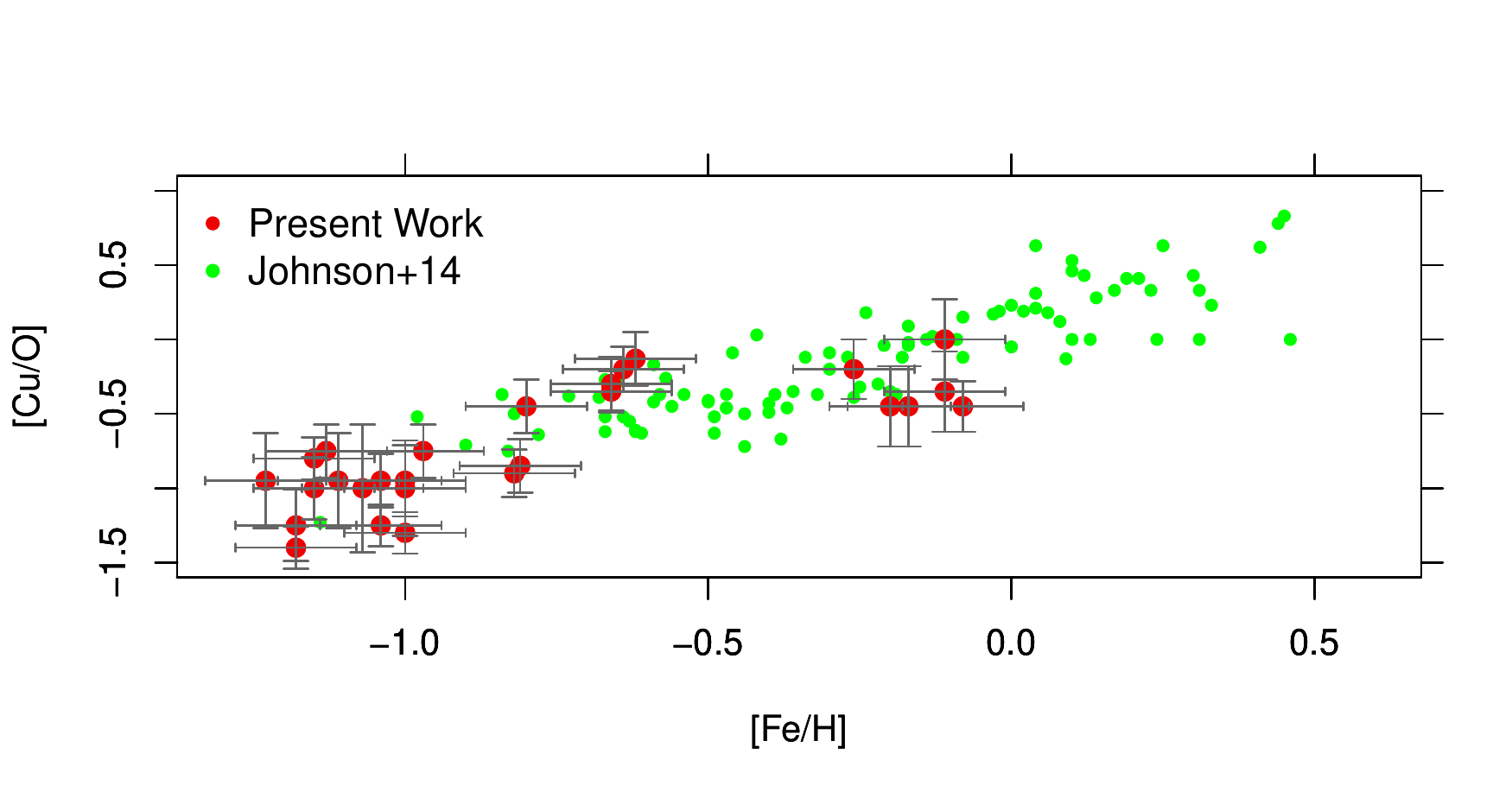}
\caption{[Cu/O] vs. [Fe/H]. Symbols: present work (red filled circles), and Johnson et al. (2014) (green filled circles).}
\label{CUOX} 
\end{figure*}

According to McWilliam (2016) [Cu/O] has much less spread than [Cu/Fe] data,
Fig. \ref{CUOX} shows [Cu/O] vs. [Fe/H], where the behaviour of the sample
cluster stars track well the Johnson et al. (2014) field stars data.
This rather straight correlation between Cu and O, 
 indicates the production of Cu and O in the same massive stars.

\subsection{Zinc}

The main isotopes of Ti, Co, Ni, Cu, Zn are 
produced only or mainly in the zone
that undergoes explosive Si burning with complete Si exhaustion (LC03).
% {\it $^{59}$Co} forms in hydrostatic core He burning and C-burning shell, 
%as well as $^{59}$Ni in complete explosive Si burning and is the decay 
%product of $^{59}$Cu, whereas  $^{63}$Cu is synthesized directly. 
%{\it $^{58}$Ni} is destroyed by both core He burning and convective C shell,
% and is largely produced during complete and incomplete Si-burning.
%$^{60,62}$Ni are produced directly and as decay of $^{60,62}$Zn, $^{60,62}$Cu.
%% $^{60,62}$Ni.
%$^{56,57}$Fe are products of radioactive decay of  $^{56,57}$Ni, for
%the electron fraction of Y$_{\rm e}$ $\sim$0.5. 
The relevant Zn isotopes
 $^{64,66,67,68}$Zn are produced in core He burning but  $^{64}$Zn is
 destroyed in convective C shell; they are also produced in 
$\alpha$-rich freeze-out layers
 in complete explosive Si-burning (LC03, WW95,
Woosley et al. 2002,  Nomoto et al. 2013).
These contributions do not explain however, the high
[Zn/Fe] observed in metal-poor stars.  
Umeda \& Nomoto (2002, 2005), Nomoto et al. (2013)
 suggested that $^{64}$Zn is produced in
energetic explosive nucleosynthesis so-called hypernovae.

\begin{figure*}[!]
\centering
\includegraphics[scale=1.0]{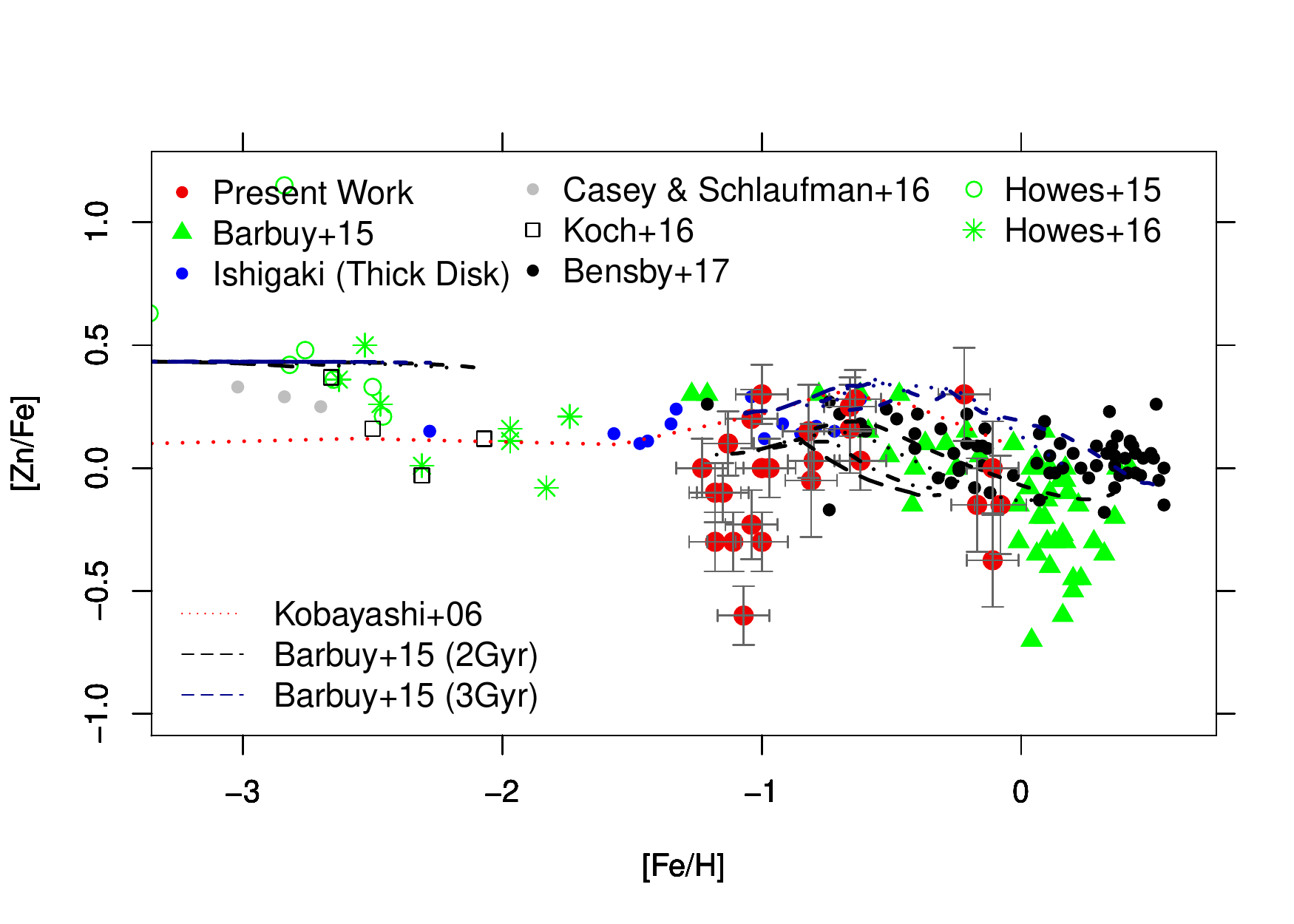}
\caption{[Zn/Fe] vs. [Fe/H]. Symbols: present work (filled red circles), Barbuy et al. (2015) (green filled triangles), Ishigaki et al. (2013) (blue filled circles), Howes et al. (2015) (green open circles), Howes et al. (2016) (green crosses), Casey \& Schlaufman (2016) (grey filled circles), Koch et al. (2016) (black open squares), Bensby et al. (2017) (black filled circles).  Chemical evolution models by Kobayashi et al. (2006) (red dotted line).
The Barbuy et al. (2015) models are shown for enrichment timescales of
2 (black) and 3 (blue) Gyr. In each case, the models for radius with respect to the Galactic
center of r$<$0.5 kpc (dashed lines), 0.5$<$r$<$1 kpc (dotted lines),
 1$<$r$<$2 kpc (dashe-dotted lines),  2$<$r$<$3 kpc (long-dashed lines)
are overplotted.
 Errors are assumed as in Fig. \ref{plotsc}.
}
\label{plotzn} 
\end{figure*}

Figure \ref{plotzn} shows [Zn/Fe] vs. [Fe/H] for the present sample,
and bulge field stars from Barbuy et al. (2015), Bensby et al. (2013, 2017),
and  metal-poor
stars from Howes et al. (2015, 2016), Casey \& Schlaufman (2016) and
Koch et al. (2016), and   for thick
disk stars data from Ishigaki et al. (2013) and Nissen et al. (2011).

A high Zn abundance is found for bulge metal-poor stars in the range 
-3.0$\simless$[Fe/H]$\simless$-0.8.
This behaviour is similar to that previously reported
 in metal-poor  halo  and disk 
stars (e.g. Sneden et al. 1991;
Nissen \& Schuster 2011). In all samples 
[Zn/Fe] decreases with increasing metallicity, 
reaching a solar value at [Fe/H]$>$-0.4.

 The nucleosynthesis taking place in  hypernovae is needed to reproduce
this Zn enhancement in metal-poor stars, as proposed by Umeda \& Nomoto (2005), 
 and Nomoto et al. (2013, and references therein). The contribution
in Zn by hypernovae in the
chemical evolution models by Barbuy et al. (2015) proved to be needed
to reproduce the data (see their Fig. 12).
As for the present results [Zn/Fe] is enhanced in the metal-poor
clusters and decreases with metallicity, following the literature data.
The exception is the globular cluster HP~1, showing
low [Zn/Fe] at its metallicity of [Fe/H]$\sim$-1.0. 
A further inspection of this cluster would be of great
interest, given that it has characteristics of being very old,
and could reveal particularities due to its early formation.
 
%[Zn/Fe]$\sim$+0.3 at -1.3$\simless$[Fe/H]$\simless$-0.5, and reaches
%it decreases
% with increasing metallicity, reaching the solar ratio at around

Figure \ref{plotzn} compares [Zn/Fe] vs. [Fe/H] for bulge stars
 with chemodynamical evolution models 
of the Galactic bulge by Barbuy et al. (2015), further described in
 Fria\c ca \& Barbuy (2017).
% A  massive dark halo of mass 
%$M_{H}$= 1.3$\times$10$^{10}$ M$_{\odot}$ is assumed, containing
% baryonic gas of mass  2$\times$10$^9$ M$_{\odot}$ that is turned into stars. 
%For the Galactic bulge, FB17 deduced a best
%fit, from oxygen abundances, of a specific star formation 
%($\nu_{\rm SF}=$ 0.5 Gyr$^{-1}$).
%Barbuy et al. (2015), and FB17 computed the chemical evolution of 
%O, Mg, Si, Ca, Ti and Zn,
%adopting in particular the hypernovae contribution to explain Zn 
%abundances in metal-poor stars.
The hypernovae yields are suitable for metallicities more
metal-poor than [Fe/H$\simless$-2.0, as adopted in these models
(Barbuy et al. 2015; Fria\c ca \& Barbuy 2017; da Silveira 2017).
For this reason these models in the range -2.0$<$[Fe/H]$<$-1.0
are interrupted.
Models by Kobayashi et al. (2006) taking into
account hypernovae also reproduce well the Zn behaviour.

For disk stars with [Fe/H]$>$0.0, 
Reddy et al. (2003, 2006) obtained [Zn/Fe]$\sim$0.0,
Bensby et al. (2003, 2005) found [Zn/Fe] essentially constant, whereas 
Allende-Prieto et al. (2004) found increasing [Zn/Fe] with
increasing metallicity.

The Bensby et al. (2013, 2017) results for microlensed dwarf bulge stars also 
give a solar [Zn/Fe] at all metallicities, differently from 
Barbuy et al. (2015), where [Zn/Fe] decreases sharply at the high
metallicity end. The present results for the metal-rich clusters also
appear to decrease with increasing metallicity, despite some spread.
 This decrease implies
the action of SNIa, and could be an evidence of differences
in the chemical enrichment of bulge giants
and a thick disk sample.
It is interesting that Duffau et al. (2017) also found decreasing
[Zn/Fe] for red giants, and constant [Zn/Fe] for dwarfs, at the 
supersolar metallicities.
They interpreted this discrepancy in terms of stellar populations, 
i.e. that their red giants should be younger than the dwarfs, and
for this reason, to contain Fe enriched from SNIa.
The age explanation does not fit 
the present data, because our sample consists of old globular clusters. Stars
in NGC~6528 have subsolar [Zn/Fe], whereas NGC~6553 has
[Zn/Fe]$\sim$+0.3 for one star, and subsolar in the other star. 
 In particular, at its location in the Galaxy, NGC 6553 has kinematical 
characteristics compatible with bulge or disk stars (Zoccali et al. 2001b), 
whereas NGC~6528 is located in the bulge, so that they might be different
from each other. In conclusion, there seems to be a trend
to have decreasing Zn-to-Fe with increasing metallicity, despite it not being
clear for NGC~6553.
Another aspect is the suggestion by Grisoni et al. (2017) that the local
metal-rich thick disk consists of stars having migrated from the inner
regions of the Galaxy. Recio-Blanco et al. (2017) has also
advanced a possibility of this population corresponding to
a dwarf galaxy that previously merged with the Milky Way in the solar
vicinity. A question that comes to mind is if
it would be possible that the metal-rich bulge stars either by
Barbuy et al. (2015), or those by Bensby et al. (2017) 
correspond to the alpha-enhanced
thick disk by Grisoni et al. (2017), and for testing this it would
be of interest to derive Zn abundances in these metal-rich thick disk stars.

 Sk\'ulad\'ottir et al. (2017) derived Zn abundances 
in stars of the
dwarf galaxy Sculptor. They also find [Zn/Fe] decreasing with increasing
metallicities, and verified that the same occurs with other dwarf galaxies
studied in the literature such as Sagittarius, Sextans, Draco and Ursa Minor.
The authors suggest that it 
 is more naturally explained by the enrichment of Fe, and no
Zn enrichment from SNIa, therefore a behaviour similar to that of 
alpha-elements, although other less likely possibilities are discussed.

Finally, a general comment is that there is a trend for the cluster stars to be deficient relative to field bulge stars
for Sc, V, and Zn. In particular at the metal-rich end, it could be attributed to noise in the spectra. For
the metal-poor clusters,  on the other hand, further inspection would be of great interest,
because it could have an impact in the interpretation of enrichment of these globular clusters.

\begin{table*}[!]
%\begin{landscape}
\caption{ Line-by-line abundance ratios of Sc, V, Mn, Cu, Zn for the sample.}             
\label{abonds} 
\scalefont{0.70}     
\centering          
%\rotatebox{90}{
\begin{tabular}{lcccccccccccccccccc}     % 12 columns 
\noalign{\vskip 0.1cm}
\noalign{\hrule\vskip 0.1cm}
\noalign{\vskip 0.1cm}    
& & \multicolumn{5}{c}{\hbox{\bf 47 Tucanae}} & \multicolumn{4}{c}{\hbox{\bf NGC 6553}} & \multicolumn{3}{c}{\hbox{\bf NGC 6528}} & \multicolumn{2}{c}{\hbox{\bf HP~1}}  \\
\noalign{\vskip 0.1cm} 
\hline\hline    
\noalign{\smallskip}
Line &$\lambda$({\rm \AA}) & M8 & M11 & M12 & M21 & M25 & II-64 & II-85 & III-8 & 267092 & I-18 & I-36 & I-42 & 2 & 3 & \\      
\noalign{\vskip 0.1cm}
\noalign{\hrule\vskip 0.1cm}
%                     M8      M11      M12      M21      M25      II-64    II85     III8     III9     I-18     I-36     I-42     2         3
ScI  & 5671.805  & + 0.00 & + 0.03 & + 0.05 & + 0.00 &$-$0.10 &$-$0.30 &$-$0.30 &$-$0.30 &$-$0.15 &$-$0.30 &$-$0.30 & + 0.00 &---      & + 0.00 & \\
ScI  & 5686.826  & + 0.00 & + 0.00 & + 0.10 &---     & + 0.00 &$-$0.30 & + 0.00 &$-$0.30 &$-$0.25 &$-$0.30 &$-$0.50 & + 0.00 & + 0.00  &---     & \\
ScI  & 6210.676  &$-$0.25 &$-$0.30 &$-$0.10 &---     &$-$0.20 &$-$0.30 &---     &$-$0.30 &$-$0.30 &$-$0.50 &$-$0.60 &$-$0.30 & + 0.00  & + 0.00 & \\
ScII & 5526.790  & + 0.00 &$-$0.30 & + 0.05 & + 0.00 & + 0.00 &$-$0.30 & + 0.00 &$-$0.30 & + 0.00 &$-$0.40 &$-$0.30 &$-$0.30 &$-$0.15  & + 0.00 & \\
ScII & 5552.224  & + 0.00 & + 0.00 &---     &---     & + 0.10 & + 0.00 & + 0.30 & + 0.00 &---     &---     &---     &---     & ---     & + 0.00 & \\ 
ScII & 5657.896  & + 0.20 & + 0.15 & + 0.30 & + 0.00 & + 0.30 & + 0.00 & + 0.30 & + 0.00 & + 0.30 &$-$0.30 & + 0.00 & + 0.00 & + 0.00  & + 0.00 & \\
ScII & 5684.202  & + 0.30 & + 0.05 & + 0.30 & + 0.00 & + 0.30 &$-$0.10 & + 0.30 &$-$0.25 & + 0.00 &$-$0.30 &$-$0.30 & + 0.00 & + 0.00  & + 0.00 & \\
ScII & 6245.637  & + 0.10 & + 0.00 & + 0.00 & + 0.00 & + 0.25 & + 0.00 & + 0.10 &$-$0.30 & + 0.00 &$-$0.30 &$-$0.30 &---     & + 0.00  & + 0.00 & \\  
ScII & 6300.698  & + 0.15 & + 0.00 & + 0.10 &---     & + 0.00 & + 0.00 & + 0.25 &$-$0.15 &---     &$-$0.30 &$-$0.40 & + 0.00 & + 0.00  & + 0.30 & \\   
ScII & 6320.851  & + 0.15 & + 0.10 & + 0.25 & + 0.00 & + 0.20 & + 0.00 &---     & + 0.00 & + 0.00 &$-$0.30 &$-$0.30 & + 0.00 & + 0.00  & + 0.00 & \\
ScII & 6604.601  & + 0.15 & + 0.00 & + 0.05 &$-$0.10 & + 0.00 &$-$0.15 &---     &$-$0.30 & + 0.00 &$-$0.35 &$-$0.45 &$-$0.30 &$-$0.20  &$-$0.15 & \\  
VI   & 5703.560  & + 0.10 & + 0.00 & + 0.10 &$-$0.20 & + 0.00 & + 0.00 & + 0.25 &$-$0.25 & + 0.00 &$-$0.12 &$-$0.10 & + 0.00 &$-$0.30   & + 0.00 & \\ 
VI   & 6081.440  &$-$0.10 & + 0.00 &$-$0.05 &$-$0.30 &$-$0.10 &$-$0.10 & ---    &$-$0.10 &$-$0.15 &$-$0.30 &$-$0.15 & + 0.00 &$-$0.30   &$-$0.25 & \\  
VI   & 6090.220  & + 0.00 & + 0.00 & + 0.15 &$-$0.05 & + 0.00 &$-$0.25 & + 0.00 &$-$0.15 &$-$0.15 &$-$0.25 &$-$0.30 & + 0.00 &$-$0.30   &$-$0.10  & \\  
VI   & 6119.520  & + 0.00 & + 0.05 & + 0.12 &$-$0.05 & + 0.00 & + 0.00 & + 0.00 & + 0.00 &$-$0.15 &$-$0.10 &$-$0.30 & + 0.00 &---      &$-$0.20  & \\ 
VI   & 6199.190  &$-$0.15 & + 0.00 &$-$0.10 &$-$0.10 & + 0.00 &$-$0.30 &---     &$-$0.10 &$-$0.30 &$-$0.30 &$-$0.30 & + 0.00 &$-$0.10   &$-$0.12 & \\ 
VI   & 6243.100  &$-$0.10 & + 0.00 & + 0.00 & + 0.00 & + 0.00 &$-$0.35 &---     &$-$0.25 &$-$0.15 &$-$0.20 &$-$0.30 & + 0.00 &$-$0.15  &$-$0.10  & \\  
VI   & 6251.820  &$-$0.10 & + 0.00 & + 0.00 &$-$0.30 & + 0.00 & + 0.00 &---     &$-$0.30 &$-$0.15 &$-$0.30 &$-$0.30 & + 0.00 & -0.30    &$-$0.05 & \\   
VI   & 6274.650  &$-$0.10 & + 0.00 &$-$0.10 &$-$0.15 & + 0.00 &$-$0.15 &---     &$-$0.30 &$-$0.15 &$-$0.20 &$-$0.15 & + 0.00 & -0.10    &---     & \\  
VI   & 6285.160  & + 0.00 &$-$0.05 & + 0.00 & + 0.00 & + 0.00 &---     &---     &---     &---     &$-$0.30 &---     &---     &---      &---     & \\
MnI  & 6013.513  &$-$0.10 &$-$0.05 & + 0.00 &$-$0.40 &$-$0.30 &$-$0.30 & + 0.00 &$-$0.45 &$-$0.30 &$-$0.30 & + 0.00 & + 0.00 &$-$0.60   &$-$0.50  & \\ 
MnI  & 6016.640  &$-$0.30 &$-$0.30 &$-$0.30 &$-$0.40 &$-$0.30 &$-$0.30 & ---     &$-$0.45 &$-$0.30 &$-$0.30 &$-$0.30 & ---     &$-$0.60   &$-$0.55 & \\
MnI  & 6021.800  &$-$0.20 &$-$0.30 &$-$0.30 &$-$0.50 &$-$0.30 &$-$0.30 & + 0.00 &$-$0.40 & ---    &$-$0.30 &$-$0.30 &$-$0.20 &$-$0.50   &$-$0.60  & \\ 
CuI  & 5105.537  & + 0.20 & + 0.00 & + 0.00 &$-$0.30 & + 0.00 &$-$0.30 &$-$0.10 &$-$0.30 & + 0.00 &$-$0.30 &$-$0.30 &$-$0.60 &---      &$-$0.60  & \\
CuI  & 5218.197  & + 0.30 & + 0.30 & + 0.30 & + 0.00 & + 0.00 &   ---  &  ---   &$-$0.15 & + 0.30 &$-$0.15 &$-$0.45 &   ---  &$-$1.00     &$-$0.30  & \\
ZnI  & 4810.529  & + 0.25 & +0.00 & + 0.02 & + 0.05  & + 0.25 &  ---  &  ---   & $-$0.05  & + 0.30 & +0.00  &$-$0.30 &   ---  & ---  & +0.00   & \\
ZnI  & 6362.339  & + 0.30 & +0.05 & + 0.30 & + 0.05  & + 0.25 & --- &---     &$-$0.00 &+ 0.30 & + 0.00 &$-$0.30 &---     &---      &+0.00    & \\
%VI & 4831.640 B        &--     &+0.3 &--- &-0.3 &+0.3 & & & & & & &  \\ % ELIMINAR=BLENDS
%VI & 4851.480=lar. &--     &+0.3 &--- &+0.3 &+0.3 & & & & & & &  \\ % ELIMINAR=BLENDSy 
%VI & 4875.480 i      &--     &blend &--- &-0.3  & & & & & & & &  \\  % blend: varia muito pouco
%VI & 4932.030gf &--     &gfruim &-- &-0.3  & & & & & & & &  \\ %gf ruim, e blend dominando
%VI & 5627.640    &+0.2   &+0.2   &+0.3 & 0.0 &0.0 & & & & & & &  \\  % blend domina
%VI & 5670.850    &-0.1   &-0.2   &-0.1 & 0.0 &-0.30 & & & & & & &  \\   % blend domina
%VI & 6216.370    & --    &blend  &--- &-0.5 &-0.6 & & & & & & &  \\blend, varia pouco e gf errado 
\hline\hline    
\noalign{\smallskip}
\noalign{\vskip 0.1cm} 
& & \multicolumn{5}{c}{\hbox{\bf HP-1}} & \multicolumn{4}{c}{\hbox{\bf NGC 6522}} & \multicolumn{4}{c}{\hbox{\bf NGC 6558}}  \\      
\noalign{\vskip 0.1cm}
\noalign{\hrule\vskip 0.1cm}
\noalign{\vskip 0.1cm}    
Line &$\lambda$({\rm \AA}) &2115 & 2461 & 2939  & 3514 & 5037    & 5485   &B-107 &B-122 &B-128 & B-130 & 283   & 364 & 1072 & 1160 &\\
\noalign{\vskip 0.1cm}
\noalign{\hrule\vskip 0.1cm}
\noalign{\vskip 0.1cm} 
%                    2115      2461       2939     3514        5037       5485      B-107    B-122     B-128      B-130     283        364     1072   1160
ScI & 5671.805    &$-$0.10   &---      &+0.12     &+0.05     &$-$0.12   &---      &---     &+0.00     &$-$0.30  &---      &---      &---     &---     &---     & \\    
ScI & 5686.826    &+0.10     &---      &+0.25     &---       &--        &---      &---     &---       &---      &---      &---      &---     &---     &---     & \\   
ScI & 6210.676    &+0.15     &---      &+0.15     &---       &$-$0.05   &---      &---     &---       &$-$0.05  &---      &---      &---     &---     &---     & \\    
ScII & 5526.790   &$-$0.15   &+0.00    &+0.00     &$-$0.30   &$-$0.18   &$-$0.30  &+0.00   &+0.00     &+0.00    &$-$0.05  &$-$0.30  &$-$0.15 &$-$0.15 &$-$0.15 & \\ 
ScII & 5552.224   &+0.30     &---      &---       &---       &---       &---      &$-$0.30 &---       &---      &+0.30    &---      &---     &---     &---     & \\    
ScII & 5657.896   &+0.00     &+0.00    &+0.30     &+0.00     &+0.10     &+0.15    &+0.00   &+0.05     &+0.00    &+0.00    &$-$0.3   &$-$0.15 &+0.30   &$-$0.10 & \\  
ScII & 5684.202   &+0.00     &+0.00    &+0.00     &$-$0.15   &+0.00     &+0.00    &$-$0.15 &+0.10     &+0.00    &+0.00    &+0.00    &$-$0.10 &+0.30   &---     & \\    
ScII & 6245.637   &+0.10     &+0.00    &+0.30     &+0.00     &+0.12     &+0.03    &$-$0.10 &+0.15     &+0.00    &+0.00    &$-$0.03  &---   &$-$0.30 &$-$0.20 & \\   
ScII & 6300.698   &+0.30     &+0.00    &---       &+0.30     &+0.30     &---      &---     &+0.00     &+0.00    &+0.00    &---      &+0.60   &---     &+0.00   & \\    
ScII & 6320.851   &+0.00     &+0.15    &+0.30     &+0.30     &+0.15     &+0.00    &$-$0.15 &+0.15     &+0.30    &+0.15    &+0.00    &---     &+0.15   &$-$0.30 & \\    
ScII & 6604.601   &+0.00     &+0.00    &+0.30     &+0.30     &+0.00     &$-$0.05  &$-$0.30 &+0.05     &+0.15    &+0.00    &+0.00    &---     &+0.10   &$-$0.05 & \\ 
VI & 5703.560     &+0.00     &$-$0.10   &+0.00     &+0.00     &+0.00     &+0.00    &---     &$-$0.20    &$-$0.10  &$-$0.05  &$-$0.05  &+0.00   &$-$0.30 &+0.00   & \\    
VI & 6081.440     &+0.00     &---      &+0.00     &+0.02     &+0.00     &+0.00    &---     &$-$0.25   &$-$0.10  &$-$0.15  &+0.05    &+0.30     &+0.00   &+0.00   & \\     
VI & 6090.220     &+0.00     &+0.00    &+0.00     &+0.00     &+0.00     &---      &+0.00   &+0.00     &$-$0.15  &$-$0.1   &$-$0.05  &+0.00   &$-$0.15 &$-$0.10 & \\    
VI & 6119.520     &+0.05     &+0.00    &+0.00     &+0.00     &+0.00     &---      &+0.00   &+0.00     &$-$0.15  &+0.0     &$-$0.05  &+0.00   &+0.00   &$-$0.10 & \\    
VI & 6199.190     &+0.00     &$-$0.05  &+0.00     &+0.05     &+0.00     &---      &---     &+0.00     &+0.00    &+0.0     &+0.00    &---     &$-$0.30 &+0.05   & \\    
VI & 6243.100     &+0.00     &+0.00    &+0.00     &+0.05     &+0.00     &+0.00    &$-$0.15 &+0.00     &$-$0.10  &+0.0     &+0.00    &---     &$-$0.20 &+0.00   & \\     
VI & 6251.820     &+0.00     &+0.00    &+0.05     &+0.00     &+0.00     &+0.00    &---     &$-$0.10   &+0.0     &$-$0.05  &+0.00    &+0.15   &$-$0.15 &+0.10   & \\     
VI & 6274.650     &---       &+0.00    &+0.00     &+0.00     &+0.00     &---      &$-$0.10 &+0.00     &$-$0.25  &$-$0.15  &---      &---     &$-$0.15 &+0.00   & \\     
VI & 6285.160     &---       &---      &---       &---       &---       &---      &---     &---       &---      &---      &---      & ---    & ---    & ---    & \\
MnI & 6013.513    &$-$0.30   &$-$0.3   &$-$0.30   &$-$0.30   &$-$0.35   &$-$0.40  &$-$0.55 &$-$0.50   &$-$0.55  &$-$0.50  &$-$0.50  &$-$0.25 &$-$0.60 &$-$0.50 & \\   
MnI  & 6016.640   &$-$0.30   &$-$0.3   &$-$0.30   &$-$0.30   &$-$0.35   &$-$0.55  &$-$0.55 &$-$0.60   &$-$0.60  &$-$0.60  &    ---  &$-$0.32 &$-$0.60 &$-$0.35 & \\     
MnI  & 6021.800   &$-$0.30   &$-$0.2   &$-$0.30   &$-$0.50   &$-$0.40   &$-$0.50  &$-$0.55 &$-$0.60   &$-$0.60  &$-$0.60  &$-$0.40  &$-$0.30 &$-$0.65 &$-$0.50 & \\  
CuI  & 5105.537   &$-$0.80   &$-$0.8   &$-$1.00   &$-$1.10   &$-$1.00   &$-$1.00  &$-$0.60 &$-$0.60   &$-$0.50  &$-$0.60  &$-$0.70  &$-$0.60 &$-$0.80 &$-$0.70 & \\   
CuI  & 5218.197   &$-$0.30   &$-$0.1   &+0.00     &$-$0.60   &$-$0.30   &---      &$-$0.30 &$-$0.30   &$-$0.30  &$-$0.30  &$-$0.30  &---     &$-$0.1  &---     & \\     
ZnI  & 4810.529   &---       &$-$0.30    &$-$0.60   &$-$0.10   &$-$0.30   &$-$0.30  &+0.05   &$-$0.25   &+0.00    &$-$0.30  &$-$0.10    &---     &+0.00   &+0.20   & \\     
ZnI  & 6362.339   &---       &---      &---       &---       &---       &---      &+0.15   &+0.15     &+0.30    & $-$  0.15  &---      &---     &---     &---     & \\    
\hline                   
\hline                  
\end{tabular}
\end{table*}

\begin{table*}[!]
\caption{Mean abundances of Sc, V, Mn, Cu, Zn for the sample.} 
\centering 
\scalefont{0.70}  
\begin{tabular}{l@{} c c c c c c c c c c c c c } 
%\toprule
 \multicolumn{14}{c}{Mean abundances} \\ 
%\cmidrule(l){2-6} 
\hline
Stars & [Fe/H] & [ScI/Fe] &$\delta_{[ScI/Fe]}$ & [ScII/Fe] &$\delta_{[ScII/Fe]}$ & [VI/Fe] &$\delta_{[VI/Fe]}$ & [MnI/Fe] &$\delta_{[MnI/Fe]}$ & [CuI/Fe] &$\delta_{[CuI/Fe]}$ & [ZnI/Fe] &$\delta_{[ZnI/Fe]}$ \\ 
\hline
\multicolumn{8}{c}{47Tuc} \\
\hline
M8  & $-$0.64  & $-$0.08 & 0.12  & +0.13  & 0.08   & $-$0.05  & 0.07     & $-$0.20 & 0.08  & +0.25 & 0.04  
& +0.28  & 0.03  \\ 
M11 & $-$0.62  & $-$0.09 & 0.15   & +0.04 & 0.06   &  +0.00    & 0.02   & $-$0.22  & 0.12   & +0.12 & 0.12  & +0.03  & 0.03  \\ 
M12 & $-$0.66  &  +0.02  & 0.08 & +0.15   & 0.12   &  +0.01    & 0.09  & $-$0.20   & 0.14 & +0.15  & 0.12   & +0.16 & 0.14  \\ 
M21 & $-$0.80  &  +0.00  & 0.08 & $-$0.02 & 0.04   &  $-$0.13  & 0.11   & $-$0.43  & 0.05   & $-$0.15& 0.12  & +0.03 & 0.03  \\ 
M25 & $-$0.66  & $-$0.10 & ?    &+0.14    & 0.12   & $-$0.01   & 0.03   & $-$0.30 & 0.00 & +0.00  & 0.00   
& +0.25 & 0.00   \\ 
\hline 
\multicolumn{8}{c}{NGC 6553} \\
\hline
II-64 & $-$0.20 & $-$0.30 & 0.00 & $-$0.06 &0.06  & $-$0.14 & 0.13    & $-$0.30 & 0.00   &+0.00 & 0.24   
&  --- & --- \\
II-85 & $-$0.26  & $-$0.15 & 0.15  & +0.21 &0.08   & +0.08  & 0.12    &+0.00    & 0.00   &+0.10  & 0.16 
 &  --- & --- \\ 
III-8 & $-$0.17 & $-$0.30 & 0.00 & $-$0.16 &0.13 & $-$0.18  & 0.10   &$-$0.43   & 0.02   &+0.15  & 0.24
  & $-$0.15 & 0.03 \\ 
III-9 & $-$0.22 & $-$0.23 & 0.06 &+0.05    &0.12  & $-$0.15  & 0.08   & $-$0.30 & 0.00   &+0.30  & 0.12 
  & +0.30  & 0.00 \\ 
\hline
\multicolumn{8}{c}{NGC 6528} \\
\hline
I-18  & $-$0.08 & $-$0.37   & 0.09 &  $-$0.32 &0.02  & $-$0.23 & 0.08   & $-$0.30 & 0.00 &$-$0.15 & 0.12
 &  $-$0.15  & 0.08  \\ 
I-36  & $-$0.11  &  $-$0.47 & 0.12 &  $-$0.29 &0.14 &  $-$0.24 & 0.08   & $-$0.20 & 0.14 &+0.00 & 0.24 
 &  $-$0.37 & 0.03   \\ 
I-42  & $-$0.11  &  $-$0.10 & 0.14 &  $-$0.10 &0.12 & +0.00    & 0.00   & $-$0.10 & 0.10 &$-$0.30 & 0.24 
 & ---   & ?   \\
\hline
\multicolumn{8}{c}{HP 1} \\
\hline 
2    & $-$1.00  &  +0.00 & ? &  $-$0.05  &0.07  &  $-$0.22   & 0.09  & $-$0.57   & 0.05  & $-$1.00 & ? 
&  +0.30 & ? \\ 
3   & $-$0.97 &  +0.00  & ?  &   $-$0.02 &0.12   &   $-$0.12 & 0.08   & $-$0.55  & 0.04  & $-$0.45 & 0.12 
& +0.00  & ?\\ 
2115 & $-$1.00 &  +0.05 & 0.11   & +0.07  &0.13  &  +0.01    & 0.02   & $-$0.30  & 0.00  & $-$0.55 & 0.20 
 & ---  & ---    \\ 
2461  & $-$1.11 &   ---  & ---   &  +0.02 &0.06   &  $-$0.02  & 0.04  & $-$0.27  & 0.05  & $-$0.45 & 0.29
& $-$0.30 & ?  \\ 
2939 & $-$1.07 &  +0.17 & 0.06   & +0.20  &0.12   &  +0.01    & 0.02  & $-$0.30  & 0.00   & $-$0.50 & 0.41 
 & $-$0.60 & ? \\ 
3514  & $-$1.18 &  +0.05 & ?     & +0.06  &0.18   &  +0.02    & 0.02  & $-$0.37  & 0.09   & $-$0.85 & 0.20 
 & $-$0.10 & ? \\ 
5037 & $-$1.00 & $-$0.09 & 0.04  &  +0.08  &0.10 &  +0.00     & 0.00  & $-$0.37  & 0.02   & $-$0.65 & 0.29
 & $-$0.30 & ? \\ 
5485 & $-$1.18 &   ---  & ---    &  $-$0.03 &0.07  &  +0.00   & 0.00  & $-$0.48  & 0.06   & $-$1.00 & ? 
& $-$0.30 & ? \\
\hline
\multicolumn{8}{c}{NGC 6522} \\
\hline
B-107  & $-$1.13 & ---  & ---     & $-$0.11 &0.12  &  $-$0.06 & 0.06 &  $-$0.55 & 0.00 & $-$0.45 & 0.12
 &  +0.10  & 0.05   \\ 
B-122  & $-$0.81 & +0.00 & ?      &  +0.07  &0.06  &  $-$0.07 & 0.10  &  $-$0.57 & 0.05 & $-$0.45 & 0.12
  &  $-$0.05 & 0.20  \\ 
B-128  & $-$0.82 & $-$0.18 & 0.13 &  +0.06  &0.11  &  $-$0.12 & 0.07  &  $-$0.58& 0.02 & $-$0.40 & 0.08  
 &  +0.15  & 0.15    \\ 
B-130  & $-$1.04  &  --- &  ---   &  +0.05  &0.11  &  $-$0.06 & 0.06  & $-$0.57 & 0.05 & $-$0.45 & 0.12
 &  $-$0.23  & 0.08   \\
\hline
\multicolumn{8}{c}{NGC6558} \\
\hline
283 & $-$1.15   & --- &  --- & $-$0.16 &0.16  &  $-$0.01 & 0.03 & $-$0.45 & 0.05 & $-$0.50 & 0.16 
& $-$0.10 & ?  \\ 
364 & $-$1.15   & --- &  --- & +0.05 &0.32  &  +0.09   & 0.12 & $-$0.29 & 0.03 & $-$0.60 & ?
 &---   & ---     \\ 
1072  & $-$1.23 & --- &  --- & +0.07   &0.22 &  $-$0.16  & 0.11 & $-$0.62 & 0.07 & $-$0.45  & 0.29
& +0.00  & ?    \\ 
1160  & $-$1.04 & --- &  --- & $-$0.13 &0.11  &  $-$0.01 & 0.06 & $-$0.45 & 0.02 & $-$0.70 & ? 
&+0.20   & ?    \\ 
\hline 
\hline 
%\bottomrule 
\end{tabular}
\label{meanabundance} 
\end{table*}

\section{Summary}

 Globular clusters of the Galactic bulge should trace the 
formation process of the central parts of the Galaxy. They are also
tracers of the  older stellar populations in the bulge.
Chemical tagging is a next big step for the understanding of the
Milky Way formation. The iron-peak elements have been little studied so far,
but their study should help understanding: 
 a) nucleosynthesis of these elements is
complex and observations can help constraining their formation;
b) Sc and V appear to vary in lockstep with Fe in the present sample, 
but Sc has been found to be alpha-like in thick disk and halo stars,
and further studies are needed; c) Mn is deficient in metal-poor stars,
and steadily increases with metallicity due to enrichment from SNIa; 
d) Cu shows a secondary-like behavior, in principle indicating its
production in a weak s-process in massive stars; d) Zn is alpha-like
in halo and thick disk stars, and also in the bulge, as concerns metal-poor
stars. For metal-rich stars there is a controversy as to whether it
decreases with increasing metallicity, or if [Zn/Fe] keeps a solar value.

We have derived abundances of the iron-peak elements Sc, V, Mn, Cu, and Zn,
in 23 red giants in the 5 bulge globular clusters
NGC~6553, NGC~6528, HP~1, NGC~6522, NGC~6558, and 5 red giants in the
reference inner halo/thick disk cluster 47 Tucanae.
The work was based on FLAMES-UVES high-resolution spectra obtained at the
VLT UT2 telescope.

Vanadium varies in lockstep with Fe. Sc behaves closely as V, not showing
a clear enhancement, that was previouly suggested by Nissen et al. (2000) for
alpha-rich halo and thick disk stars. Both [Sc,V/Fe] seem to decrease
with increasing metallicity at the high metallicity end.

 Mn is deficient in metal-poor stars and increases to solar values for the more
metal-rich stars, indicating that it is underproduced in massive stars,
and later produced in SNIa.

 Cu show a behaviour as secondary elements, having low values
at low metallicities, and steadily increasing with increasing metallicity,
 indicating an enrichment through a weak-s process in massive stars, and
in good agreement with chemical evolution models.

Zn is enhanced in metal-poor stars, likewise an alpha-element, 
and decreases with
increasing metallicity. At the high metallicity end the
behaviour of the present data is different from that found by
Bensby et al. (2013, 2017), that show solar ratios at the high metallicities.
This could be a discriminator of having the contribution of SNIa or not.
This is made less clear given the difference in [Zn/Fe] found by Duffau et al.
(2017) for red giants and dwarfs.   Sk\'ulad\'ottir et al. (2017) found
[Zn/Fe] decreasing with metallicity for the dwarf galaxy Sculptor, and
point out that the same is true for other dwarf galaxies.
 It is of great interest to pursue 
abundance derivation of iron-peak elements, and in particular
Sc in all stellar populations, and Zn in bulge stars.

\begin{acknowledgements}

We acknowledge grants and fellowships from FAPESP, CNPq and CAPES.
%This publication makes use of data products from the Two Micron All Sky
%Survey, which is a joint project of the University of Massachusetts and the
%Infrared Processing and Analysis Caltech, funded by the NASA and the NSF.

\end{acknowledgements}

%===============================================================================
%                           References
%===============================================================================

%\section{References}

%\section*{Appendix A: Atomic data}

\begin{appendix}

\section{Atomic data}

In Table \ref{sclines1} are given the hyperfine structure constants of Sc, V and Cu lines.

\begin{table*}
\begin{flushleft}
\scalefont{0.87}
\caption{Atomic constants for ScI and ScII used to compute hyperfine structure:
A and B constants from Mansour, N. B., Dinneen, T. P., Young, L. 1989, NIMPB, 40-252M , Villemoes et al. 1992, PhRvA, 45-6241V for ScII and Biehl (1976) for ScI.  
For VI the A and B constants are from 
1 UBDE      Unkel, P., Buch, P., Dembczynski, J., Ertmer, W.. and Johan, U. 1989, Z. Phys. D 11, 259-271.
2 CPGC      Childs, W.J., Poulsen, O., Goodman, L.S., and Crosswhite, H. 1979, Phys. Rev. A 19, 168-176.
3 PBAG      Palmeri, P., Biemont, E., Aboussaid, A,, and Godefroid, M. 1995, J.Phys.B 28, 3741-3752.
 4 CBFG      Cochrane, E.C.A., Benton, D.M., Foreset, D.H., and Griffith, J.A.R. 1998, J.Phys.B 31, 2203-2213.
5 LGB       Lefebre, P-H, Garnir, H-P, Biemont, E 2002, Physica Scripta 66, 363-366.
B constants not available in the literature are assumed as null.}             
\label{sclines1}      
\centering          
\begin{tabular}{lc@{}c@{}c@{}c@{}c@{}c@{}c@{}c@{}c@{}c@{}c@{}c@{}ccc} 
\noalign{\smallskip}
\hline\hline    
\noalign{\smallskip}
\noalign{\vskip 0.1cm} 
species & $\lambda$ ({\rm \AA}) & \phantom{-}Lower level 
& \phantom{-}J &\phantom{-}A(mK)& \phantom{-}A(MHz) 
&\phantom{-}B(mK)& \phantom{-}B(MHz) & \phantom{-}Upper level 
& \phantom{-}J &\phantom{-}A(mK)& 
\phantom{-}A(MHz) &\phantom{-}B(mK)& \phantom{-}B(MHz)  \\
\noalign{\vskip 0.1cm}
\noalign{\hrule\vskip 0.1cm}
\noalign{\vskip 0.1cm}
$^{45}$ScI & 5671.805 & 3d$^2$($^3$F)4s $4$F &  9/2 &+9.5 & 284.8029 & -0.4& -11.9917 &  3d$^2$($^3$F)4p $^4$G & 11/2 & +1.5 & 44.9689  & --- & ---&   \\
$^{45}$ScI & 5686.826 &  3d$^2$($^3$F)4s 4F &  7/2 &+8.3& 248.8278 & -0.3& -8.9938 & 3d$^2$($^3$F)4p 4G & 7/2 & +4.9 & 146.8983 & --- & ---&  \\
%$^{45}$ScI & 5717.30 &  3d$^2$(3F)4s 4F &  7/2 &+8.3& 248.8278 & -0.3& -8.9938 &  3d$^2$(3F)4p 4G & 7/2 & +4.9&  146.8983 & --- & --- &  \\
$^{45}$ScI & 6210.676 &  3d4s$^2$ 2D &  3/2 &+8.98& 269.2137 & -0.88& -26.3817 & 3d4s(1D)4p 2D & 3/2 & -11.5 & -344.7614 & --- & ---&  \\
\noalign{\vskip 0.1cm}
\noalign{\hrule\vskip 0.1cm}
\noalign{\vskip 0.1cm}
%$^{45}$ScII & 5239.813 & 3p$^6$4s$^2$ 1S & 0.0 &--- & --- & --- & --- & 3p$^6$3d4p 1P$^o$ & 1.0 & --- & --- & --- & --- & \\
%$^{45}$ScII & 5357.199 & 3p$^6$3d$^{2}$ 3P & 2.0 &---& -27.2 & --- & 26.0 & 3p$^6$3d4p 1P$^o$ & 1.0 & --- & --- & --- & --- & \\
$^{45}$ScII & 5526.790 &  3p$^6$3d$^{2}$ 1G &  4.0 & --- &M 135.232 & --- &M -63.44 &  3p$^6$3d4p 1F$^o$ & 3.0 & --- & 193.1 & --- & -65 &  \\
$^{45}$ScII & 5552.224 & 3p$^6$4s$^2$ 1S & 0.0 &--- & --- & --- & --- & 3p$^6$3d4p 3P$^o$ & 1.0 & --- & 258.0 & --- & 12.0 & \\
$^{45}$ScII & 5657.896 &  3p$^6$3d$^2$ 3P & 2.0 & --- &M -27.732 &--- &M 22.13 &  3p$^6$3d4p 3P$^o$ & 2.0 & --- & 105.6 & --- & -21 & \\
$^{45}$ScII & 5684.202 & 3p$^6$3d$^2$ 3P & 2.0 & --- & -27.2 & --- & 26.0 & 3p$^6$3d4p 3P$^o$ & 1.0 & --- & 258.0 & --- & 12.0 & \\
$^{45}$ScII & 6245.637 &  3p$^6$3d$^2$ 3P &  2.0 & --- & M -27.732 & --- & 22.13 &  3p$^6$3d4p 3D$^o$ & 3.0 & --- & 101.8 & --- & 24&  \\
$^{45}$ScII & 6300.698 & 3p$^6$3d$^2$ 3P & 2.0 & --- & -27.2 & --- & 26.0 & 3p$^6$3d4p 3D$^o$ & 2.0 & --- & 125.7 & --- & 6.0 & \\
$^{45}$ScII & 6320.84  & 3p$^6$3d$^2$ 3P & 1.0 & --- & -108.1 & --- & -13.0 & 3p$^6$3d4p 3D$^o$ & 1.0 & --- & 307.0 & --- & 1.0 & \\
$^{45}$ScII & 6604.601 &  3p$^6$3d$^2$ 1D & 2.0 & --- & 149.361 & --- & 7.818 &  3p$^6$3d4p 1D$^o$ & 2.0 & --- & 215.7 & --- & 18 & \\
%$^{45}$ScII & 4831.640   & 3d${^3}$4s${^2}$  a${^4}$F &  5/2 & --- & 321.238$^{1}$ &&   3.715$^{1}$ & & 3d${^3}$(${^4}$F)4s4p($^{3}$P$^{0}$) z${^4}$D${^°}$ &  5/2 && 611.07$^{3}$ && 0.0$^{3}$ &  \\
\noalign{\vskip 0.1cm}
\noalign{\hrule\vskip 0.1cm}
\noalign{\vskip 0.1cm}
%$^{51}$VI & 4851.480   & 3d${^3}$4s${^2}$    a${^4}$F &  3/2 &  560.062$^{1}$ &&   4.077$^{1}$ & & 3d$^{3}$($^{4}$F)4s4p($^{3}$P$^{0}$) z${^4}$D${^°}$ &  1/2 && -185.9$^{3}$ && 0.0$^{3}$ &  \\
%$^{51}$VI &4864.740   & 3d${^3}$4s${^2}$    a$^{4}$F &  5/2 &  321.238$^{1}$ &&   3.715$^{1}$ & & 3d$^{3}$($^{4}$F)4s4p($^{3}$P$^{0}$) z${^4}$D${^°}$ &  3/2 && 544.81$^{3}$ && 0.0$^{3}$ &  \\
%$^{51}$VI &4875.480   & 3d${^3}$4s${^2}$    a$^{4}$F &  7/2 &  249.748$^{1}$ &&   5.423$^{1}$ & & 3d$^{3}$($^{4}$F)4s4p($^{3}$P$^{0}$) z${^4}$D${^°}$ &  5/2 && 611.07$^{3}$ && 0.0$^{3}$ &  \\
%$^{51}$VI &4932.030   & 3d${^3}$4s${^2}$    a${^4}$P &  5/2 &  112.834$^{1}$ && -11.095$^{1}$ & & 3d${^3}$(${^4}$P)4s4p($^{3}$P$^{0}$) y${^4}$P${^°}$ &  3/2 &&   687.0$^{5}$ && 0.0$^{5}$ &  \\
%$^{51}$VI &5627.640   & 3d${^4}$(${^5}$D)4s a$^{4}$D &  7/2 & -160.172$^{1}$ &&  15.256$^{1}$ & & 3d$^{4}$($^{5}$)D)4p              y${^4}$D${^°}$ &  7/2 &&  -17.1$^{3}$ && 0.0$^{3}$ &  \\
%$^{51}$VI &5670.850   & 3d$^{4}$($^{5}$)D)4s a$^{4}$D &  7/2 & -160.172$^{1}$ &&  15.256$^{1}$ & & 3d$^{3}$($^{4}$F)4s4p($^{3}$P$^{0}$) z${^2}$G${^°}$ &  9/2 &&  94.64$^{3}$ && 0.0$^{3}$ &  \\
$^{51}$VI &5703.560   & 3d$^{4}$($^{5}$)D)4s a$^{4}$D &  3/2 &    7.558$^{1}$ &---&   2.075$^{1}$ &--- & 3d$^{4}$($^{5}$)D)4p              y${^4}$F${^°}$ &  5/2 &---&   216.0$^{5}$ &---& 0.0$^{5}$ &  \\
%$^{51}$VI &5727.030   & 3d$^{4}$($^{5}$)D)4s a$^{4}$D &  7/2 & -160.172$^{1}$ &---&  15.256$^{1}$ &--- & 3d$^{4}$($^{5}$)D)4p              y${^4}$F${^°}$ &  9/2 &---&   89.0$^{3}$ &---& 0.0$^{3}$ &  \\
$^{51}$VI &6081.440   & 3d$^{4}$($^{5}$)D)4s a$^{4}$D &  3/2 &    7.558$^{1}$ &---&   2.075$^{1}$ &--- & 3d$^{4}$($^{5}$)D)4p              z${^4}$P${^°}$ &  3/2 &---& -286.4$^{2}$ &---& -6.0$^{2}$ &  \\
$^{51}$VI &6090.220   & 3d$^{4}$($^{5}$)D)4s a$^{4}$D &  7/2 & -160.172$^{1}$ &---&  15.256$^{1}$ &--- & 3d$^{4}$($^{5}$)D)4p              z${^4}$P${^°}$ &  5/2 &---&  -89.8$^{2}$ &---& 8.0$^{2}$ &  \\
$^{51}$VI &6119.520   & 3d$^{4}$($^{5}$)D)4s a$^{4}$D &  5/2 & -143.367$^{1}$ &---&   1.067$^{1}$ &--- & 3d$^{4}$($^{5}$)D)4p              z${^4}$P${^°}$ &  3/2 &---& -286.4$^{2}$ &---& -6.0$^{2}$ &  \\
$^{51}$VI &6199.190   & 3d$^{4}$($^{5}$)D)4s a$^{6}$D &  7/2 &  382.368$^{1}$ &---&   2.220$^{1}$ &--- & 3d$^{3}$($^{4}$F)4s4p($^{3}$P$^{0}$) z${^6}$D${^°}$ &  9/2 &---& 503.46$^{4}$ &---& 3.3$^{4}$ &  \\
%$^{51}$VI &6216.370   & 3d$^{4}$($^{5}$)D)4s a$^{6}$D &  5/2 &  373.595$^{1}$ &&  -2.575$^{1}$ & ---& 3d$^{3}$($^{4}$F)4s4p($^{3}$P$^{0}$) z${^6}$D${^°}$ &  7/2 && 514.35$^{4}$ && -1.2$^{4}$ &  \\
$^{51}$VI &6243.100   & 3d$^{4}$($^{5}$)D)4s a$^{6}$D &  9/2 &  406.854$^{1}$ &---&  14.721$^{1}$ & ---& 3d$^{3}$($^{4}$F)4s4p($^{3}$P$^{0}$) z${^6}$D${^°}$ &  9/2 &---& 503.46$^{4}$ &---& 3.3$^{4}$ &  \\
$^{51}$VI &6251.820   & 3d$^{4}$($^{5}$)D)4s a$^{6}$D &  7/2 &  382.368$^{1}$ &---&   2.220$^{1}$ &--- & 3d$^{3}$($^{4}$F)4s4p($^{3}$P$^{0}$) z${^6}$D${^°}$ &  7/2 &---& 514.35$^{4}$ &---& -1.2$^{4}$ &  \\
$^{51}$VI &6274.650   & 3d$^{4}$($^{5}$)D)4s a$^{6}$D &  3/2 &  405.605$^{1}$ &---&  -8.060$^{1}$ & ---& 3d$^{3}$($^{4}$F)4s4p($^{3}$P$^{0}$) z${^6}$D${^°}$ &  1/2 &---& 939.94$^{4}$ &---& 0.0$^{4}$ &  \\
$^{51}$VI &6285.160   & 3d$^{4}$($^{5}$)D)4s a$^{6}$D &  5/2 &  373.595$^{1}$ &---&  -2.575$^{1}$ &--- & 3d$^{3}$($^{4}$F)4s4p($^{3}$P$^{0}$) z${^6}$D${^°}$ &  3/2 &---& 594.69$^{4}$ &---& -4.4$^{4}$ &  \\

\noalign{\vskip 0.1cm}
\noalign{\hrule\vskip 0.1cm}
\noalign{\vskip 0.1cm}

$^{63}$CuI & 5105.537 & 4p 2P [case e] & 1.5 &6.5& 194.865 &-0.96& -28.78 &  4s2 2D [case b] & 
2.5  & 24.97& 748.582 &6.20&  185.871   \\ 
    & 5218.197 & 4p 2P [case e]   & 1.5  &6.5& 194.865 &-0.96& -28.78  & 4d 2D [---] & 2.5 & 
0.0* & 0.0* & 0.0* & 0.0* \\
%    & 5782.127 & 4p 2P [case d]&0.5 &16.93& 507.549 &0.0*& 0.0* & 4s2 2D [case c]   & 1.5  &59.7& 1789.76 &0.14& 4.197  \\
\noalign{\vskip 0.1cm}
\noalign{\hrule\vskip 0.1cm}
\noalign{\vskip 0.1cm}
$^{65}$CuI & 5105.537 & 4p 2P [case e] & 1.5 &6.96& 208.66 &-0.86& -25.78 & 4s2 2D [case b] & 
2.5  & 26.79& 803.14 &5.81&  174.18  \\ 
    & 5218.197 & 4p 2P [case e]   & 1.5  &6.96& 208.66 &-0.86& -25.78  & 4d 2D [---] & 2.5 & 
0.0* & 0.0* & 0.0* & 0.0* \\

\noalign{\vskip 0.1cm}
\noalign{\hrule\vskip 0.1cm}
\noalign{\vskip 0.1cm}  
\hline                  
\end{tabular}
\end{flushleft}
\end{table*}

\begin{table*}
\begin{flushleft}
\scalefont{0.87}
\caption{Central wavelengths from NIST or Kur\'ucz line lists and total oscillator strengths
from line lists by Kur\'ucz, NIST and VALD, literature, and adopted
values. In column 7, literature oscillator strength values are from the following references: 1 Ram\'{i}rez \& Allende Prieto 2011; 2 Lawler et al. (2014).}             
\label{sclines2}      
\centering 
\small         
\begin{tabular}{l@{}c@{}c@{}c@{}c@{}c@{}c@{}c@{}c@{}ccccc}     % 12 columns 
\noalign{\smallskip}
\hline\hline    
\noalign{\smallskip}
\noalign{\vskip 0.1cm} 
species & {\rm $\lambda$} ({\rm \AA}) & \phantom{-}{\rm $\chi_{ex}$ (eV)} & 
\phantom{-}{\rm gf$_{Kurucz}$} &
 \phantom{-}{\rm gf$_{NIST}$} & \phantom{-}{\rm gf$_{VALD}$} &
 \phantom{-}{\rm gf$_{literature}$} & \phantom{-}{\rm gf$_{adopted}$}&    \\
\noalign{\vskip 0.1cm}
\noalign{\hrule\vskip 0.1cm}
\noalign{\vskip 0.1cm}
%EuII     & 6173.029 & 1.319712 & $-$0.854 &---  & $-$0.860 &$-$0.86$^3$ & $-$0.86 &
%& ---    & --- & --- & ---  \\

$^{45}$ScI & 5671.805/828N & 1.447908     & +0.640      & +0.495    & +0.495   & ---        & +0.495 &  \\
$^{45}$ScI & 5686.826/856 & 1.439588      & +0.530      & +0.376    & +0.376   & ---       & +0.276 &  \\
%$^{45}$ScI & 5717.300/314 & 1.439588      & $-$0.410    & $-$0.530  & $-$0.532 &---       & ---    &  \\
$^{45}$ScI & 6210.676/658 & 0.000000      & $-$1.570    & $-$1.53   & $-$1.529 & ---       & $-$1.53 &  \\
\noalign{\vskip 0.1cm}
\noalign{\hrule\vskip 0.1cm}
\noalign{\vskip 0.1cm}
%$^{45}$ScII & 5239.813/811N & 1.455221    & $-$0.77     & $-$0.77   & $-$0.765 & ---       & $-$0.765 &  \\
%$^{45}$ScII & 5357.199/202 & 1.507058     & $-$2.210    & $-$2.11   & $-$2.111 & $-$2.11$^1$ & $-$2.21 &  \\
$^{45}$ScII & 5526.790/785 & 1.768298     & +0.13       & +0.02     & +0.024   & ---        & $-$0.28 &  \\
$^{45}$ScII & 5552.224/235 & 1.455221     & $-$2.270    & ---       & $-$2.119 & $-$2.28$^1$ & $-$2.27 &  \\
$^{45}$ScII & 5657.896/907 & 1.507058     & $-$0.50     & $-$0.60   & $-$0.603 & ---        & $-$0.60 &  \\
$^{45}$ScII & 5684.202/214 & 1.507508     & $-$1.050    & $-$1.07   & $-$1.074 & $-$1.07$^1$ & $-$1.07 &  \\
$^{45}$ScII & 6245.637/641 & 1.507508     & $-$0.98     & ---       & $-$1.030 & $-$1.04$^1$ & $-$1.18 &  \\
$^{45}$ScII & 6300.698/746 & 1.507508     & $-$1.840    & ---       & $-$1.887 & $-$1.95$^1$ & $-$1.99 &  \\
$^{45}$ScII & 6320.851/843 & 1.500496     & $-$1.770    & ---       & $-$1.819 & $-$1.92$^1$ & $-$1.97 &  \\
$^{45}$ScII & 6604.601/578 & 1.357044     & $-$1.48     & $-$1.31   & $-$1.309 & $-$1.31$^1$ & $-$1.41 &  \\
\noalign{\vskip 0.1cm}
\noalign{\hrule\vskip 0.1cm}
\noalign{\vskip 0.1cm}
%$^{51}$VI & 4831.640  & 0.017033  && -1.38   & -1.380  & &-1.88 & \\
%$^{51}$VI & 4851.480  & 0.00000   && -1.138  & -1.139  & &-1.138 & \\
%$^{51}$VI & 4875.480  & 0.040104  && -0.81   & -0.810  & &-0.810 & \\
%$^{51}$VI & 4932.030  & 1.218096  && -1.17   & -1.170  & &-1.32 voltei a esse valor & \\
%$^{51}$VI & 5627.640  & 1.080616  && -0.363  & -0.363  & &-0.363 & \\
%$^{51}$VI & 5670.850  & 1.080616  && -0.42   & -0.420  & &-0.42 & \\
$^{51}$VI & 5703.560  & 1.050919  &---& -0.211  & -0.211  & -0.21$^2$ &-0.211 & \\
$^{51}$VI & 6081.440  & 1.050919  &---& -0.578  & -0.579  & -0.61$^2$ &-0.578 & \\
$^{51}$VI & 6090.220  & 1.080616  &---& -0.062  & -0.062  & -0.07$^2$ &-0.162 & \\
$^{51}$VI & 6119.520  & 1.063602  &---& -0.320  & -0.320  & -0.36$^2$ &-0.47 & \\
$^{51}$VI & 6199.190  & 0.286572  &---& -1.28   & -1.300  & -1.46$^2$ &-1.48 & \\
%$^{51}$VI & 6216.370  & 0.275259  &---& -1.29   & -1.290  & &-0.95 & \\
$^{51}$VI & 6243.100  & 0.300634  &---& -0.98   & -0.980  & -0.94$^2$ &-0.88 & \\
$^{51}$VI & 6251.820  & 0.286572  &---& -1.34   & -1.340  & -1.37$^2$ &-1.44 & \\
$^{51}$VI & 6274.650  & 0.266964  &---& -1.67   & -1.670  & -1.70$^2$ &-1.72 & \\
%$^{51}$VI & 6285.160  & 0.275259  && -1.51   & -1.510  & -1.54$^2$ &-1.56 & \\
%$^{51}$VI & 4864.740  & 0.017033  && -0.96   & -0.960  & & & \\
%$^{51}$VI & 5727.030  & 1.080616  && -0.011  & -0.012  & & & \\
   
\noalign{\vskip 0.1cm}
\noalign{\hrule\vskip 0.1cm}
\noalign{\vskip 0.1cm}

CuI & 5105.537 & 1.389035 & $-$1.516 & $-$1.50 & $-$1.542  & --- & $-$1.52 \\ 
CuI & 5218.197 & 3.816948 & $+$0.476 & $+$0.26 & $+$0.364  & ---& $+$0.0  \\
 
\noalign{\vskip 0.1cm}
\noalign{\hrule\vskip 0.1cm}
\noalign{\vskip 0.1cm}  
\hline                  
\end{tabular}
\end{flushleft}
\end{table*}

%------------Table hfsMn------------------------------------------------------------
\begin{table*}
\caption{Hyperfine structure for \ion{Cu}{I} lines. }
\label{hfsCu}
\centering
\begin{tabular}{ccccccccccc}
\hline
\noalign{\smallskip}
\cline{1-3} \cline{5-7} \\
\multicolumn{3}{c}{5105.50$\rm \AA$;  $\chi$=1.39 eV} && \multicolumn{3}{c}{5218.20$\rm \AA$; $\chi$=3.82 eV}  & \\
\multicolumn{3}{c}{log gf(total) = $-$1.520} && \multicolumn{3}{c}{log gf(total) = +0.0}  & \\
\noalign{\smallskip}
\cline{1-3} \cline{5-7}  \\
\noalign{\smallskip}
\cline{1-3} \cline{5-7}  \\
$\lambda$ ($\rm \AA$) & log gf & iso && $\lambda$ ($\rm \AA$) & log gf &iso \\
\noalign{\smallskip}
\cline{1-3} \cline{5-7}  \\
 5105.562 & $-$2.8856 & 63 && 5218.195 & $-$1.2041 & 63 & \\
 5105.563 & $-$2.9314 & 63 && 5218.197 & $-$1.2499  &63 & \\ 
 5105.554 & $-$2.5634 & 63 && 5218.197 & $-$0.8819 &63 & \\
 5105.567 & $-$3.8856 & 63 && 5218.201 & $-$2.2041  &63 & \\
 5105.558 & $-$2.8187 & 63 && 5218.201 & $-$1.1372 &63 & \\
 5105.540 & $-$2.3135 & 63 && 5218.203 & $-$0.6320 &63 & \\ 
 5105.562 & $-$4.0617 & 63 && 5218.206 & $-$2.3802 &63 & \\
 5105.544 & $-$2.9156 & 63 && 5218.206 & $-$1.2341 &63 & \\
%\multicolumn{3}{c}{5782.14 $\rm \AA$;  $\chi$=1.64 eV}  \\ 
 5105.516 & $-$2.1075 & 63  && 5218.206 & 0.0 &63 & \\
%\multicolumn{3}{c}{log gf(total) = $-$1.720$^b$} \\ 
 5105.565 & $-$3.3619 & 65  && 5218.194 &$-$1.2041 &65 & \\ %  5782.032 & $-$3.632 & 65 &  \\       
 5105.566 & $-$3.4077 & 65  && 5218.196 &$-$1.2499 &65 & \\ % 5782.042 & $-$3.9349 & 65 & \\ 
 5105.555 & $-$3.0397 & 65  && 5218.196 &$-$0.8819 &65 & \\ % 5782.054 & $-$3.234 & 65 & \\      
 5105.570 & $-$4.3619 & 65  && 5218.201 &$-$2.2041 &65 & \\ %  5782.064 & $-$3.285 & 63 & \\     
 5105.559 & $-$3.2950 & 65  && 5218.201 &$-$1.1372 &65 & \\ %   5782.073 & $-$3.589 & 63 & \\    
 5105.540 & $-$2.7898 & 65  && 5218.201 &$-$0.6320 &65 & \\ %   5782.084 & $-$2.888 & 63 & \\     
 5105.564 & $-$4.5380 & 65  && 5218.206 &$-$2.3802 &65 & \\ % 5782.086 & $-$3.234  & 65  &   \\
 5105.545 & $-$3.3919 & 65  && 5218.206 &$-$1.2341 &65 & \\ % 5782.098 & $-$3.234 & 65 &   \\
 5105.514 & $-$2.5838 & 65  && 5218.206 &$-$0.4260 &65 & \\ % 5782.113 & -2.888 & 63 &   \\
%          &           &     &&  & & & \\ % 5782.153 & -2.787 & 65 &   \\
%          &           &     &&  & & & \\ % 5782.173 & -2.441 & 63 &   \\
\noalign{\smallskip}
\hline
\end{tabular}
\end{table*}
%$-$$-$$-$$-$$-$$-$$-$$-$$-$$-$$-$$-$$-$$-$$-$$-$$-$$-$$-$$-$$-$$-$$-$$-$$-$$-$$-$$-$$-$$-$$-$$-$$-$$-$$-$$-$$-$$-$$-$$-$$-$$-$$-$$-$$-$$-$$-$$-$$-$$-$$-$$-$$-$$-$$-$$-$$-$$-$$-$$-$$-$$-$$-$$-$$-$$-$$-$$-$$-$$-$$-$$-$$-$$-$$-$$-$$-$$-$$-$

\end{appendix}


\begin{thebibliography}{}

%\bibitem[]{} Allen, D.M., Porto de Mello, G.F. 2011, A\&A, 525, A63

\bibitem[Alonso et al.(1999)]{Alonso} 
Alonso, A., Arribas, S., Mart\'{\i}nez-Roger, C. 1999, A\&AS, 140, 261 

%\bibitem[Alonso]{Alonso} 
%Alonso, A., Arribas, S., Mart\'{\i}nez-Roger, C. 2001, A\&A, 376, 1039

\bibitem[]{} Allende Prieto, C., Lambert, D.L., Asplund, M. 2001,
      ApJ, 556, L63

%\bibitem[]{} Allende Prieto, C., ... 2004

\bibitem[Alves-Brito et al.(2005)]{alvesbrito05} 
Alves-Brito, A., Barbuy, B., Ortolani, S., Momany, Y., Hill, V.,
Zoccali, M., Renzini, A., Minniti, D., Pasquini, L., Bica, E. \& Rich, R.M.
2005, A\&A, 435, 657

\bibitem[Alves-Brito et al.(2006)]{alvesbrito06} 
Alves-Brito, A., Barbuy, B., Zoccali, M., Minniti, D., Ortolani, S., Hill, V.,
Renzini, A., Pasquini, L., Bica, E., Rich, R.M. \&  Mel\'endez, J. 
A\&A, 2006, 460, 269.

%\bibitem[Arnett et al.(1971)]{arnett71} 	
%Arnett, W. D., Truran, J. W., Woosley, S. E. 1971, ApJ, 165, 87

%\bibitem[Asplund et al.(2005)]{asplund05} Asplund, M., Grevesse, N., Sauval, A.
%J. 2005, ASPC, 336, 25
\bibitem[Asplund et al.(2009)]{asplund09}  Asplund, M., Grevesse, N., Sauval, A.J., Scott, P. 2009,
ARA\&A, 47, 481

\bibitem[Ballester et al. (2000)]{ballester00}
Ballester, P., Modigliani, A., Boitquin, O., Cristiani, S.,
Hanuschik, R., Kaufer, A., Wolf, S. 2000, in {\it The Messenger}, 101, 31.

\bibitem[Barbuy et al.(1998)]{barbuy98}
Barbuy, B., Bica, E., Ortolani, S. 1998, A\&A, 333, 117

\bibitem[]{} 
Barbuy, B., Perrin, M.-N., Katz, D., Coelho, P., Cayrel, R.,
Spite, M., van't Veer-Menneret, C. 2003, A\&A, 404, 661

\bibitem[]{} Barbuy, B., Zoccali, M., Ortolani, S., et al. 2006, A\&A, 449, 349
%\bibitem[]{} Barbuy, B., Zoccali, M., Ortolani, S., et al.             2007, AJ, 134, 1613 

\bibitem[]{} Barbuy, B.,  Zoccali, M., Ortolani, S., et al. 2009, A\&A, 507, 405
\bibitem[]{} Barbuy, B., Chiappini, C., Cantelli, E. et al. 2014, A\&A, 
      570, A76

\bibitem[Barbuy et al.(2013)]{barbuy13} 
Barbuy, B., Hill, V., Zoccali, M., %Minniti D, Renzini A, 
et~al. 2013, A\&A, 559, A5

\bibitem[Barbuy et al.(2014)]{barbuy14} 
Barbuy, B., Chiappini, C., Cantelli, E., %Depagne E, Pignatari M, 
et~al. 2014, A\&A, 570, A76

\bibitem[Barbuy et al.(2015)]{barbuy15} 
Barbuy, B., Fria\c ca, A., da Silveira, C.R., %Hill V, Zoccali M.,
 et~al. 2015, A\&A, 580, A40

\bibitem[Barbuy et al.(2016)]{barbuy16} 
Barbuy, B., Cantelli, E., Vemado, A., %Ernandes H, Ortolani S,
 et~al. 2016, A\&A, 591, A53

\bibitem[Barbuy et al.(2017)]{barbuy17} 
Barbuy, B., Siqueira-Mello, C., Muniz, L.I., %Ernandes H, Ortolani S,
 et~al. 2017, A\&A, in preparation

\bibitem[Barbuy et al.(2018)]{barbuy18}
Barbuy, B., Chiappini, C., Gerhard, O. 2018, ARA\&A, in press


\bibitem[Battistini \& Bensby(2015)]{battistini15}
Battistini, C., Bensby, T., 2015, A\&A, 577, A9


\bibitem[Bensby et al.(2003)]{bensby03} 
Bensby, T., Feltzing, S., Lundstr\"om, I., 2003, A\&A, 410, 527

%\bibitem[Bensby et al.(2004)]{bensby04} 
%Bensby T, Feltzing S, Lundstr\"om I. 2004. \textit{A\&A} 415:155--170

\bibitem[Bensby et al.(2005)]{bensby05} 
Bensby, T., Feltzing, S., Lundstr\"om, I., Ilyin, I. 2005, 
A\&A, 433, 185

\bibitem[]{} Bensby, T., Yee, J.C., Feltzing, S. et al. 2013, A\&A, 549, A147

\bibitem[Bensby et al.(2017)]{bensby17} 
Bensby, T., Feltzing, S., Gould, A., %Yee JC, Johnson JA,
 et~al. 2017, {\bf A\&A, 605, A89}
%\textit{A\&A} 549:147

\bibitem[Bergemann \& Gehren(2008)]{bergemann08}
Bergemann, M., Gehren, T. 2008, A\&A, 492, 823

\bibitem[]{}
Biehl, H. 1976, University of Kiel

\bibitem[Bielski(1975)]{Bi75}
Bielski, A. 1975, JQSRT, 15, 463

\bibitem[Casey \& Schlaufman(2015)]{casey15}
Casey, A.R., Schlaufman, K.C. 2015, ApJ,  809, 110

\bibitem[]{} Cayrel, R. 1988, in IAU Symposium, Vol. 132, The Impact of Very High
S/N Spectroscopy on Stellar Physics, ed. G. Cayrel de Strobel \& M. Spite, 345

\bibitem[]{} Cayrel, R., Depagne, E., Spite, M.,
Hill, V., Spite, F. et al. 2004, A\&A, 416, 1117

\bibitem[Cescutti et al.(2008)]{cescutti08}  
Cescutti, G., Matteucci, F., Lanfranchi, G. 2008, A\&A, 491, 401

%\bibitem[Brodzinski et al.(1987)]{brod87}
%Brodzinski T., Kronfeldt H.-D., Kropp J.-R., Winkler R., Z. 1987, Phys. D, 7,
%161

%\bibitem[]{}
%Chiappini, C., Matteucci, F., Gratton, R. 1997, ApJ 477, 765C

\bibitem[]{} Coelho, P., Barbuy, B., Mel\'endez, J., Schiavon, R.P.,
 Castilho, B.V.  2005, A\&A, 443, 735

%\bibitem[]{} 
%Carretta, E., Gratton, R. G., Bragaglia, A., Bonifacio, P., Pasquini, L. 2004,
%A\&A, 416, 925

%\bibitem[]{} 	
%Gallino, R., Arlandini, C., Busso, M., Lugaro, M., Travaglio, C., Straniero,
%O., Chieffi, A., Limongi, M. 1998, ApJ, 497, 388

%\bibitem[Gratton]{gratton04} 
%Gratton, R., Sneden, C., Carretta, E. 2004, ARA\&A, 42, 385.

\bibitem[]{}
da Silveira, C.R. 2017, PhD thesis, Universidade de S\~ao Paulo

\bibitem[]{}
Duffau, S., Caffau, E., Sbordone, L. et al. 2017, {\bf A\&A, 605, A128}

\bibitem[]{}
Feltzing, S., Fohlman, M., Bensby, T. 2007, A\&A, 467, 665

\bibitem[]{}
Fishlock, C.K., Yong, D., Karakas, A.I. et al. 2017, MNRAS, 466, 4672

\bibitem[Freeman \& Bland-Hawthorn(2002)]{freemanBH02}
Freeman K, Bland-Hawthorn J. 2002, ARA\&A, 40, 487

\bibitem[Fria\c ca \& Barbuy(2017)]{friaca17}
Fria\c ca A.C.S., Barbuy B. 2017. A\&A, 598, 121

%\bibitem[Fr\"ohlich et~al.(2006)]{frohlich06}
%Fr\"ohlich C, Mart\'{\i}nez-Pinedo G, Liebend\"orfer M,
% Thielemann F-K, Bravo E, Langanke K, Zinner NT.	
% et~al. 2006. \textit{PhRvL} 96:142502


\bibitem[Fulbright et~al.(2007)]{fulbright07}
Fulbright, J.P., McWilliam, A., Rich, R.M. 2007, ApJ, 661, 1152

\bibitem[]{}
Gratton, R.G., Sneden, C. 1990, A\&A, 234, 366

\bibitem[]{}
Gratton, R.G., Carretta, E., Bragaglia, A. 2012, A\&ARv, 20, 50

\bibitem[Grevesse et~al.(1996)]{grevesse96}
Grevesse, N., Noels, A., Sauval, J. 1996,
ASP Series, 99, 117
	
\bibitem[Grevesse \& Sauval(1998)]{grevesse98} 
Grevesse, N., Noels, A. \& Sauval, J.N. 1996, ASPC, 99, 117

\bibitem[Grisoni et al.(2017)]{grisoni17}
Grisoni, V., Spitoni, E., Matteucci, F. et al. 2017, MNRAS, 472, 3637

\bibitem[Guarnieri et al.(1998)]{guarnieri98}
Guarnieri, M.D., Ortolani, S., Montegriffo, P., et al. 1998, A\&A, 331, 70

\bibitem[]{} Gustafsson, B., Edvardsson, B., Eriksson, K. et al.
% J{\o}rgensen, U. G., Nordlund, {\AA}.; Plez, B.
 2008, A\&A, 486, 951

\bibitem[Harris(1996)]{harris96} 
Harris W. 1996, AJ, 112, 1487

%\bibitem[]{} 	
%Heger, A., Woosley, S. E. 2002, ApJ, 567, 532

\bibitem[Hansen et al.(2013)]{hansen13}
Hansen, B.M.S., Kalirai, J.S., Anderson, J. et al. 2013, Nature, 500, 51 

\bibitem[Hinkle et al.(2000)]{hink00}
Hinkle, K., Wallace, L., Valenti, J., Harmer, D. 2000, Visible and Near
Infrared Atlas of the Arcturus Spectrum 3727-9300 A, ed. K. Hinkle, L. Wallace,
J. Valenti, and D. Harmer (San Francisco: ASP) 

\bibitem[]{} 
Houdashelt, M.L., Bell, R.A., Sweigart, A.V. 2000, AJ, 119, 1448

\bibitem[Howes et~al.(2014)]{howes14} 
Howes, L.M., Asplund, M., Casey, A.R., Keller, S.C., Yong, D.,
 et~al. 2014, MNRAS, 445, 4241

\bibitem[Howes et~al.(2015)]{howes15} 
Howes, L.M., Casey, A.R., Asplund, M., Keller, S.C., Yong, D.,
  et~al. 2015, Nature, 527, 484

\bibitem[Howes et~al.(2016)]{howes16} 
Howes, L.M., Asplund, M., Keller, S.C., Casey, A.R., Yong, D.,
 et~al. 2016, MNRAS, 460, 884

\bibitem[]{} Ishigaki, M.N., Aoki, W., Chiba, M. 2013, ApJ, 771, 67

\bibitem[]{}
Iwamoto, K., Brachwitz, F., Nomoto, K., Kishimoto, N., Umeda, H., Hix, W. R.,
Thielemann, F. 1999, ApJS, 125, 439


\bibitem[Johnson et~al.(2014)]{johnson14} 
Johnson, C.I., Rich, R.M., Kobayashi, C., Kunder, A., Koch, A. 2014,
 AJ, 148, 67

\bibitem[Kobayashi et~al.(2006)]{kobayashi06}
Kobayahi, C., Umeda, H.,  Nomoto, K., Tominaga, N.,
Ohkubo, T.  2006, ApJ, 653, 1145

\bibitem[Koch et~al.(2016)]{koch16}
Koch, A., McWilliam, A., Preston, G.W., Thompson, I.B. 2016, A\&A, 587, 124

\bibitem[Kraft(1983)]{kraft83}
Kraft, R.P. 1983, in Highlights of Astronomy, 6, 129

\bibitem[Kruijssen(2015)]{kruijssen15}
Kruijssen, J.M.D. 2015, MNRAS, 454, 1658

%\bibitem[]{}
%Kroupa, P., Tout, C.A., Gilmore, G. 1993, MNRAS, 262, 545K

\bibitem[]{} Kur\' ucz, R. 1993, CD-ROM 23

\bibitem[Lecureur et~al.(2007)]{lecureur07} 
Lecureur, A., Hill, V., Zoccali, M., Barbuy, B., G\'omez, A.,
 et~al. 2007, A\&A, 465, 799


\bibitem[Limongi \& Chieffi(2003)]{limongi03}
Limongi, M., Chieffi, A. 2003. ApJ, 592, 404

\bibitem[Martell et al.(2011)]{martell11}
Martell, S.L., Smolinski, J.P., Beers, T.C., Grebel, E.K. 2011,
A\&A, 534, A136

\bibitem[Martin et al.(2002)]{mart02}Martin, W.C., Fuhr, J.R., Kelleher, D.E.,
et al. 2002, NIST Atomic Database (version 2.0), http://physics.nist.gov/asd.
    National Institute of Standards and Technology, Gaithersburg, MD.

%\bibitem[McWilliam (1997)]{mcwill97}
%McWilliam, A. 1997, ARA\&A, 35, 503

\bibitem[McWilliam]{mcwilliam16}
McWilliam, A. 2016, PASA, 33, 40

\bibitem[McWilliam]{}
McWilliam, A., Rich, R. M., Smecker-Hane, T. A. 2003, ApJ, 592, L21

%\bibitem[McWilliam]{}
%McWilliam, A., Rich, R. M., Smecker-Hane, T. A. 2003b, ApJ, 593L, 145  
	
%\bibitem[McWilliam]{}
%McWilliam, A., Smecker-Hane, T. A. 2005, ApJ, 622L, 29

\bibitem[McWilliam]{}
McWilliam, A., Wallerstein, G., Mottini, M. 2013, ApJ, 778, 149


\bibitem[]{} Mel\'endez, J., Barbuy, B., Bica, E. et al. 2003, A\&A, 411, 417

\bibitem[Nissen]{}
Nissen, P.E., Chen, Y.Q., Schuster, W.J., Zhao, G. 2000, A\&A, 353, 722

\bibitem[Nissen]{}
Nissen, P.E., Schuster, W.J., 2011, A\&A, 530, 15


\bibitem[Nomoto et~al.(2013)]{nomoto13} 
Nomoto, K., Kobayashi, C., Tominaga, N. 2013, ARA\&A, 51, 457


%\bibitem[Ortolani et al.(1995)]{ortolani95} 
%Ortolani, S., Renzini, A., Gilmozzi, R., Marconi, G., Barbuy, B., Bica, E.,
%\& Rich, R. M. 1995, Nature 377, 701

\bibitem[Ortolani et al.(2007)]{ortolani07}
Ortolani, S., Barbuy, B., Bica, E., Zoccali, M., Renzini, A. 2007,
A\&A, 470, 1043

\bibitem[Ortolani et al.(2011)]{ortolani11}
Ortolani, S., Barbuy, B., Momany, Y., Saviane, I., Bica, E. et al.
2011, ApJ, 737, 31

\bibitem[]{}
Pignatari, M., Gallino, R., Heil, M., Wiescher, M., Käppeler, F., Herwig, F., Bisterzo, S. 2010,ApJ, 710, 1557P

\bibitem[Piskunov et al.(1995)]{pisk95}
Piskunov, N., Kupka, F., Ryabchikova, T., Weiss, W., Jeffery, C., 1995, A\&AS,
112, 525

\bibitem[]{} 
Pomp\'eia, L., Hill, V., Spite, M. et al. 2008, A\&A, 480, 379

\bibitem[]{} 
Ram\'{\i}rez, I., Allende-Prieto, C. 2011, ApJ, 743, 135

\bibitem[]{}
Recio-Blanco, A. et al. 2017, IAU Symposium 334, in press

\bibitem[]{}
Reddy, B.E., Tomkin, J., Lambert, D.L., Allende Prieto, C. 2003, MNRAS, 340,
304

\bibitem[]{}
Reddy, B.E., Lambert, D.L., Allende Prieto, C. 2006, MNRAS, 367, 1329

\bibitem[]{}
Renzini, A. 2017, MNRAS, 469, L63

\bibitem[Rossi et al.(2015)]{rossi15}
Rossi, L., Ortolani, S., Barbuy, B., Bica, E., Bonfanti, A.
2015, MNRAS, 450, 3270

%\bibitem[Romano et~al.(2010)]{romano10}
%Romano, D., Karakas, A.I., Tosi, M., Matteucci, F. 2010, A\&A, 522, 32

\bibitem[Schiavon et al.(2017)]{schiavon17}
Schiavon, R.P., Zamora, O., Carrera, R., et al. 2017, MNRAS, 465, 501

\bibitem[Schultheis et~al.(2017)]{schultheis17}
Schultheis, M., Rojas-Arriagada, A., Garc\'{\i}a-P\'erez, A.E., 
J\"onsson, H., Hayden, M.,  et~al. 2017, A\&A, 600, A14
 
\bibitem[Sk\'ulad\'ottir et al.(2017)]{}
Sk\'ulad\'ottir, \'A., Tolstoy, E., Salvadori, S., Hill, V., Pettini, M.
2017, A\&A, in press

\bibitem[]{}
Smith, V.V., Cunha, K., Shetrone, M.D. et al. 2013, ApJ, 765, 16

\bibitem[]{}
Sneden, C., Gratton, R. G., Crocker, D. A. 1991, A\&A, 246, 354S 

\bibitem[]{}	
Sobeck, J.S., Ivans, I. I., Simmerer, J. A., Sneden, C., Hoeflich, P.,
Fulbright, J. P., Kraft, R. P., 2006, AJ, 131, 2949

\bibitem[]{}
Steffen, M., Prakapavicius, D., Caffau, E. et al. 2015, A\&A, 583, A57

%\bibitem[Tsujimoto \& Shigeyama(1998)]{tsujimoto98} 
%Tsujimoto, T., Shigeyama, T., 1998, ApJ, 508, L151
\bibitem[Terndrup(1988)]{terndrup88}
Terndrup, D.M. 1988, AJ, 96, 884	

\bibitem[]{}
Umeda, H., Nomoto, K., 2002, ApJ, 565, 385

\bibitem[Umeda \& Nomoto(2003)]{umedanomoto03} 
Umeda, H., Nomoto, K., 2003, Nature,  422, 871

\bibitem[Umeda \& Nomoto(2005)]{umedanomoto05} 
Umeda, H., Nomoto, K., 2005, ApJ, 619, 427 

%\bibitem[Walther(1962)]{walther62}
%Walther H., 1962, Z. Phys., 170, 507

%\bibitem[]{}
%Wolfe, A. M., Gawiser, E., Prochaska, J. X. 2005, ARA\&A, 43, 861

%\bibitem[Woodgate \& Martin(1957)]{wood57}
%Woodgate G.K., Martin J.S., 1957, Proc. Phys. Soc. Lon. A, 70, 485

\bibitem[Woosley \& Weaver]{ww95} 
Woosley, S. E., Weaver, T. A. 1995, ApJS, 101, 181

\bibitem[Woosley et~al.(2002)]{Woosley2002}
Woosley, S., Heger, A., Weaver, T.A. 2002. Rev. Mod. Phys. 74, 1015

%\bibitem[Yoshida et~al.(2008)]{yoshida08}
%Yoshida T, Umeda H, Nomoto K. 2008. ApJ, 672:1043

\bibitem[Zoccali et al.(2001a)]{zoccali01a}
Zoccali, M., Renzini, A., Ortolani, S., Bragaglia, A., Bohlin, R. et al.
2001a, ApJ, 553, 733

\bibitem[Zoccali et al.(2001b)]{zoccali01b} 
Zoccali, M., Renzini, A., Ortolani, S., Bica, E., Barbuy, B. 2001b, 
AJ, 121, 2638

\bibitem[Zoccali et al.(2004)]{zoccali04} 
Zoccali, M., Barbuy, B., Hill, V., Ortolani, S., Renzini, A., Bica, E., Momany,
Y., Pasquini, L., Minniti, D., \& Rich, R.M. 2004, A\&A, 423, 507

\bibitem[Zoccali et~al.(2006)]{Zoccali2006} 
Zoccali, M., Lecureur, A., Barbuy, B., Hill, V., Renzini, A.,
 et~al. 2006, A\&A, 457, L1

\bibitem[Zoccali et~al.(2008)]{Zoccali2008} 
Zoccali, M., Hill, V., Lecureur, A., Barbuy, B., Renzini, A.,
 et~al. 2008, A\&A, 486, 177



\end{thebibliography}
\end{document}